\documentclass{article}
\pdfoutput=1
\usepackage{authblk}
\usepackage{geometry}
\usepackage[utf8]{inputenc}
\usepackage[english]{babel}
\usepackage[T1]{fontenc}
\usepackage{amsmath}
\usepackage{xcolor}
\colorlet{myPurple}{blue!40!red}
\colorlet{myPurplee}{blue!10!red}
\colorlet{myCyan}{cyan!60!gray}
\colorlet{myRed}{red!66!black}
\usepackage{tikz}
\usepackage{pgfplots}
\pgfplotsset{compat=1.14}
\usepackage{hyperref}
\hypersetup{colorlinks=true,citecolor=myRed,urlcolor=myRed,linkcolor=myRed}
\usepackage{exscale}
\usepackage{bbm}
\usepackage{graphicx}
\usepackage{latexsym}
\usepackage{amsfonts}
\usepackage{amssymb}
\usepackage{amsthm}
\usepackage{enumerate}
\usepackage{bbold}
\usepackage{color}
\usepackage{nicefrac}
\newcommand{\sket}[1]{{\ensuremath{\lvert#1\rangle}}}
\newcommand{\lket}[1]{{\ensuremath{\left\lvert#1\right\rangle}}}
\newcommand{\ket}[1]{\if@display\lket{#1}\else\sket{#1}\fi}
\newcommand{\tp}{\otimes}
\newcommand{\sbra}[1]{{\ensuremath{\langle#1\rvert}}}
\newcommand{\lbra}[1]{{\ensuremath{\left\langle#1\right\rvert}}}
\newcommand{\bra}[1]{\if@display\lbra{#1}\else\sbra{#1}\fi}

\newcommand{\sbraket}[2]{{\ensuremath{\langle#1\rvert#2\rangle}}}
\newcommand{\lbraket}[2]{{\ensuremath{\left\langle#1\!\left\rvert\vphantom{#1}#2\right.\!\right\rangle}}}
\newcommand{\braket}[2]{\if@display\lbraket{#1}{#2}\else\sbraket{#1}{#2}\fi}

\newcommand{\sketbra}[2]{{\ensuremath{\lvert #1\rangle\!\langle #2\rvert}}}
\newcommand{\lketbra}[2]{{\ensuremath{\left\lvert #1\right\rangle\!\!\left\langle #2\right\rvert}}}
\newcommand{\ketbra}[2]{\if@display\lketbra{#1}{#2}\else\sketbra{#1}{#2}\fi}
\usepackage{tcolorbox}
\usepackage{mathtools}

\newcommand{\proj}[1]{\ketbra{#1}{#1}}

\newcommand{\sx}{\sigma_{\textrm{x}}}
\newcommand{\sz}{\sigma_{\textrm{z}}}
\newcommand{\sy}{\sigma_{\textrm{y}}}
\newcommand{\tr}{\mathrm{tr}}

\newcommand{\idd}{\mathds{1}}

\newcommand{\rP}{\text{P}}

\newcommand{\A}{\mathrm{A}}
\newcommand{\B}{\mathrm{B}}
\newcommand{\C}{\mathrm{C}}
\newcommand{\Pp}{\mathrm{P}}

\usepackage{calrsfs}
\DeclareMathAlphabet{\pazocal}{OMS}{zplm}{m}{n}
\usepackage{tikz}
\usepackage{lipsum}
\usepackage{mathrsfs}
\theoremstyle{plain}

\usepackage{graphicx}
\usepackage{bm}
\usepackage{dsfont}
\usepackage{tikz}
\usepackage{amsthm}
\usepackage{array}
\usepackage{amssymb}
\usepackage{amsfonts}
\usepackage{cancel}
\usepackage[toc,page]{appendix}
\usepackage{multirow}
\usepackage{color}
\usepackage{calrsfs}
\usetikzlibrary{backgrounds,decorations.pathreplacing,calc}

\usepackage{tkz-euclide}

\newcommand{\ba}{\begin{eqnarray}}
\newcommand{\be}{\begin{equation}}
\newcommand{\ee}{\end{equation}}

\newcommand{\bn}{\begin{equation*}}
\newcommand{\en}{\end{equation*}}
\newcommand{\ea}{\end{eqnarray}}
\newcommand{\ban}{\begin{eqnarray*}}
\newcommand{\ean}{\end{eqnarray*}}

\newcommand{\junk}{\ket{\xi}}
\newcommand{\junkz}{\ket{\xi_0}}
\newcommand{\junko}{\ket{\xi_1}}

\newcommand{\pur}{\ket{\psi}^{\A\B\C\Pp}}

\newcommand{\ot}{\otimes}

\newcommand{\aj}{\vec{a}^j}

\newcommand{\refstate}{\ket{\Psi}}
\newcommand{\refstatestar}{\ket{{\Psi'}^*}}
\newcommand{\physstate}{\ket{\psi}}

\newcommand{\aA}{\mathbf{a}}
\newcommand{\xX}{\mathbf{x}}
\newtheorem{theorem}{Theorem}
\newtheorem{lemma}[theorem]{Lemma}

\definecolor{main}{HTML}{5989cf}  
\definecolor{sub}{HTML}{cde4ff}   
\definecolor{main1}{HTML}{f2acaf}  
\definecolor{sub1}{HTML}{cc141d}   
\newtheorem{defin}{Definition}
\newcommand{\la}{\lambda}
\usepackage{tcolorbox}
\newtcolorbox{boxI}{colback = sub,colframe = main,boxrule = 0pt,toprule = 6pt}

\begin{document}

\title{All pure multipartite entangled states of qubits can be self-tested up to complex conjugation}

\author[1]{Maria Balanz\'o-Juand\'o}
\author[2]{Andrea Coladangelo}
\author[3]{Remigiusz Augusiak}
\author[1,4]{Antonio Ac\'in}
\author[5]{Ivan \v{S}upi\'c}
\affil[1]{\small ICFO – Institut de Ciències Fotòniques, The Barcelona Institute of Science and Technology, 08860 Castelldefels, Spain}
\affil[2]{\small Paul G. Allen School of Computer Science and Engineering, University of Washington, Seattle, USA}
\affil[3]{\small Center for Theoretical Physics, Polish Academy of Sciences, Aleja Lotnik\'ow 32/46, 02-668 Warsaw, Poland}
\affil[4]{\small ICREA-Institució Catalana de Recerca i Estudis Avançats, Lluís Companys 23, 08010 Barcelona, Spain
}
\affil[5]{\small LIP6, Sorbonne Universit\'e, CNRS, 4 Place Jussieu, 75005 Paris, France}

\date{}

\maketitle

\begin{abstract}

Self-testing refers to the certification of quantum states and measurements based entirely on the correlations exhibited by measurements on separate subsystems. In the bipartite case, self-testing of states has been completely characterized, up to local isometries, as there exist protocols that self-test arbitrary pure states of any local dimension. Much less is known in the multipartite case, where an important difference with respect to the bipartite case appears: there exist multipartite states that are not equivalent, up to local isometries, to their complex conjugate. Thus, any self-testing characterization must in general be complete up to not only local unitaries, but also complex conjugation. Under these premises, in this work, we give a complete characterization of self-testing in the multipartite qubit case.

\end{abstract}

\maketitle

\section{Introduction}

Bell's celebrated theorem~\cite{bell} from 1964 proposed an experimental test to address the question of locality and determinism in quantum mechanics. 
More concretely, Bell's theorem established that, for systems that admit a description in terms of a local hidden variable model, the correlations exhibited by measurements on separate subsystems must obey certain constraints, usually stated in the form of an inequality ---typically referred to as a Bell inequality---. Correlations that violate such an inequality defy an explanation in terms of local hidden variables and are commonly referred to as Bell nonlocal, or just nonlocal. Quantum theory allows for the existence of nonlocal correlations, implying that quantum phenomena cannot be reproduced by local-hidden variables. In fact, Bell inequality violations have been demonstrated convincingly in several experiments~\cite{hensen,loophole2,loophole3}.

Bell nonlocality has a long tradition in the foundations of quantum theory. In the last three decades, the topic has received even more attention, as it has proven to be a crucial ingredient in several information processing tasks. The key insight is that the violation of a Bell inequality witnesses entanglement shared among the devices involved, \emph{without making any assumption} about the inner workings of the devices themselves. Such an approach to certifying properties of quantum systems is referred to as \emph{device-independent}. Its appeal is that it removes trust in the behavior of the quantum devices involved. The approach has been adopted in various cryptographic protocols, such as quantum key distribution~\cite{Barrett2005,Acin2006,Scarani2006,Acin2007}, randomness generation~\cite{colbeck,pironio2010random,colbeckkent}, and verifiable delegated computation protocols~\cite{ruv, coladangelo2019verifier, natarajan2023bounding}.

Remarkably, certain nonlocal correlations suffice to single out a unique state that is compatible with them, up to certain degrees of freedom. Such correlations are said to \emph{self-test} the quantum state. It has been known for a long time that the Clauser--Horne--Shimony--Holt (CHSH) Bell inequality can be maximally violated only by using a maximally entangled pair of qubits, and measuring anticommuting observables~\cite{Tsirelson1987,Summers1987,Popescu1992,BMR}. In fact, the concept of self-testing, which was formally introduced in the works of Mayers and Yao~\cite{Mayers98,Mayers2004}, applies not only to states, but also to quantum measurements and channels~\cite{Sekatski,chen2024all}. These works already pointed out that certain physical transformations, namely local isometries, do not affect the observed measurement correlations. In addition to local isometries, complex conjugation of the state and measurement operators also preserves the measurement correlations. Consequently, the certification of states and measurements provided by a self-testing correlation is modulo all of these transformations. 

The field of self-testing has progressed substantially since its inception (see~\cite{STreview} for a detailed review). In the bipartite case, we know that all pure entangled states can be self-tested~\cite{Coladangelo2017}. However, in multipartite case, only a limited number of results are known. Notable examples include, in the qubit case, self-testing of Dicke states~\cite{Wstate,Ivan,Fadel2017,marginal}, GHZ-states~\cite{Pal} and graph states~\cite{McKague2014,Flavio}, and, for arbitrary dimension, all states that admit a Schmidt decomposition~\cite{Ivan}. Recently, Hardy-type paradoxes have been used to show that some genuine multipartite nonlocal states can be self-tested~\cite{adhikary2024networkassist,adhikary2024selftesting}. Finally, it has been shown that all pure entangled states can be self-tested when incorporated into a quantum network~\cite{Networks_ST_all}. Although this protocol can self-test any pure entangled state, it also necessitates several auxiliary maximally entangled states to facilitate the self-testing process. The question of whether all pure multipartite entangled states can be self-tested in the standard Bell scenario, thus, remains an open challenge. A crucial challenge arises from the fact that, while in the bipartite scenario all pure entangled state are equivalent to their complex conjugates up to a local unitary transformation, in the tripartite scenario there are already states for which this equivalence does not hold~\cite{Acin2001,Kraus2010}. 

In this manuscript, we completely solve the problem of self-testing multipartite entangled states of an arbitrary number of qubits in the standard Bell scenario. We show that a state $\ket{\psi}$ can be self-tested up to an uncertainty in the complex conjugation. More precisely, the best that one can achieve is to self-test the state up to a ``flag'' qubit, i.e.\ self-test the state 
\begin{equation}
 \sqrt{p}\ket{\Psi} \ket{\textbf{0}}+ \sqrt{1-p}\ket{\Psi^*}\ket{\textbf{1}}\,, 
\end{equation}
where $\ket{\mathbf{0}}$ and $\ket{\mathbf{1}}$ are orthogonal flag states (see below for a precise definition).

This manuscript is organized as follows. In Section~\ref{sec:preliminaries}, we introduce the standard Bell scenario, along with several technical lemmas that are essential for proving our main result. Section~\ref{sec:tripartite} outlines our method for self-testing any three-qubit state. In Section~\ref{sec:multipartite}, we demonstrate how the results from the tripartite scenario can be generalized to the case of an arbitrary number of qubits, proving our main result. In Appendix~\ref{app:proof1}, we provide a detailed explanation of how to self-test partially entangled pairs of qubits. Appendix~\ref{appML} contains the proof of the measurement lemma, which is the key technical result that our approach is built on, and Appendix~\ref{app:tripartite} contains a detailed proof of self-testing of any tripartite state. Finally, Appendix~\ref{app:thmMulti} provides the detailed proof of the main theorem from Section~\ref{sec:multipartite}.

\bigskip
\section{Preliminaries and useful lemmas\label{sec:preliminaries}}

Within this section, we establish the notation that will be consistently employed throughout the manuscript. We also provide several technical statements that are pivotal to proving our results.

We consider a Bell scenario with $n$ parties that can run $m$ binary measurements. The measurement input for each party is labeled by $x_i=0,\dots,m-1$, and the obtained outcome by $a_i=0,1$, with $i=1,\dots,n$. We introduce shortened notation $\textbf{a} = (a_1,\dots,a_n)$ and $\textbf{x} = (x_1,\dots,x_n)$. The correlations obtained in this quantum experiment are described by the joint conditional probabilities 
\begin{equation}\label{eq:correlation}
  p(\textbf{a}|\textbf{x})=\tr\left(\rho^{[n]} M_{a_1|x_1}^{(1)}\otimes \cdots \otimes M_{a_n|x_n}^{(n)}\right),
\end{equation}
where $\rho^{[n]}$ as a density operator on the Hilbert space $\bigotimes_{i=1}^n\pazocal{H}^{(i)}$, denotes the state shared by the $n$ parties whereas $M_{a_i|x_i}^{(i)}$ are the positive semi-definite operators defining the local measurements performed by them. Given that in our framework all measurements are binary, having as a result $0$ or $1$, they can be characterized by specifying the measurement observables $A_{x_j}^{(j)} = \sum_{a_j}(-1)^{a_j}M_{a_j|x_j}^{(j)}$. The measurement correlations given in Eq.~\eqref{eq:correlation} can be equivalently phrased in terms of the correlators:
\begin{equation}
  \langle A_{x_1}^{(1)}A_{x_2}^{(2)}\cdots A_{x_n}^{(n)}\rangle = \tr\left(A_{x_1}^{(1)}\otimes A_{x_2}^{(2)} \otimes\cdots\otimes A_{x_n}^{(n)}\rho_n\right).
\end{equation}
These correlators may involve an observable for a subset of the parties, in which case, the operator acting on the remaining parties is the identity, e.g., $\langle A_{x_1}^{(1)}A_{x_2}^{(2)}\rangle = \tr(A_{x_1}^{(1)}\otimes A_{x_2}^{(2)} \otimes\mathds{1}\otimes\cdots\otimes \mathds{1}\rho_n)$. 

A self-testing protocol aims to establish equivalence between two experiments, which are customarily called the physical and the reference experiment. By experiment, we consider a set consisting of the shared state and measurements applied by the different parties. The reference experiment $\{\ket{\Psi}^{[n']},{M}_{a|x}^{(j')}\}$ is the specification or a blueprint with which we want to compare the physical experiment $\{\rho^{[n]},M_{a|x}^{(j)}\}$ producing the observed correlations\footnote{Note that $n'$ does not represent an integer distinct from $n$. In this context, integers in the superscripts of Hilbert spaces indicate the party to which the corresponding share of the state belongs. Primes, double primes, or subscripts accompanying the integer serve to distinguish between different Hilbert spaces associated with the same party.}. The reference state is pure and it is a vector in $n$-qubit Hilbert space $\pazocal{H}^{[n']} = \bigotimes_{i=1'}^{n'}\pazocal{H}^{(i)}$. Contrary the physical state shared among $n$ parties may not be pure, and in general it is a density operator on a Hilbert space of unknown dimension, but with known tensor product structure $\pazocal{H}^{[n]} = \bigotimes_{i=1}^n\pazocal{H}^{(i)}$. For ease of computation, we absorb the purification of the physical state $\physstate^{[n],P}$, ensuring, however, that none of the parties can perform any operation on the purification Hilbert space. As highlighted previously, it is imperative to acknowledge that the device-independent certification of the identical nature of two experiments is unattainable. Instead, the achievable goal is to certify their equivalence, signifying that they are connected through a set of permissible transformations comprising local isometries and complex conjugations. In line with this perspective, we introduce the following definition. 

\begin{defin}[Self-testing quantum states]\label{stDef}
The correlations $p(\aA \vert \xX)$ \emph{self-test} the state $\refstate^{[n']}$ if for all states $\rho^{[n]}$ compatible with $p(\aA \vert \xX)$ via Eq.~\eqref{eq:correlation}, there exists a local unitary $U =\bigotimes_{i=1}^n U^{(i,i')}$ 
such that for any purification $\physstate^{[n],\rP}$ of $\rho^{[n]}$ 
\begin{equation}
(U\otimes\mathds{1}^{\rP})\left(\physstate^{[n],\rP}\bigotimes_{i=1}^n\ket{0}^{(i')}\right)=\sqrt{p}\refstate^{[n']}\otimes\junkz^{[n],\rP} + \sqrt{1-p}\refstatestar^{[n']}\otimes\junko^{[n],\rP} ,\label{st_pure} 
\end{equation}
for some orthogonal states $\junkz$ and $\junko$ and a positive number $0\leq p < 1$. 
\end{defin}
This definition resembles that of self-testing of subspaces put forward in~\cite{subspaces}, however, here the state $\ket{\Psi}$ and its complex conjugation are not orthogonal in general. In fact, the method we introduce here allows one to certify a subspace spanned by the state $\ket{\Psi}$ one aims to certify and its complex conjugate $\ket{\Psi^{*}}$.

Let us emphasize that there are several approaches to mathematically formalize a self-testing statement. Since the native Hilbert space of the physical state always has an unknown dimension, the relationship between the physical and reference states can be described using various constructs, such as a local isometry~\cite{McKague_thesis}, an extraction channel~\cite{Jed1}, or, as in this work, a unitary transformation. In the unitary approach, however, the input to the unitary must include not only the physical state but also an auxiliary state that resides in the same Hilbert space as the reference state. In certain cases, self-testing proofs establish that the Hilbert space of the physical state decomposes as the tensor product of the Hilbert space of the reference state and an additional Hilbert space. When this decomposition is guaranteed, there is no need to include an auxiliary state as input to the self-testing unitary. This is the scenario for the self-testing proofs employed in Lemmas~\ref{lemma:selftesting} and~\ref{lemma:measurementlemma} later in this section.

Finally, we briefly address the assumptions underlying the physical experiment in our self-testing framework. It is important to note that self-testing results, which may hold under specific assumptions about the physical setup, do not necessarily generalize, as highlighted in~\cite{baptista2023mathematical}. First, we do not assume the purity of the state. While we use the purification of the state for mathematical convenience, we emphasize that none of the parties have access to the purification Hilbert space. Second, our analysis avoids assuming that the measurements are projective; instead, all measurements are modeled as positive operator-valued measures (POVMs). A common assumption in many self-testing frameworks is that the physical experiment is support-preserving, meaning that the support of the physical state remains invariant under the measurement operators. In this work, we do not make this assumption outright. Instead, we demonstrate that when the physical experiment reproduces self-testing correlations, it necessarily satisfies the support-preserving condition.

In this manuscript, we formulate self-testing frameworks for multipartite states employing a modular approach that involves various subroutines for self-testing bipartite states. This methodology was initially introduced in~\cite{Wstate} and further refined in~\cite{Ivan}. Here, we rely on a protocol presented in~\cite{erik} for self-testing any partially entangled pair of qubits and all the three Pauli measurements by one qubit. This self-testing scheme is recurrently applied throughout our construction.

To set the stage, we revisit the one-parameter class of Bell inequalities often referred to in the literature as tilted CHSH Bell inequalities~\cite{amp},
\begin{equation}\label{tiltedCHSHin}
I_{\alpha} :=
\alpha\langle A_0^{(j)}\rangle+\langle A_0^{(j)}A_0^{(k)}\rangle+\langle A_0^{(j)}A_1^{(k)}\rangle\ + \langle A_1^{(j)}A_0^{(k)}\rangle -\langle A_1^{(j)}A_1^{(k)}\rangle \leq 2+\alpha,
\end{equation}
where $0\leq\alpha\leq 2$, and $A_i^{(j)}$ and $A_i^{(k)}$ are binary observables applied by the $j$-th and $k$-th parties, respectively. The maximal quantum violation of this inequality is $2\sqrt{2}\sqrt{1+\alpha^2/4}$, and by taking 
$\alpha=2\cos2\theta/\sqrt{1+\sin^22\theta}$,
it is achieved by applying the one-qubit measurements
\begin{align}\label{idealmeasurements}
A_0^{(j)}=\sz,&\qquad A_0^{(k)}=\cos\mu \sz+\sin\mu \sx,
\\ \nonumber A_1^{(j)}=\sx,
&\qquad A_1^{(k)}=\cos\mu \sz-\sin\mu \sx,
\end{align} 
on the two-qubit partially entangled state shared by the $j$-th and $k$-th parties
\begin{equation}\label{partentangled}
\ket{\psi_{\theta}}=\cos\theta\ket{00}+\sin\theta\ket{11},
\end{equation}
where $\theta\in(0,\pi/4]$ is such that $\tan\mu=\sin2\theta$, and $\sx,\sz$ are the standard Pauli matrices. Let us further introduce two more Bell expressions
\begin{align}\label{tilted2}
J_{\alpha} &:=
\alpha\langle A_0^{(j)}\rangle+\langle A_0^{(j)}A_2^{(k)}\rangle+\langle A_0^{(j)}A_3^{(k)}\rangle+\langle A_2^{(j)}A_2^{(k)}\rangle-\langle A_2^{(j)}A_3^{(k)}\rangle,\\
\label{chsh3}
L &:=\langle A_1^{(j)}A_4^{(k)}\rangle+\langle A_1^{(j)}A_5^{(k)}\rangle+\langle A_2^{(j)}A_4^{(k)}\rangle-\langle A_2^{(j)}A_5^{(k)}\rangle .
\end{align}
The first, $J_{\alpha}$, defines another tilted CHSH inequality, characterized by the same parameter $\alpha$ as $I_{\alpha}$, while $L$ is the standard CHSH inequality. When writing Eqs.~\eqref{tiltedCHSHin}, \eqref{tilted2} and \eqref{chsh3}, we slightly abuse the notation anticipating what comes below in Lemma~\ref{lemma:selftesting}: there, we consider a Bell scenario consisting of three measurements for party $j$, $A_0^{(j)},A_1^{(j)},A_2^{(j)}$, and six for party $k$, $A_0^{(k)},\ldots,A_5^{(k)}$, and the value of the previous Bell expressions is computed for the different settings as denoted in these equations. For example, observable $A_0^{(j)}$ is used to compute both $I_{\alpha}$ and $J_{\alpha}$, see Eqs.~\eqref{tiltedCHSHin} and \eqref{tilted2}. The following lemma states the self-testing properties of these three Bell expressions.

\begin{lemma}[Self-testing partially entangled pair of qubits~\cite{erik}]\label{lemma:selftesting}
Let there be a state $\physstate^{(j,k)} \in \pazocal{H}^{(j)}\otimes \pazocal{H}^{(k)}$ and measurement observables $A_x^{(j)}$ and $A_{y}^{(k)}$ with $x=0,1,2$ and $y=0,\ldots,5$, such that
\begin{align*}
I_{\alpha} = J_{\alpha} = 2\sqrt{2}\sqrt{1+\alpha^2/4}\,, \qquad L = 2\sqrt{2}\sin{\theta}.
\end{align*}
Then
\begin{itemize}
\item Hilbert spaces $\pazocal{H}^{(j)}$ and $\pazocal{H}^{(k)}$ have tensor product structure $\pazocal{H}^{j} = \pazocal{H}^{(j_q)}\otimes \pazocal{H}^{(j'')}$ and $\pazocal{H}^{k} = \pazocal{H}^{(k_q)}\otimes \pazocal{H}^{(k'')}$, where $\pazocal{H}^{(j_q)}$ and $\pazocal{H}^{(k_q)}$ are isometric to $\mathbb{C}^2$.
\item There exist unitaries $U^{(j)}$ and $U^{(k)}$ such that 
\begin{align}
\label{eq:xz_meas}
U^{(j)}\otimes U^{(k)}\physstate^{(j,k)} &= \ket{\psi_\theta}^{(j_q,k_q)} \otimes \junk^{(j'',k'')},\\
U^{(j)}\, A_0^{(j)}\, {U^\dagger}^{(j)} = \sz^{(j_q)}\otimes \mathds{1}^{(j'')}, \qquad U^{(j)}\, A_1^{(j)}\, {U^\dagger}^{(j)} &= \sx^{(j_q)}\otimes \mathds{1}^{(j'')}, \qquad 
\label{eq:y_meas}
U^{(j)}\, A_2^{(j)}\, {U^\dagger}^{(j)} = \sy^{(j_q)}\otimes A_\mathrm{Y}^{(j'')},
\end{align}
where $\theta$ and $\mu$ are as defined above, $A_\mathrm{Y}$ is a Hermitian $\pm 1$-eigenvalue observable and $\junk$ is a normalized quantum state. 
\end{itemize}
\end{lemma}

The proof of this lemma can be found in Appendix~\ref{app:proof1}. The lemma provides us with correlations that self-test a partially entangled pair of qubits and a tomographically complete set of qubit measurements performed by party $(j)$. Although not explicitly stated in the lemma, relations~\eqref{eq:xz_meas} and \eqref{eq:y_meas} imply that the unitary $U^{(k)}$ maps the measurements of party $(k)$ to a tomographically complete set of qubit measurements, as proven in Appendix~\ref{app:proof1}. 

The second important technical lemma, which we will repeatedly use in what follows, again concerns two arbitrary parties, $j$ and $k$, and gives a way to self-test an arbitrary measurement by party $k$ if the parties already self-tested a partially entangled pair of qubits through Lemma~\ref{lemma:selftesting} and the corresponding measurements by party $j$, see Eqs.~\eqref{eq:xz_meas}-\eqref{eq:y_meas}. This lemma is an instance of the so-called ``post-hoc'' measurement self-testing, introduced in~\cite{STreview} and formalized in~\cite{chen2024all}.

\begin{lemma}[Measurement lemma]\label{lemma:measurementlemma}
Let $\physstate$ be a bipartite state and $A_0^{(j)}, A_1^{(j)}, A_2^{(j)}$ dichotomic measurement observables. Suppose there exist local unitaries $U^{(j)}$ and $U^{(k)}$ such that:
\begin{align}\label{state}
U^{(j)}\otimes U^{(k)}\physstate^{(j,k)} &= \ket{\psi_{\theta}}^{(j_q,k_q)} \otimes \junk^{(j'',k'')}, \\ 
U^{(j)}A_0^{(j)}{U^\dagger}^{(j)} = \sz^{(j_q)} \otimes \mathds{1}^{(j'')}, \qquad
U^{(j)}A_1^{(j)}{U^\dagger}^{(j)} &= \sx^{(j_q)} \otimes \mathds{1}^{(j'')},\qquad
U^{(j)}A_2^{(j)}{U^\dagger}^{(j)} = \sy^{(j_q)} \otimes A_\mathrm{Y}^{(j'')}, \label{eq: A2}
\end{align}
where $A_\mathrm{Y}$ is a Hermitian $\pm 1$-eigenvalue operator. Suppose moreover that
\begin{align}
\label{02}
\bra{\psi}A_0^{(j)}\otimes A^{(k)}\ket{\psi} = \alpha_z,&\qquad
\bra{\psi}A_1^{(j)}\otimes A^{(k)}\ket{\psi} = \alpha_x \sin{2\theta}, \\
\bra{\psi}A_2^{(j)}\otimes A^{(k)}\ket{\psi} = \alpha_y \sin{2\theta}, &\qquad 
\bra{\psi}\mathds{1}^{(j)}\otimes A^{(k)}\ket{\psi} = \alpha_z\cos{2\theta}, \label{22}
\end{align}
for some real numbers $\alpha_z$, $\alpha_x$ and $\alpha_y$ such that $\alpha_z^2 + \alpha_x^2 + \alpha_y^2 = 1$. Then, we have
\begin{align*}
  U^{(k)}\, A^{(k)}{U^\dagger}^{(k)} &= \alpha_z\, \sz^{(k_q)}\otimes\mathds{1}^{(k'')} + \alpha_x\,\sx^{(k_q)}\otimes\mathds{1}^{(k'')} + \alpha_y\, \sy^{(k_q)}\otimes \bar{B}_\mathrm{Y}^{(k'')}\\
  &= \left(\alpha_z\, \sz+ \alpha_x\,\sx + \alpha_y\, \sy\right)^{(k_q)}\otimes \left(\frac{\idd+\bar{B}_\mathrm{Y}}{2}\right)^{(k'')} + \left(\alpha_z\, \sz^*+ \alpha_x\,\sx^* + \alpha_y\, \sy^*\right)^{(k_q)}\otimes \left(\frac{\idd-\bar{B}_\mathrm{Y}}{2}\right)^{(k'')}
\end{align*}
where $\bar{B}_\mathrm{Y}$ is a Hermitian $\pm 1$-eigenvalue operator.
\end{lemma}

The proof of this lemma can be found in Appendix \ref{appML}. 

\bigskip~\section{Self-testing tripartite states\label{sec:tripartite}}
Let us now consider the tripartite scenario, where the objective is to self-test the shared state between Alice, Bob, and Charlie. This shared state is of the form
\begin{equation}\label{eq:tripartitestate}
  |\Psi\rangle^{\A'\B'\C'}=\sum_{i,j,k=0}^1 \lambda_{ijk}\ket{i}^{\A'}\otimes\ket{j}^{\B'}\otimes\ket{k}^{\C'}.
\end{equation}
A pure state is said to be genuinely multipartite entangled (GME) whenever it does not have a tensor-product form across any bipartition. Conversely, if the state is not GME, it can be decomposed into a tensor product of two states. In the tripartite case, the two states consist of one and two parties and, therefore, self-testing immediately follows from the self-testing of arbitrary bipartite qubit states, first shown in~\cite{Bamps}. Therefore, in what follows we restrict our considerations to GME tripartite states.

Before presenting the self-testing recipe for tripartite states, we first define the parameters that characterize the state and the measurements to be self-tested. Specifically, we introduce the notation for the states resulting from Alice and Bob measuring their qubits in the computational basis:
\begin{align}\label{psiplus}
  \ket{\psi_{a+}}^{\B'\C'} \propto \sum_{j,k}\lambda_{0jk}\ket{jk}^{\B'\C'}, \qquad 
  \ket{\psi_{a-}}^{\B'\C'} \propto \sum_{j,k}\lambda_{1jk}\ket{jk}^{\B'\C'},
\end{align}
where $\ket{\psi_{a+}}$ and $\ket{\psi_{a-}}$ are the post-measurement states shared by Bob and Charlie corresponding to 
the outcome $0$ and $1$ of Alice's measurement respectively. Analogous notation is used for the post-measurement state shared by Alice and Charlie followed by Alice obtaining $0$ when measuring in the computational basis:
\begin{equation}\label{psiminus}
  \ket{\psi_{b+}}^{\A'\C'} \propto \sum_{j,k}\lambda_{j0k}\ket{jk}^{\A'\C'}
\end{equation}

Since self-testing is agnostic to local unitary rotations, we have the flexibility to select the most convenient definition of computational bases. In this context, we opt for bases such that when Alice and Bob perform measurements in the basis $\{\ket{0},\ket{1}\}$, the resulting projection of the state onto the two remaining parties is entangled. It is noteworthy that such bases, ensuring entanglement in the projected state, exist for any GME state, as established in~\cite{zwerger2019device}. Furthermore, as it will become evident during the derivations, we select a basis in which the complex phases of $\lambda_{000}$ and $\lambda_{001}$ differ, as do those of $\lambda_{100}$ and $\lambda_{101}$. This can be achieved by applying a unitary operation of the form $\proj{0} + e^{i\varphi}\proj{1}$ to Charlie's qubit, with a convenient choice of $\varphi$. Additionally, to ensure that none of the coefficients $\lambda_{000}, \lambda_{001}, \lambda_{100}$ and $\lambda_{101}$ vanish, we can, for instance, apply a Hadamard gate to Charlie's qubit. Hence, for $t \in \{a,b\}$ there exist parameters $\phi_{t^\pm} \in (0,\frac{\pi}{4}]$ and Schmidt bases $\{\ket{0_{t^\pm}},\ket{1_{t^\pm}}\}$ such that 
\begin{align} \label{psia+}
   V_{a^\pm}^{\B'}\otimes W_{a^\pm}^{\C'}\ket{\psi_{a\pm}}^{\B'\C'} &= \cos\phi_{a^\pm}\ket{0_{a^\pm}0_{a^\pm}}^{\B'\C'} + \sin\phi_{a^\pm}\ket{1_{a^\pm}1_{a^\pm}}^{\B'\C'}  \equiv \ket{\Psi_{a\pm}}^{\B'\C'},\\ \label{psib+}
  T_{b^+}^{\B'}\otimes W_{b^+}^{\C'}\ket{\psi_{b+}}^{\A'\C'} &= \cos\phi_{b^+}\ket{0_{b^+}0_{b^+}}^{\A'\C'} + \sin\phi_{b^+}\ket{1_{b^+}1_{b^+}}^{\A'\C'} \equiv \ket{\Psi_{b+}}^{\A'\C'}.
\end{align}
Pauli matrices in these newly defined bases are denoted in the following way: $\sigma_{\textrm{z}t^\pm} = \sum_{l}(-1)^l\proj{l_{t^\pm}}$, $\sigma_{\textrm{x}t^\pm} = \proj{+_{t^\pm}} - \proj{-_{t^\pm}}$, where $\ket{\pm_{t^\pm}} = (\ket{0_{t^\pm}}\pm \ket{1_{t^\pm}})/\sqrt{2}$, and $\sigma_{\textrm{y}t^\pm} = \sqrt{-1}\sigma_{\textrm{z}t^\pm}\sigma_{\textrm{x}t^\pm}$. Pauli matrices in the computational basis are denoted simply as $\boldsymbol{\sigma} = (\sz,\sx,\sy)$.

With all the parameters defined, we can state our self-testing result.
\begin{theorem}[Self-testing of tripartite states]
  Any GME three-qubit state can be self-tested.\label{thmTri}
\end{theorem}

The detailed proof can be found in Appendix \ref{app:tripartite}. The key idea behind this self-testing method is to use three sub-tests, each designed to self-test one of the three states described in~\eqref{psia+} and~\eqref{psib+}, as outlined in Lemma~\ref{lemma:selftesting}. In general, the Schmidt bases self-tested in different sub-tests vary and are only determined up to unknown rotations. To address this, each sub-test employs Lemma~\ref{lemma:measurementlemma} to self-test two additional qubit observables: one used to project the other two parties in different sub-tests, corresponding to $\sz$, and another corresponding to $\sx$. For each party, the measurements corresponding to $\sz$ and $\sx$ are denoted by $\lozenge$ and $\blacklozenge$, respectively.

\begin{figure}
 \centerline{
  \begin{tikzpicture}[thick]
  %
  \tikzstyle{operator} = [draw,fill=white,minimum size=1.5em] 
  \tikzstyle{phase} = [fill,shape=circle,minimum size=5pt,inner sep=0pt]
\node at (-2,-4) (ST) {$\ket{\psi^{\A\B\C\Pp}}$};
  \node at (0,0) (q1) {$\ket{+}^{\mathrm{A}'}$};
  \node at (0,-1) (q2) {};
  \draw[-] (ST) -- (q2);
  \node at (0,-3) (qB') {$\ket{+}^{\mathrm{B}'}$};
  \node at (0,-4) (qB) {};
  \draw[-] (ST) -- (qB);
  \node at (0,-6) (qC') {$\ket{+}^{\mathrm{C}'}$};
  \node at (0,-7) (qC) {};
\draw[-] (ST) -- (qC);
  \node[phase] (phase11) at (1,0) {} edge [-] (q1);
  \node[operator] (phase12) at (1,-1) {$A_\lozenge$} edge [-] (q2);
  \draw[-] (phase11) -- (phase12);
  \node[phase] (phase1B) at (1,-3) {} edge [-] (qB');
  \node[operator] (phase1B2) at (1,-4) {$B_\lozenge$} edge [-] (qB);
  \draw[-] (phase1B) -- (phase1B2);
  \node[phase] (phase1C) at (1,-6) {} edge [-] (qC');
  \node[operator] (phase1C2) at (1,-7) {$C_\lozenge$} edge [-] (qC);
  \draw[-] (phase1C) -- (phase1C2);
  %
  \node[operator] (phase111) at (2,0) {H} edge [-] (phase11);
\node[operator] (phase111B) at (2,-3) {H} edge [-] (phase1B);
  \node[operator] (phase111C) at (2,-6) {H} edge [-] (phase1C);
  \node[phase] (phase21) at (3,0) {} edge [-] (phase111);
  \node[operator] (phase22) at (3,-1) {$A_\blacklozenge$} edge [-] (phase12);
  \draw[-] (phase21) -- (phase22); 
\node[phase] (phase21B) at (3,-3) {} edge [-] (phase111B);
  \node[operator] (phase22B) at (3,-4) {$B_\blacklozenge$} edge [-] (phase1B2);
  \draw[-] (phase21B) -- (phase22B);
\node[phase] (phase21C) at (3,-6) {} edge [-] (phase111C);
  \node[operator] (phase22C) at (3,-7) {$C_\blacklozenge$} edge [-] (phase1C2);
  \draw[-] (phase21C) -- (phase22C);
  \node (end1) at (4,0) {} edge [-] (phase21);
  \node (end2) at (4,-1) {} edge [-] (phase22);
  \node (end1B) at (4,-3) {} edge [-] (phase21B);
  \node (end2B) at (4,-4) {} edge [-] (phase22B);
  \node (end1C) at (4,-6) {} edge [-] (phase21C);
  \node (end2C) at (4,-7) {} edge [-] (phase22C);
  \end{tikzpicture}
 }
 \caption{
SWAP isometry for self-testing the tripartite states. It takes as input the physical state $\physstate$ and three auxiliary qubits initialized in the state $\ket{+} = (\ket{0}+\ket{1})/\sqrt{2}$. $H$ denotes the Hadamard gate. 
 }
 \label{isoTel}
\end{figure}
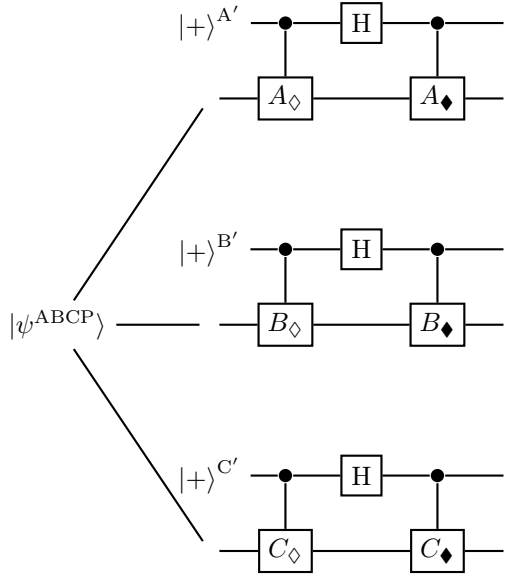

This approach enables us to self-test the forms of the three states given in Eqs.~\eqref{psiplus} and~\eqref{psiminus}, though only up to complex conjugation. The final self-testing result is achieved through the construction of a SWAP isometry, illustrated in Fig.~\ref{isoTel}. This isometry operates on the state $\pur$ and three auxiliary qubits. If $\pur$ represents a qubit state, and if the $\lozenge$
 observables correspond to the local $\sz$
operator, while the $\blacklozenge$ observables correspond to the local $\sx$ operator, then the SWAP isometry maps 
$\pur$ to the Hilbert space associated with the auxiliary qubits. Crucially, since the $\lozenge$ and $\blacklozenge$ observables for each party anti-commute, the isometry extracts $\pur$ into the auxiliary Hilbert spaces in a form expressed in the eigenbasis of the $\lozenge$ observables.

To make things more concrete let us analyze the conclusions obtained from the three sub-tests. As explained above, the first two self-test states~\eqref{psia+}, after Alice measures the physical observable $A_\lozenge$. Such self-tests as stipulated by Lemma~\ref{lemma:selftesting} ensure the existence of unitaries $U^\B_{a_\pm}$ and $U^\C_{a_\pm}$ such that:
\begin{align}\nonumber
  \left(U^\B_{a_\pm}\otimes U^\C_{a_\pm}\right)&\tr_\A\left[\frac{\idd \pm A_\lozenge}{2}\proj{\psi}^{\A\B\C\Pp}\right] \left({U^\B_{a_\pm}}\otimes {U^\C_{a_\pm}}\right)^\dagger \propto \\ &\propto\ketbra{{{\psi}}_{a\pm}}{{{\psi}}_{a\pm}}^{\B_q\C_q}\ot\frac{\idd+{B}_{Ya^\pm}}{2}\xi_{a^\pm}^{\B''\C''P} + \ketbra{{\psi^*_{a\pm}}}{\psi^*_{a\pm}}^{\B_q\C_q}\ot \frac{\idd-{B}_{Ya^\pm}}{2}\xi_{a^\pm}^{\B''\C''\Pp} \nonumber \\ &= \ketbra{{\psi}_{a\pm}}{{\psi}_{a\pm}}^{\B_q\C_q}\ot\frac{\idd+{C}_{Ya^\pm}}{2}\xi_{a^\pm}^{\B''\C''\Pp} + \proj{{\psi^*_{a\pm}}}^{\B_q\C_q}\ot\frac{\idd-{C}_{Ya^\pm}}{2}\xi_{a^\pm}^{\B''\C''\Pp},\label{stateafterAlice}
\end{align}
where operators ${B}_{Ya^\pm}$ and ${C}_{Ya^
\pm}$ result from applying Lemma~\ref{lemma:measurementlemma} to self-test the measurements $B_
\lozenge$ and $C_\lozenge$:
\begin{align}\label{b0apm}
   U^\B_{a_\pm}\,B_\lozenge\,{U^\B_{a_\pm}}^{\dagger} &=  V_{a^\pm}^\dagger{\sigma}_{\textrm{z}{a^\pm}}V_{a^\pm} \otimes \frac{\idd+{B}_{Ya^\pm}}{2} + {V_{a^\pm}^*}^\dagger{\sigma}_{\textrm{z}{a^\pm}}V_{a^\pm}^* \otimes \frac{\idd-{B}_{Ya^\pm}}{2},\\
   U^\C_{a_\pm}\,C_\lozenge\,{U^\C_{a_\pm}}^{\dagger} &= W_{a^\pm}^\dagger{\sigma}_{\textrm{z}{a^\pm}}W_{a^\pm} \otimes \frac{\idd+{C}_{Ya^\pm}}{2} + {W_{a^\pm}^*}^\dagger{\sigma}_{\textrm{z}{a^\pm}}W_{a^\pm}^* \otimes \frac{\idd-{C}_{Ya^\pm}}{2},\label{c0apm}\end{align}
   as well as $B_\blacklozenge$ and $C_\blacklozenge$
   \begin{align}
   \label{b1apm}
U^\B_{a_\pm}\,B_\blacklozenge\,{U^\B_{a_\pm}}^{\dagger} &=  V_{a^\pm}^\dagger{\sigma}_{\textrm{x}{a^\pm}}V_{a^\pm} \otimes \frac{\idd+{B}_{Ya^\pm}}{2} + {V_{a^\pm}^*}^\dagger{\sigma}_{\textrm{x}{a^\pm}}V_{a^\pm}^* \otimes \frac{\idd-{B}_{Ya^\pm}}{2},\\   U^\C_{a_\pm}\,C_\blacklozenge\,{U^\C_{a_\pm}}^{\dagger} &= W_{a^\pm}^\dagger{\sigma}_{\textrm{x}{a^\pm}}W_{a^\pm} \otimes \frac{\idd+{C}_{Ya^\pm}}{2} + {W_{a^\pm}^*}^\dagger{\sigma}_{\textrm{x}{a^\pm}}W_{a^\pm}^* \otimes \frac{\idd-{C}_{Ya^\pm}}{2},\label{c1apm}
\end{align}
where unitaries $V_{a^\pm}$ and $W_{a^\pm}$ are those given in Eq.~\eqref{psia+}.

The described self-testing statements ensure that the state of Bob and Charlie, after Alice measures observable $A_\lozenge$, is equivalent, up to a local isometry, to the reference state Alice would prepare for them by measuring in the computational basis. This, however, cannot establish the correct entanglement structure between Alice on one side and Bob and Charlie on the other. Indeed, a state that is separable across the bipartition $A|BC$ would also pass the test. 

As discussed above, in the third sub-test we use Lemma~\ref{lemma:selftesting} to self-test the projected state of Alice and Charlie following Bob's measurement of observable $B_\lozenge$. The test checks only the cases when Bob obtains the outcome $0$ and if successful ensures the existence of unitaries $U^\A_{b_+}$ and $U^\C_{b_+}$ such that:
\begin{align}
\left(U^\A_{b_+}\otimes U^\C_{b_+}\right)&\tr_\B\left[\frac{\idd + B_\lozenge}{2}\proj{\psi}^{\A\B\C\Pp}\right]\left({U^\A_{b_+}}\otimes {U^\C_{b_+}}\right)^\dagger \propto \nonumber \\ &\propto \ketbra{{{\psi}}_{b+}}{{{\psi}}_{b+}}^{\A_q\C_q}\ot\frac{\idd+{A}_{Yb^+}}{2}\xi_{b^+}^{\A''\C''\Pp} + \ketbra{{\psi^*_{b+}}}{{\psi^*_+}}^{\A_q\C_q}\ot \frac{\idd-{A}_{Yb^+}}{2} \xi_{b^+}^{\A''\C''\Pp} \nonumber \\ &= \ketbra{{\psi}_{b+}}{{\psi}_{b+}}^{\A_q\C_q}\ot\frac{\idd+{C}_{Yb^+}}{2}\xi_{b^+}^{\A''\C''\Pp} + \proj{{\psi_{b+}^*}}^{\A_q\C_q}\ot\frac{\idd-{C}_{Yb^+}}{2}\xi_{b^+}^{\A''\C''\Pp},\label{stateafterBob}
\end{align}
with again operators ${A}_{Yb^+}$ and ${C}_{Yb^+}$ stemming from using Lemma~\ref{lemma:measurementlemma} to self-test operators $A_\lozenge$, and $C_\lozenge$:
\begin{align}
  U^\A_{b_+}\,A_\lozenge\,{U^\A_{b_+}}^{\dagger} &=  T_{b^+}^\dagger{\sigma}_{\textrm{z}{b^+}}T_{b^+} \otimes \frac{\idd+{A}_{Yb^+}}{2} + {T_{b^+}^*}^\dagger{\sigma}_{\textrm{z}{b^+}}T_{b^+}^* \otimes \frac{\idd-{A}_{Yb^+}}{2},\\ 
  U^\C_{b_+}\,C_\lozenge\,{U^\C_{b_+}}^{\dagger} &= W_{b^+}^\dagger{\sigma}_{\textrm{z}{b^+}}W_{b^+} \otimes \frac{\idd+{C}_{Yb^+}}{2} + {W_{b^+}^*}^\dagger{\sigma}_{\textrm{z}{b^+}}W_{b^+}^* \otimes \frac{\idd-{C}_{Yb^+}}{2}, \label{c0bpm} \end{align}
and $A_\blacklozenge$, $C_\blacklozenge$:
  \begin{align}
  U^\A_{b_+}\,A_\blacklozenge\,{U^\A_{b_+}}^{\dagger} &=  T_{b^+}^\dagger{\sigma}_{\textrm{x}{b^+}}T_{b^+} \otimes \frac{\idd+{A}_{Yb^+}}{2} + {T_{b^+}^*}^\dagger{\sigma}_{\textrm{x}{b^+}}T_{b^+}^* \otimes \frac{\idd-{A}_{Yb^+}}{2},\\
  \label{c1bpm}
  U^\C_{b_+}\,C_\blacklozenge\,{U^\C_{b_+}}^{\dagger} &= W_{b^+}^\dagger{\sigma}_{\textrm{x}{b^+}}W_{b^+} \otimes \frac{\idd+{C}_{Yb^+}}{2} + {W_{b^+}^*}^\dagger{\sigma}_{\textrm{x}{b^+}}W_{b^+}^* \otimes \frac{\idd-{C}_{Yb^+}}{2},
\end{align}
where the unitaries $T_{b^\pm}$ and $W_{b^\pm}$ are those given in Eq.~\eqref{psib+}.
The results from the third sub-test show that the state of Alice and Charlie conditioned on Bob measuring $B_\lozenge$ and obtaining result $0$ is entangled and, up to complex conjugation has the same form as in the reference experiment. 

We employ the SWAP isometry to combine the three sub-tests and demonstrate their sufficiency for proving the main theorem, with the technical details provided in Appendix~\ref{sec:SWAP}. The output of the SWAP isometry is 
\begin{equation}
  \Phi\left(\ket{\psi}^{\A\B\C\Pp}\otimes\ket{+++}^{\A'\B'\C'}\right) = \sum_{ijk \in \{0,1\}}\ket{\xi_{ijk}}^{\A\B\C\Pp}
\otimes\ket{ijk}^{\A'\B'\C'},
\end{equation}
where \begin{equation}\ket{\xi_{ijk}} = A_\blacklozenge^i\frac{\idd + (-1)^iA_\lozenge}{2}\otimes B_\blacklozenge^j\frac{\idd + (-1)^jB_\lozenge}{2}\otimes C_\blacklozenge^k\frac{\idd + (-1)^kC_\lozenge}{2}\pur.\end{equation}
The results of the first two sub-tests allow us to simplify states $\ket{\xi_{ijk}}$. The states $\ket{\xi_{0jk}}$ correspond those examined in the first sub-test because they have the form $(\idd + A_\lozenge)/2\otimes f_{jk}(B_\lozenge,B_\blacklozenge,C_\lozenge,C_\blacklozenge)\pur$, where $f_{jk}(\cdots)$ is a linear functional. Similarly, the states $\ket{\xi_{1jk}}$ correspond to the second sub-test, taking the form $A_\blacklozenge (\idd -A_\lozenge)/2\otimes f_{jk}(B_\lozenge,B_\blacklozenge,C_\lozenge,C_\blacklozenge)\pur$. Using the self-testing results from Eqs.~\eqref{stateafterAlice}, \eqref{b0apm}, \eqref{c0apm}, \eqref{b1apm}, and \eqref{c1apm}, we find $\ket{\xi_{0jk}} = \lambda_{0jk}\ket{\xi_0} + \lambda_{0jk}^*\ket{\xi_1}$, where $\braket{\xi_0}{\xi_1} = 0$, and $\ket{\xi_{1jk}} = \lambda_{1jk}\ket{\xi'_0} + \lambda_{1jk}^*\ket{\xi'_1}$, where $\braket{\xi'_0}{\xi'_1} = 0$. Substituting these into the output of the SWAP isometry, we obtain
\begin{multline}
  \Phi\left(\ket{\psi}^{\A\B\C\Pp}\otimes\ket{+++}^{\A'\B'\C'}\right) = \ket{\xi_{0}}^{\A\B\C\Pp}\otimes \sum_{jk}\lambda_{0jk}\ket{0jk}^{\A'\B'\C'} + \ket{\xi_1}^{\A\B\C\Pp}\otimes \sum_{jk}\lambda_{0jk}^*\ket{0jk}^{\A'\B'\C'} + \\ + \ket{\xi'_0}^{\A\B\C\Pp}\otimes \sum_{jk}\lambda_{1jk}\ket{1jk}^{\A'\B'\C'} + \ket{\xi'_1}^{\A\B\C\Pp}\otimes \sum_{jk}\lambda_{1jk}^*\ket{1jk}^{\A'\B'\C'}.
\end{multline}
If $\ket{\xi_0} = \ket{\xi'_0}$ and $\ket{\xi_1} = \ket{\xi'_1}$, this equation proves the main theorem. The results of the third sub-test verify this by analyzing the states $\ket{\xi_{i0k}}$. Using Eqs.~\eqref{stateafterBob}–\eqref{c1bpm}, we find $\ket{\xi_{i0k}} = \lambda_{i0k}\ket{\xi''_0} + \lambda^*_{i0k}\ket{\xi''_1}$, where $\braket{\xi''_0}{\xi''_1} = 0$. By comparing two distinct decompositions of $\ket{\xi_{00k}}$ and $\ket{\xi_{10k}}$ we conclude that $\ket{\xi_0} = \ket{\xi'_0} = \ket{\xi''_0}$ and $\ket{\xi_1} = \ket{\xi'_1} = \ket{\xi''_1}$. This establishes the final output of the SWAP isometry as
\begin{equation}
  \Phi\left(\pur\otimes\ket{+++}^{\A'\B'\C'}\right) = \ket{\xi_{0}}^{\A\B\C\Pp}\otimes \ket{\Psi}^{\A'\B'\C'} + \ket{\xi_1}^{\A\B\C\Pp}\otimes \ket{\Psi^*}^{\A'\B'\C'},
\end{equation}
completing the proof of the main theorem.

\bigskip
\section{Recipe for self-testing multipartite states\label{sec:multipartite}}

\begin{figure}[htbp]
  \centering
  \includegraphics[width=\textwidth]{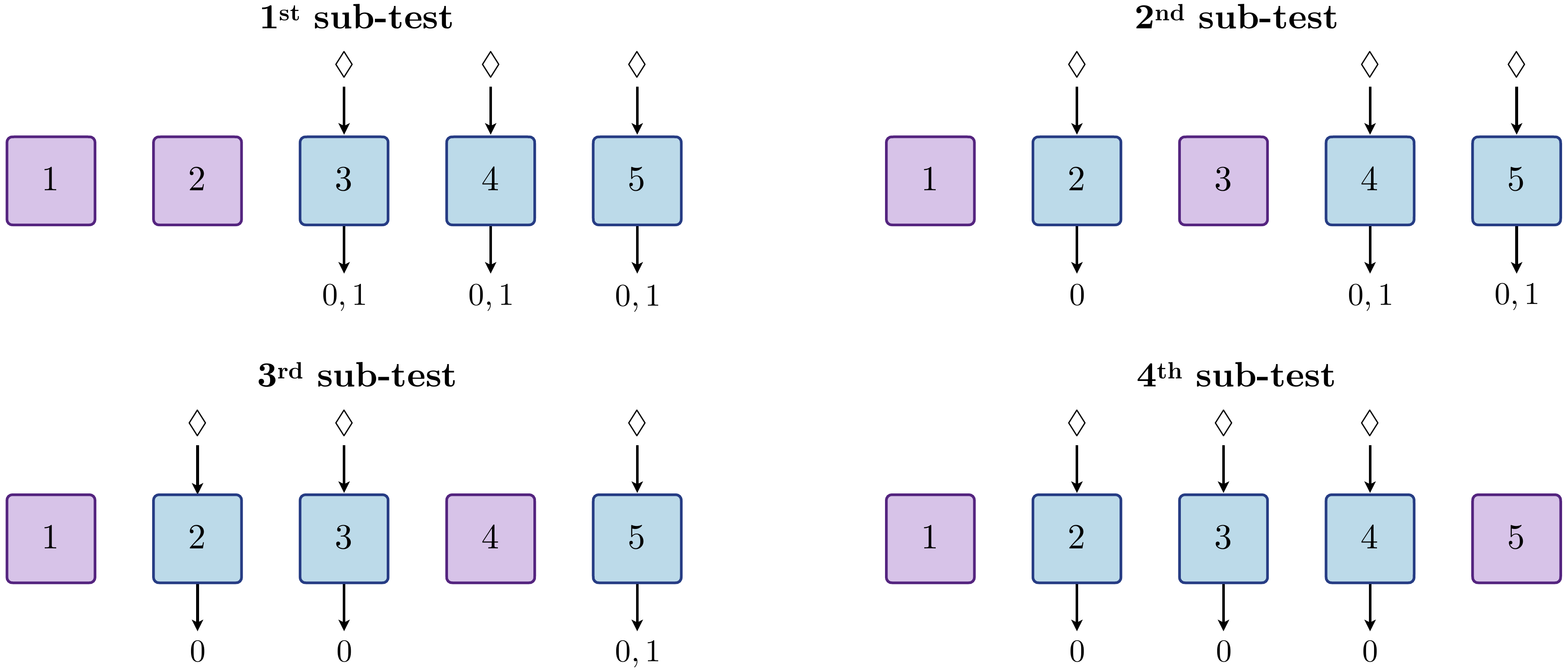}
  \caption{A schematic representation of the self-testing scenario for five-partite states. In each sub-test, three projecting parties (depicted with blue boxes) receive the input $\lozenge$, and the resulting projected states of the two tested parties (shown with purple boxes) are then self-tested. The correlations corresponding to the output $1$ of a projecting party contribute to the self-testing of the tested parties’ states only in sub-tests that occur before the projecting party itself assumes the role of a tested party. In this case, the first sub-test self-tests $2^3$
 distinct states, corresponding to $2^3$
 different global outputs from the projecting parties. In the second sub-test, this number reduces to $2^2$, then to $2$ in the third sub-test, and finally, the last sub-test self-tests a single state corresponding to one specific global output.}
  \label{fig:MultiST}
\end{figure}

We now show how the ideas discussed in the previous section can be extended to achieve self-testing of any $n$-partite qubit state. An arbitrary state has the form
\begin{equation}
\label{multistate}
\ket{\Psi} = \sum_{\vec{a}\in \{0,1\}^n} \lambda_{\vec{a}}\ket{\vec{a}}
\end{equation}
in a product basis $\{\ket{\vec a}\}$ and for normalized coefficients $\{\lambda_{\vec{a}}\}$. 

Our main self-testing result states the following.
\begin{theorem}\label{theoremMulti}
For any $n$-qubit pure state there exists a self-testing protocol requiring at most $9\cdot2^{n-2}-4$ two-outcome measurements per party.
\end{theorem}

To begin, note that self-testing a GME state of $n$ parties inherently implies the self-testing of any non-GME state of $n+1$ parties. Consequently, our analysis focuses on GME states. The self-testing procedure generalizes the approach previously developed for tripartite states, extending it to the $n$-party scenario.

We assign a unique number to each party, ranging from $1$ to $n$, and structure the self-testing procedure into $n-1$ sub-tests. In the $j$-th sub-test, the parties $\{1,\cdots, n\}/\{1,j+1\}$ perform measurements corresponding to the input $\lozenge$. These measurements prepare states for parties $1$ and $j+1$ that are then self-tested. In the $j$-th sub-test we name parties
$1$ and $j+1$ tested parties, while all the remaining are called projecting parties. Party $1$ serves as the tested party in every sub-test, while each of the other parties acts as a projecting party in all sub-tests except the one where it is itself being tested. In each sub-test, the $n-2$ projecting parties collectively produce $2^{n-2}$ different combinations of measurement outputs, resulting in $2^{n-2}$ distinct post-measurement states for the two tested parties. However, not all these post-measurement states are self-tested within a given sub-test. The subset of states that undergo self-testing depends on the sub-test in question. Specifically, in sub-test $k$, the procedure self-tests the post-measurement states corresponding to all possible outputs of parties $\{k+2,\cdots,n\}$, but only those states corresponding to output $0$ for parties $\{2,\cdots,k\}$. Using the recipe provided in Lemma~\ref{lemma:selftesting}, in every sub-test the post-measurement states are self-tested, and using Lemma~\ref{lemma:measurementlemma} also the observables corresponding to the inputs $\lozenge$ and $\blacklozenge$ of the tested parties.

The total number of measurement settings required for this self-testing procedure depends on the party. Let us analyze this, focusing on the number of different inputs needed by each party beyond the $\lozenge$ and $\blacklozenge$ measurements:\begin{itemize}
\item
Party $n$ is tested only in the final sub-test. Its projected sub-state is self-tested only when all projecting parties obtain the outcome $0$ for their $\lozenge$-measurements. To reproduce the required correlations (as specified in Lemma~\ref{lemma:selftesting}), party $n$ needs $6$ additional measurements, while party $1$ requires $3$ measurements.
\item
Party $n-1$ is a tested party only in a sub-test where its shared state with party $1$ is self-tested for both outcomes of party $n$’s $\lozenge$-measurement (and only outputs $0$ of the other projecting parties). This means that two different states are self-tested, requiring $6$ measurements for one state and $3$ for the other, totaling $9$ measurements for both tested parties.
\item Continuing this analysis, party $n-j$ requires $9\cdot2^{(j-1)}$ measurements because its shared state with party $1$ is tested for $2^j$ global outcomes of the remaining parties. Each tested state alternates between requiring $6$ and $3$ measurements.
\item The total number of measurements required by party $1$ is the sum of its measurements across all sub-tests discussed above, which equals $3 + 9(2^{n-2}-1)$
\end{itemize}

The full proof of Theorem~\ref{theoremMulti}, including constructive details of the procedure, is provided in Appendix~\ref{app:thmMulti}. Here, we provide a brief outline. Observables corresponding to dichotomic $\lozenge$ and $\blacklozenge$ measurements are used to construct a SWAP isometry. The output of the SWAP isometry is of the form:\begin{equation*}\sum_{\vec{a}\in (0,1)^n}(\lambda_{\vec{a}}\ket{\xi_{\vec{a}}} + \lambda^*_{\vec{a}}\ket{\xi_{\vec{a}}'})\otimes \ket{\vec{a}}.\end{equation*} The goal is to demonstrate that for every pair $\vec{a}_1,\vec{a}_2$, the equalities $\ket{\xi_{\vec{a}_1}} = \ket{\xi_{\vec{a}_2}}$ and $\ket{\xi'_{\vec{a}_1}} = \ket{\xi'_{\vec{a}_2}}$ hold, thereby establishing that all $2^n$ vectors are identical. The first sub-test demonstrates that vectors $\ket{\xi_{\vec{a}}}$ corresponding to $\vec{a}$ with identical last $n-2$ components are the same. Each subsequent sub-test reduces the number of distinct vector groups by half, until, by the ($n-1$)-th sub-test, it is proven that all $\ket{\xi_{\vec{a}}}$ are identical and equal to $\ket{\xi}$, and likewise for $\ket{\xi'_{\vec{a}}}$, being all equal to $\ket{\xi'}$. This yields the final form for the state resulting from the isometry:
\begin{equation*}
  \ket{\xi}\otimes\ket{\Psi} + \ket{\xi'}\otimes\ket{\Psi^*},
\end{equation*}
which completes the proof of the theorem, c.f. Eq.\eqref{st_pure} with $p=\langle{\xi}\ket{\xi}$ and $1-p=\langle{\xi'}\ket{\xi'}$.

\bigskip
\section{Discussion\label{sec:discussion}}
In this work, we have developed a self-testing procedure capable of certifying any pure entangled state of $n$ qubits for an arbitrary number of parties. The protocol is general and applies universally to all such states, fully certifying them or doing so up to complex conjugation, depending on whether the target state is locally equivalent to its complex conjugate. While the number of measurements required in our general construction grows exponentially with the number of parties, this increase reflects the universality of the approach rather than a fundamental limitation. For specific states, the relationships between their amplitudes could allow for a reduction in the number of measurements. For example, alternative methods for self-testing partially entangled qubit pairs, such as those introduced in~\cite{Barizien2024custombell} and~\cite{barizien2024quantumstatisticsminimalscenario}, could be employed to streamline the process. Furthermore, for highly structured states, the number of measurements per party can be reduced to a constant, as demonstrated for W-states and graph states in~\cite{Ivan}. An important direction for future research is to explore how different families of multipartite states can be self-tested with a number of measurements that scales more favorably with system size, either through refined protocols or leveraging the structure of the family in question.

\bigskip
\section*{Acknowledgements}
We thank Laura Man\v{c}inska for useful suggestions and Mariami Gachechiladze for insightful discussions. This project was funded within the QuantERA II Programme (VERIqTAS project) that has received funding from the European Union’s Horizon 2020 research and innovation programme under Grant Agreement No 101017733 and from the Polish National Science Center (projects No. 2021/03/Y/ST2/00175. We acknowledge support from the Government of Spain (Severo Ochoa CEX2019-000910-S, FUNQIP and European Union NextGenerationEU PRTR-C17.I1), Fundaci\'o Cellex, Fundaci\'o Mir-Puig, Generalitat de Catalunya (CERCA program), European Union (QSNP 101114043 and Quantera project Veriqtas), the ERC AdG CERQUTE, the AXA Chair in Quantum Information Science. MBJ acknowledges funding from the European Union's Horizon 2020 research and innovation programme under the Marie Sk\l{}odowska-Curie grant agreement No 847517. I\v{S} acknowledges the support from French national quantum initiative managed by Agence Nationale de la Recherche in the framework of France 2030 with the reference EPIQ ANR-22-PETQ-0007.

\bibliographystyle{unsrturl}

\appendix
\setcounter{equation}{0}
\setcounter{figure}{0}
\setcounter{table}{0}
\setcounter{page}{1}
\setcounter{section}{0}
\makeatletter
\renewcommand{\theequation}{A\arabic{equation}}
\renewcommand{\thefigure}{A\arabic{figure}}

\appendixpage
\addappheadtotoc

\section{Proof of Lemma \ref{lemma:selftesting}\label{app:proof1}}

In this Appendix, we provide a detailed proof of all steps in Lemma \ref{lemma:selftesting}, which establishes a method for self-testing partially entangled states of two qubits and three Pauli observables for the $j$-th party. For reference, we recall the form of the tilted CHSH Bell inequality, as presented in the main text (cf. Eq.~\eqref{tiltedCHSHin}):
\begin{equation*} I_{\alpha} := \alpha\langle A_0^{(j)}\rangle + \langle A_0^{(j)}A_0^{(k)}\rangle + \langle A_0^{(j)}A_1^{(k)}\rangle + \langle A_1^{(j)}A_0^{(k)}\rangle - \langle A_1^{(j)}A_1^{(k)}\rangle \leq 2+\alpha. \end{equation*}
We follow the proof presented in \cite{erik}, proceeding in four parts to align with the progressive structure of the argument.
\paragraph{Qubit States.}
In the first part of the proof, we consider pure two-qubit states. Any pure two-qubit state can be expressed in the form
\begin{equation*} \ket{\psi} = \cos(\tau/2)\ket{00} + \sin(\tau/2)\ket{11}, \end{equation*}
where $0 \leq \tau \leq \pi/2$. The corresponding density matrix for this state is given by
\begin{equation*} \proj{\psi} = \frac{1}{4}\left[\mathds{1}\otimes\mathds{1} + \cos\tau(\sz\otimes\mathds{1} + \mathds{1}\otimes \sz) + \sin\tau(\sx\otimes \sx - \sy\otimes \sy) + \sz\otimes \sz\right]. \end{equation*}
Since maximal quantum violations of Bell inequalities require extremal measurements, which for qubits and two outcomes correspond to projective measurements, we assume the following:
\begin{align*} A_l^{(j_q)} = \mathbf{a}^{(j_q)}_l\cdot\mathbf{\sigma}, &\qquad A_l^{(k_q)} = \mathbf{a}^{(k_q)}_l\cdot\mathbf{\sigma}, \end{align*}
where $\sigma = (\sz,\sx,\sy)$ represents the vector of Pauli matrices, and $\|\mathbf{a}^{(j_q)}_l\| = 1$ for all $l$.

By substituting the state and measurement operators into the expression for the Bell inequality, we obtain
\begin{equation}\label{be1} I_{\alpha} = \alpha\cos{\tau}a^{(j_q)}_{0,z} + S,
\end{equation}
where
\begin{equation}\label{sv1} S = \mathbf{a}^{(j_q)}_0\cdot \mathbf{T}\left(\mathbf{a}^{(k_q)}_0 + \mathbf{a}^{(k_q)}_1\right) + \mathbf{a}^{(j_q)}_1\cdot \mathbf{T}\left(\mathbf{a}^{(k_q)}_0 - \mathbf{a}^{(k_q)}_1\right), \end{equation}
and the correlation matrix 
$\mathbf{T}$ is defined as
\begin{equation*} \mathbf{T} = \left[ {\begin{array}{ccc} \sin{\tau} & 0 & 0\\ 0 & -\sin{\tau} & 0\\ 0 & 0 & 1\ \end{array}} \right]. \end{equation*}
Let us introduce the parameter $\nu \in \left[0,\pi/2\right]$, defining the vectors $\mathbf{a}^{(k_q)}_{\pm}$ as follows:
\begin{equation}\label{Raimat} \mathbf{a}^{(k_q)}_0 + \mathbf{a}^{(k_q)}_1 = 2\cos\left(\frac{\nu}{2}\right)\mathbf{a}^{(k_q)}_{+}, \qquad \mathbf{a}^{(k_q)}_0 - \mathbf{a}^{(k_q)}_1 = 2\sin\left(\frac{\nu}{2}\right)\mathbf{a}^{(k_q)}_{-}. \end{equation}
By construction, these vectors are normalized and orthogonal. Using this parametrization, the expression for $S$, as defined in \eqref{sv1}, can be rewritten as:
\begin{align} S &= \label{in0} 2\cos\left(\frac{\nu}{2}\right)\mathbf{a}^{(j_q)}_0 \cdot \mathbf{T}\mathbf{a}^{(k_q)}_{+} + 2\sin\left(\frac{\nu}{2}\right)\mathbf{a}^{(j_q)}_1 \cdot \mathbf{T}\mathbf{a}^{(k_q)}_{-} \\
&\leq \label{in1} 2\cos\left(\frac{\nu}{2}\right) \left|\mathbf{T}\mathbf{a}^{(k_q)}_{+}\right| + 2\sin\left(\frac{\nu}{2}\right)\left| \mathbf{T}\mathbf{a}^{(k_q)}_{-}\right| \\
&\leq \label{in15} 2\sqrt{\left|\mathbf{T}\mathbf{a}^{(k_q)}_{+}\right|^2 + \left|\mathbf{T}\mathbf{a}^{(k_q)}_{-}\right|^2} \\
&= \label{in33}\sqrt{\mathbf{a}^{(k_q)}_{+} \cdot \mathbf{T}^2\mathbf{a}^{(k_q)}_{+} + \mathbf{a}^{(k_q)}_{-} \cdot \mathbf{T}^2\mathbf{a}^{(k_q)}_{-}} \\
&\leq \label{in3} 2\sqrt{1+\sin^2\tau}. \end{align}
The transitions between these lines are justified as follows:\begin{itemize}
\item From \eqref{in0} to \eqref{in1}: The Cauchy-Schwarz inequality is applied to each term, with the norms of $\mathbf{a}^{(j_q)}_0$ and $\mathbf{a}^{(j_q)}_1$ being at most 1.
\item 
From \eqref{in1} to \eqref{in15}: The terms in \eqref{in1} can be interpreted as an inner product between the vectors $2\left(\cos(\frac{\nu}{2}),\sin(\frac{\nu}{2})\right)$ and $\left(\mathbf{T}\,\mathbf{a}^{(k_q)}_+, \mathbf{T}\,\mathbf{a}^{(k_q)}_-\right)$. Applying the Cauchy-Schwarz inequality to this inner product yields \eqref{in15}.
\item
From \eqref{in15} to \eqref{in33}: The expression in \eqref{in15} is rewritten using the definition of the norm, $\left|\mathbf{T}\mathbf{a}\right| = \mathbf{a}\cdot \mathbf{T}^2\mathbf{a}$
\item
Final bound in \eqref{in3}: Since $\mathbf{a}^{(k_q)}_\pm$ are orthonormal, the sum in \eqref{in3} is bounded by the sum of the squares of the two largest eigenvalues of $\mathbf{T}$, which equals 
$1+\sin^2\tau$
 \end{itemize}
Returning to the Bell expression in \eqref{be1}, we have
\begin{align} \nonumber I_{\alpha} &\leq \alpha\cos{\tau}a^{(j_q)}_{0,z} + 2\sqrt{1 + \sin^2\tau} \\ \nonumber &\leq \alpha\cos{\tau} + 2\sqrt{1 + \sin^2\tau} \\ \label{infin} &\leq 2\sqrt{2}\sqrt{1 + \alpha^2/4}. \end{align}
The transition to the second line follows from the fact that $a^{(j_q)}_{0,z} \leq 1$. The third line results from optimizing the expression in the second line over $\tau$. The maximum value is achieved when
\begin{equation*} 2\cos\tau = \alpha\sqrt{1 + \sin^2\tau}. \end{equation*}
If $I_{\alpha}$	reaches the quantum bound, all inequalities in \eqref{in1} through \eqref{infin} are saturated. From the saturation of \eqref{infin}, we deduce that
\begin{equation*}a^{(j_q)}_{0,z} = 1, \qquad \Rightarrow \qquad \mathbf{a}^{(j_q)}_0 = \begin{pmatrix} 0 \ 0 \ 1 \end{pmatrix}^T. \end{equation*}
Similarly, the saturation of \eqref{in3} implies
\begin{equation*} \mathbf{a}^{(k_q)}_{+} = \begin{pmatrix} 0 \ 0 \ 1 \end{pmatrix}^T, \qquad \mathbf{a}^{(k_q)}_{-} = \begin{pmatrix} 1 \ 0 \ 0 \end{pmatrix}^T. \end{equation*}
Using these results and Eq.~\eqref{Raimat}, the vectors 
$\mathbf{a}^{(k_q)}_l$ are obtained as
\begin{equation*}\label{sat2-l} \mathbf{a}^{(k_q)}_l = \begin{pmatrix} \cos\frac{\nu}{2}\quad \ 0\quad \ (-1)^l \sin\frac{\nu}{2} \end{pmatrix}^T, \qquad (l = 0, 1). \end{equation*}
To turn the inequality in \eqref{in0} into equality, the vectors $\mathbf{a}^{(j_q)}_l$ must be parallel to $\mathbf{T}\mathbf{a}^{(k_q)}_{\pm}$. This condition implies
\begin{equation*}\mathbf{a}^{(j_q)}_0 = \mathbf{a}^{(k_q)}_{+} = \begin{pmatrix} 0 \ 0 \ 1 \end{pmatrix}^T, \qquad \mathbf{a}^{(j_q)}_1 = \mathbf{a}^{(k_q)}_{-} = \begin{pmatrix} 1 \ 0 \ 0 \end{pmatrix}^T. \end{equation*}
Finally, for the expressions in \eqref{in1} and \eqref{in15} to be equal, the angles $\nu$ and $\tau$ must satisfy the following relation:
\begin{equation*} \cos\left(\frac{\nu}{2}\right)\sin\tau = \sin\left(\frac{\nu}{2}\right). \end{equation*}
With all the above, we conclude that the maximal violation of the tilted CHSH inequality, $I_{\alpha} = \sqrt{8+2\alpha^2}$, imposes the following constraints on the qubit strategy that achieves this violation:
\begin{align*} &\ket{\psi} = \cos\left(\frac{\theta_\alpha}{2}\right)\ket{00} + \sin\left(\frac{\theta_\alpha}{2}\right)\ket{11},\qquad \sin\theta_{\alpha} = \sqrt{\frac{1 - \alpha^2/4}{1 + \alpha^2/4}},\\ &A^{(j_q)}_0 = \sz, \qquad A^{(j_q)}_1 = \sx,\\ &A^{(k_q)}_0 = \cos\left(\frac{\mu_\alpha}{2}\right) \sz + \sin\left(\frac{\mu_\alpha}{2}\right)\sx, \qquad A^{(k_q)}_1 = \cos\left(\frac{\mu_\alpha}{2}\right) \sz - \sin\left(\frac{\mu_\alpha}{2}\right)\sx,\\ &\sin\left(\frac{\mu_\alpha}{2}\right) = \sqrt{\frac{1 - \alpha^2/4}{2}}. \end{align*}
\paragraph{Arbitrary Dimension and Mixed States.}
The tilted CHSH inequality involves two binary observables on each side, allowing us to apply the Jordan lemma, which shows that, given two operators, there is a choice of basis in which they have a diagonal form with blocks of size at most $2\times 2$. Consequently, the measurement observables can be expressed as:
\begin{align*} A^{(j)}_x = \sum_l A^{(j_q)}_{x,l} \otimes \ket{l}\bra{l}^{j''} \oplus {A^{(j)}_x}^\perp, \qquad A^{(k)}_y = \sum_l A^{(k_q)}_{y,l} \otimes \ket{l}\bra{l}^{(k'')} \oplus {A^{(k)}_y}^\perp, \end{align*}
where $x,y = 0,1$, $A_{x,l}^{(j_q)}$ and $A_{y,l}^{(k_q)}$ are $2\times 2$ operators, and the operators with the $\perp$ superscript correspond to the remaining $1\times 1$ blocks. The latter are irrelevant in our analysis as they cannot contribute to Bell violations. As a result, the operators ${A_0^{(j)}}^\perp$ and ${A_1^{(j)}}^\perp$, and ${A_0^{(k)}}^\perp$ and ${A_1^{(k)}}^\perp$ trivially commute.

In this basis, any pure state can be written as:
\begin{equation*} \ket{\Psi} = \bigoplus_{m,n} \sqrt{p_{mn}} \ket{\psi_{mn}}, \end{equation*}
where the indices $m,n$ can be of two types, denoting whether the sub-state $\ket{\psi_{mn}}$ resides in a qubit-qubit subspace, or a one-dimensional subspace. Using this representation, the tilted CHSH violation takes the form:
\begin{equation*} I_\alpha = \sum_{mn} p_{mn} I_{\alpha,mn}. \end{equation*}
The quantum bound $\beta_Q$ is achieved if, and only if, for all $m$ and $n$ such that $p_{mn} \neq 0$, one has $I_{\alpha,mn} = \beta_Q$. However, unless both $m$ and $n$ enumerate two-dimensional subspaces, the value $I_{\alpha,mn}$ cannot exceed the classical bound because in these cases either Alice's or Bob's measurements necessarily commute within the corresponding subspace. This implies that $\ket{\Psi}$ must be confined to the tensor product of two Hilbert spaces containing only $2\times 2$ blocks on both sides. Therefore, the state can be written as:
\begin{equation} \ket{\Psi} = \sum_{mn} \sqrt{q_{mn}} \ket{\varphi_{mn}}^{(j_q,k_q)} \ket{m}^{(j'')} \ket{n}^{(k'')}. \end{equation}
The tilted CHSH violation can be expressed as
\begin{align*}
I_{\alpha} &= \sum_{mn}q_{mn}\bra{\phi_{mn}}^{(j_q,k_q)}\left(\alpha A_{0,m}^{(j_q)}\tp\idd^{(k_q)}_n+ A^{(j_q)}_{0,m}\tp\left(A^{(k_q)}_{0,n} + A^{(k_q)}_{1,n}\right) + A^{(j_q)}_{1,m}\tp \left(A^{(k_q)}_{0,n} - A^{(k_q)}_{1,n}\right)\right)\ket{\phi_{mn}}^{(j_q,k_q)} \\&= \sum_{mn}q_{mn}I_{\alpha,mn},
\end{align*}
where $\idd^{(k_q)}_n$ is the identity operator on the $n$-th $2\times 2$ block in Hilbert space $\pazocal{H}^{(k)}$. 
For the quantum bound $\beta_Q$ to be achieved, it must hold that $I_{\alpha,mn} = \beta_Q$ for all $m$ and $n$ such that $q_{mn} \neq 0$. Thus, for every $m$ with at least one $n$ such that $q_{mn}\neq 0$ (and vice versa), the following conditions must be satisfied:
\begin{align*}
\ket{\psi_{mn}} &= \cos\left(\frac{\theta_\alpha}{2}\right)\ket{00} + \sin\left(\frac{\theta_\alpha}{2}\right)\ket{11},\\
A^{(j_q)}_{0,m} &= \sz,\qquad A^{(j_q)}_{1,m} = \sx,\\
A^{(k_q)}_{0,n} &= \cos\left(\frac{\mu}{2}\right) \sz + \sin\left(\frac{\mu}{2}\right)\sx, \qquad A^{(k_q)}_{1,n} = \cos\left(\frac{\mu}{2}\right) \sz-\sin \left(\frac{\mu}{2}\right)\sx.
\end{align*}
This implies
\begin{equation*}
\ket{\Psi}^{(j,k)} = \ket{\psi_\theta}^{(j_q,k_q)}\otimes \ket{junk}^{(j'',k'')},
\end{equation*}
where
\begin{equation}
\ket{junk}^{(j'',k'')} = \sum_{mn}\sqrt{q_{mn}}\ket{m}^{(j'')}\ket{n}^{(k'')}.
\end{equation}
The measurement operators can be written as:
\begin{align*}
A^{(j)}_0 &= \sz^{(j_q)}\otimes \mathds{1}^{(j'')}\oplus {A^{j}_0}^\perp,\\ A^{(j)}_1 &= \sx^{(j_q)}\otimes \mathds{1}^{(j'')}\oplus {A^{(j)}_1}^\perp,\\ 
A^{(k)}_0 &= \left(\cos\left(\frac{\mu}{2}\right) \sz^{(k_q)} + \sin\left(\frac{\mu}{2}\right)\sx^{(k_q)}\right) \otimes \mathds{1}^{(k'')} \oplus {A^{(k)}_0}^\perp, \\ A^{(k)}_1 &= \left(\cos\left(\frac{\mu}{2}\right) \sz^{(k_q)} - \sin\left(\frac{\mu}{2}\right)\sx^{(k_q)}\right) \otimes \mathds{1}^{(k'')} \oplus {A^{(k)}_1}^\perp,
\end{align*}
where operators with the superscript $\perp$ do not act on the support of $\ket{\Psi}^{(j,k)}$. \\

This result generalizes straightforwardly to mixed states, as any mixed state $\rho$ can be expressed as a convex combination of pure states: $\rho = \sum_sp_s\proj{\Psi_s}$.
If the tilted CHSH inequality is maximally violated, every pure state \ket{\Psi_s} in the decomposition must also achieve the quantum bound. Following the same reasoning as above, each \ket{\Psi_s} can be written as:
\begin{equation*}
\ket{\Psi_s}^{(j,k)} = \sum_{mn}\sqrt{q_{mn}^s}\ket{\phi_{mn}^s}^{(j_q,k_q)}\ket{m}^{(j'')}\ket{n}^{(k'')}.
\end{equation*}
Thus
\begin{equation*}
\ket{\Psi_s}^{(j,k)} = \ket{\psi_\theta}^{(j_q,k_q)}\otimes \ket{junk_s}^{(j'',k'')}, \qquad \ket{junk_s}^{(j'',k'')} = \sum_{mn}q_{mn}^s\ket{m}^{(j'')}\otimes \ket{n}^{(k'')}.
\end{equation*}
Consequently:
\begin{equation*}
\rho^{(j,k)} = \proj{\psi_\theta}^{(j_q,k_q)}\otimes \sigma_{junk}^{(j'',k'')},
\end{equation*}
where
\begin{equation*}
\sigma_{junk}^{(j'',k'')} = \sum_sp_s\proj{junk_s}^{(j'',k'')}.
\end{equation*}

For the maximal violation of $J_{\alpha}$, the condition $\{A_{0}^{(j)},A_{2}^{(j)}\} = 0$ must hold. The general form of $A_{2}^{(j)}$ is
\begin{equation*}
  A_{2}^{(j)} = \mathds{1}^{(j_q)}\tp A_{2,\mathds{1}}^{(j'')} + \sx^{(j_q)}\tp A_{2,\mathrm{X}}^{(j'')} + \sy^{(j_q)}\tp A_{2,\mathrm{Y}}^{(j'')} + \sz^{(j_q)}\tp A_{2,\mathrm{Z}}^{(j'')},
\end{equation*}
where $A_{2}^{(j)}$ satisfies $\left(A_{2}^{(j)}\right)^2=\mathds{1}$ and anti-commutes with $\sz^{(j_q)}\tp\mathds{1}^{(j'')}$. This implies that $A_{2,\mathds{1}}^{(j'')} =A_{2,\mathrm{Z}}^{(j'')} = 0$, reducing $A_{2}^{(j)}$ to
\begin{equation}\label{2meas}
  A_{2}^{(j)} = \sx^{(j_q)}\tp A_{2,\mathrm{X}}^{(j'')} + \sy^{(j_q)}\tp A_{2,\mathrm{Y}}^{(j'')},
\end{equation}
where $[A_{2,\mathrm{X}}^{(j'')},A_{2,\mathrm{Y}}^{(j'')}] = 0$ and ${A_{2,\mathrm{X}}^{(j'')}}^2 + {A_{2,\mathrm{Y}}^{(j'')}}^2 = \mathds{1}$. 
In a similar manner we can write the operators of party $k$ in the following form
\begin{equation*}
  A_{l}^{(k)} = \mathds{1}^{(k_q)}\tp A_{l,\mathds{1}}^{(k'')} + \sx^{(k_q)}\tp A_{l,\mathrm{X}}^{(k'')} + \sy^{(k_q)}\tp A_{l,\mathrm{Y}}^{(k'')} + \sz^{(k_q)}\tp A_{l,\mathrm{Z}}^{(k'')},
\end{equation*}
such that
\begin{equation*}
  {A_{l,\mathds{1}}^{(k'')}}^2 + {A_{l,\mathrm{X}}^{(k'')}}^2 + {A_{l,\mathrm{Y}}^{(k'')}}^2 + {A_{l,\mathrm{Z}}^{(k'')}}^2 = \mathds{1}. 
\end{equation*}

To account for violations such as $L = 2\sqrt{2}\sin\theta$, the following expectations must be considered:
\begin{align*}
  \langle A_1^{(j)}A_l^{(k)}\rangle &= \sin\theta\,\tr\left[\mathds{1}^{(j'')}\tp A_{l,\mathrm{X}}^{(k'')}\sigma_{junk}^{j'',k''}\right],\\
  \langle A_2^{(j)}A_l^{(k)}\rangle &= \sin\theta\,\tr\left(\tr\left[A_{2,\mathrm{X}}^{(j'')}\tp A_{l,\mathrm{X}}^{(k'')}\sigma_{junk}^{(j'',k'')}\right] - \tr\left[A_{2,\mathrm{Y}}^{(j'')}\tp A_{l,\mathrm{Y}}^{(k'')}\sigma_{junk}^{(j'',k'')}\right]\right), 
\end{align*}
for $l = 4,5$. The conditions~\eqref{2meas} allow choosing a common basis for the commuting operators $A_{2,\mathrm{Y}}^{(j'')}$ and $A_{2,\mathrm{X}}^{(j'')}$, which also square to identity
\begin{equation*}
  \mathds{1}^{(j)''} = \sum_m\ketbra{m}{m}, \qquad A_{2,\mathrm{X}}^{(j'')} = \sum_mx_m\ketbra{m}{m}, \qquad A_{2,\mathrm{Y}}^{(j'')} = \sum_my_m\ketbra{m}{m},
\end{equation*}
such that $x_m^2 + y_m^2 = 1$ for all values of $m$. Let us denote $\sigma_m = \tr_{(j'')}\left[\ketbra{m}{m}^{(j'')}\tp\mathds{1}^{(k'')}\sigma_{junk}^{j'',k''}\right]$. Now we can write the following set of relations 
\begin{align*}
\frac{L}{\sin\theta} &= \tr\left[(\mathds{1}^{(j'')}+A_{2,\mathrm{X}}^{(j'')})\tp A_{4,\mathrm{X}}^{(k'')}\sigma_{junk}^{j'',k''}\right] - \tr\left[A_{2,\mathrm{Y}}^{(j'')}\tp A_{4,\mathrm{Y}}^{(k'')}\sigma_{junk}^{j'',k''}\right] + \\ & \qquad\qquad\qquad + \tr\left[(\mathds{1}^{(j'')}-A_{2,\mathrm{X}}^{(j'')})\tp A_{5,\mathrm{X}}^{(k'')}\sigma_{junk}^{j'',k''}\right] + \tr\left[A_{2,\mathrm{Y}}^{(j'')}\tp A_{5,\mathrm{Y}}^{(k'')}\sigma_{junk}^{j'',k''}\right] \\
&= \sum_m\left((1+x_m)\tr\left[A_{4,\mathrm{X}}^{(k'')}\sigma_m^{k''}\right] - y_m\tr\left[A_{4,\mathrm{Y}}^{(k'')}\sigma_m^{k''}\right]  +  (1-x_m)\tr\left[A_{5,\mathrm{X}}^{(k'')}\sigma_m^{k''}\right] + y_m\tr\left[A_{5,\mathrm{Y}}^{(k'')}\sigma_m^{k''}\right]   \right)\\
&\leq \sum_m\sqrt{2(1+x_m)}\sqrt{\left(\tr\left[A_{4,\mathrm{X}}^{(k'')}\sigma_m^{(k'')}\right]\right)^2 + \left(\tr\left[A_{4,\mathrm{Y}}^{(k'')}\sigma_m^{(k'')}\right]\right)^2} + \\ & \qquad\qquad\qquad + \sum_m\sqrt{2(1-x_m)}\sqrt{\left(\tr\left[A_{5,\mathrm{X}}^{(k'')}\sigma_m^{(k'')}\right]\right)^2 + \left(\tr\left[A_{5,\mathrm{Y}}^{(k'')}\sigma_m^{(k'')}\right]\right)^2}   \\
&\leq \sum_m \left(\sqrt{2(1+x_m)}+ \sqrt{2(1-x_m)}\right)\tr[\sigma_m^{(k'')}]\\
&\leq \sum_m 2\sqrt{2}\tr[\sigma_m^{(k'')}]\\
&= 2\sqrt{2}.
\end{align*}
The fourth line is obtained from the Cauchy-Schwarz inequality aided with the relation $x_m^2+y_m^2 = 1$. The sixth line is obtained from the Cauchy-Schwarz inequality of different kind: $\tr\left[A\sigma_m\right] \leq \sqrt{\tr\left[A^2\sigma_m\right]}\sqrt{\tr{\sigma_m}}$ and $\tr\left[\left({A_{l,\mathrm{X}^{(k)}}}^2 + {A_{l,\mathrm{Y}^{(k)}}}^2\right)\sigma_m\right] \leq \tr{\sigma_m}$. The last inequality is saturated if $(1+x_m) = (1-x_m)$, 
implying $x_m = 0$ and thus $y_m = \pm 1$. This implies $A_{2,\mathrm{X}}^{(j'')} = 0$ and $A_{2}^{(j)} = \sy^{(j_q)}\tp A_{2,\mathrm{Y}}^{(j'')}$, such that ${A_{2,\mathrm{Y}}^{(j'')}}^2 = \mathds{1}$.
What is left is to characterize operators $A_{2}^{(k)}$ and $A_3^{(k)}$. Given that (i) these measurements are used to obtain the maximal violation of $J_\alpha$, (ii) the underlying state is $\rho = \proj{\psi_{\theta}}^{(j_q,k_q)}\tp\sigma_{junk}^{j'',k''}$ and (iii) the $j$-th's party measurements are $A_0^{(j)} = \sz^{(j_q)}\tp\mathds{1}^{(j'')}$ and $A_2^{(j)} = \sy^{(j_q)} \tp A_{2,\mathrm{Y}}^{(j'')}$, it must be that
\begin{equation}\label{aix}
  A_{2}^{(k)} + A_3^{(k)} = \cos\left(\frac{\mu}{2}\right)\sz^{(k_q)}\tp A_{\mathrm{Z}}^{(k'')}, \qquad A_{2}^{(k)} - A_3^{(k)} = \sin\left(\frac{\mu}{2}\right)\sy^{(k_q)}\tp A_{\mathrm{Y}}^{(k'')},
\end{equation}
where $\tr\left[A_{\mathrm{Z}}^{(k'')}\sigma_{junk}^{(j'',k'')}\right] = 1$ and $\tr\left[A_{\mathrm{Y}}^{(j'')}\tp A_{\mathrm{Y}}^{(k'')}\sigma_{junk}^{(j'',k'')}\right] = 1$. Hence,
\begin{equation*}
  A_{2}^{(k)} = \cos\left(\frac{\mu}{2}\right)\sz^{(k_q)}\tp A_{\mathrm{Z}}^{(k'')} + \sin\left(\frac{\mu}{2}\right)\sy^{(k_q)}\tp A_{\mathrm{Y}}^{(k'')}, \qquad 
  A_{3}^{(k)} = \cos\left(\frac{\mu}{2}\right)\sz^{(k_q)}\tp A_{\mathrm{Z}}^{(k'')} - \sin\left(\frac{\mu}{2}\right)\sy^{(k_q)}\tp A_{\mathrm{Y}}^{(k'')},
\end{equation*}
and $[A_{\mathrm{Z}}^{(k'')},A_{\mathrm{Y}}^{(k'')}] = 0$ and ${A_{\mathrm{Z}}^{(k'')}}^2 = {A_{\mathrm{Y}}^{(k'')}}^2 = \mathds{1}$. Since $A_{\mathrm{Z}}^{(k'')}$ and $A_{\mathrm{Y}}^{(k'')}$ commute, we can write them in a common basis:
\begin{equation*}
  \mathds{1}^{(k'')} = \sum_n\ketbra{n}{n}, \qquad A_{\mathrm{Z}}^{(k'')} = \sum_nz_n\ketbra{n}{n}, \quad A_{\mathrm{Y}}^{(k'')} = \sum_ny'_n\ketbra{n}{n},
\end{equation*}
where $\|z_n\| = \|y'_n\| = 1$. The conditions given below Eq.~\eqref{aix} imply
\begin{align*}
  \sum_{m,n}z_n\tr\left[\sigma_{junk}^{(j'',k'')}
|m\rangle\langle m|^{(j'')}\tp|n\rangle\langle n|^{(k'')}\right] &= 1,\\
  \sum_{m,n}y_my'_n\tr\left[\sigma_{junk}^{(j'',k'')}|m\rangle\langle m|^{(j'')}|n\rangle\langle n|^{(k'')}\right] &= 1.
\end{align*}
Since $\sum_{n,m}\tr\left[\sigma_{junk}^{(j'',k'')}|m\rangle\langle m|^{(j'')}|n\rangle\langle n|^{(k'')}\right]=1$ there are two possibilities:
\begin{enumerate}
  \item $\sigma_{junk}^{(j'',k'')} = \proj{\psi_{junk}}^{(j'')}\tp \proj{\psi_{junk}}^{(k'')}$, and $\ket{\psi_{junk}}^{(k'')}$ is the common eigenstate of $A_{\mathrm{Z}}^{(k'')}$ and $A_{\mathrm{Y}}^{(k'')}$ corresponding to the eigenvalue $+1$, while $\ket{\psi_{junk}}^{(j'')}$ is the $+1$ eigenstate of $A_{\mathrm{Y}}^{(j'')}$. In that case, all these operators act as the identity operator, $\mathds{1}$, on the support of $\sigma_{junk}^{(j'',k'')}$;
  \item $\sigma_{junk}^{(j'',k'')}$ is entangled, $A_{\mathrm{Z}}^{(k'')} = \mathds{1}$ and $A_{\mathrm{Y}}^{(j'')}\tp A_{\mathrm{Y}}^{(k'')}\sigma_{junk}^{(j'',k'')} = \sigma_{junk}^{(j'',k'')}$.
\end{enumerate}
Since the second condition is more general, it will be used onward:
\begin{equation*}
  A_{2}^{(k)} = \cos\left(\frac{\mu}{2}\right)\sz^{(k_q)}\tp \mathds{1}^{(k'')} + \sin\left(\frac{\mu}{2}\right)\sy^{(k_q)}\tp A_{\mathrm{Y}}^{(k'')}, \qquad 
  A_{3}^{(k)} = \cos\left(\frac{\mu}{2}\right)\sz^{(k_q)}\tp \mathds{1}^{(k'')} - \sin\left(\frac{\mu}{2}\right)\sy^{(k_q)}\tp A_{\mathrm{Y}}^{(k'')}.
\end{equation*}

This completes the proof of Lemma~\ref{lemma:selftesting}, and it is done for the most general class of physical experiments: there is no assumption that the state is pure or that the measurements are projective. The deduced form of measurements and state imply that the physical experiment must be support-preserving if it reproduces correlations stipulated in Lemma~\ref{lemma:selftesting}. Important to note, since we were free to chose first the Schmidt basis in the first iteration, later the basis in which the measurements are block-diagonal, to be fully rigorous we have to involve unitaries mapping the state and measurement to these bases, so the final self-testing results take the form:
\begin{align*}
  U^{(j)}\ot U^{(k)}\rho^{(j,k)}\left(U^{(j)}\ot U^{(k)}\right)^\dagger &= \proj{\psi_\theta}^{(j_q,k_q)}\otimes \sigma_{junk}^{(j'',k'')},\\
  U^{(j)}A^{(j)}_0{U^{(j)}}^\dagger &= \sz^{(j_q)}\otimes \mathds{1}^{(j'')},\\ {U^{(j)}}A^{(j)}_1{U^{(j)}}^\dagger &= \sx^{(j_q)}\otimes \mathds{1}^{(j'')},\\ {U^{(j)}}A_2^{(j)}{U^{(j)}}^\dagger &= \sy^{(j_q)} \tp A_{2,\mathrm{Y}}^{(j'')}\\
U^{(k)}A^{(k)}_0{U^{(k)}}^\dagger &= \left(\cos\left(\frac{\mu}{2}\right) \sz^{(k_q)} + \sin\left(\frac{\mu}{2}\right)\sx^{(k_q)}\right) \otimes \mathds{1}^{(k'')} , \\ U^{(k)}A^{(k)}_1{U^{(k)}}^\dagger &= \left(\cos\left(\frac{\mu}{2}\right) \sz^{(k_q)} - \sin\left(\frac{\mu}{2}\right)\sx^{(k_q)}\right) \otimes \mathds{1}^{(k'')}, \\
U^{(k)}A_{2}^{(k)}{U^{(k)}}^\dagger &= \cos\left(\frac{\mu}{2}\right)\sz^{(k_q)}\tp \mathds{1}^{(k'')} + \sin\left(\frac{\mu}{2}\right)\sy^{(k_q)}\tp A_{\mathrm{Y}}^{(k'')}, \\ 
  U^{(k)}A_{3}^{(k)}{U^{(k)}}^\dagger &= \cos\left(\frac{\mu}{2}\right)\sz^{(k_q)}\tp \mathds{1}^{(k'')} - \sin\left(\frac{\mu}{2}\right)\sy^{(k_q)}\tp A_{\mathrm{Y}}^{(k'')},\end{align*}
where unitaries $U^{(j)}$ and $U^{(k)}$ act on the support of $\rho^{(j,k)}$, whence the omition of the parts of the operators acting outside of that support. 

\section{Proof of Lemma \ref{lemma:measurementlemma}}\label{appML}
\setcounter{equation}{0}
\makeatletter
\renewcommand{\theequation}{B\arabic{equation}}

Without loss of generality, we assume that $\alpha_x,\alpha_y,\alpha_z \ge 0$, as the derivations for the other cases can be made analogously. To proceed, we express $A^{(k)}$ in terms of the Pauli operators acting on qubit Hilbert spaces $\pazocal{H}^{(k_q)}$ and Hermitian operators acting on the Hilbert space $\pazocal{H}^{(k'')}$. Specifically, we write:
\begin{equation*}
U_kA^{(k)}U_k^\dagger = \sum_{P}P^{(k_q)}\otimes A_P^{(k'')}\,,
\end{equation*}
where $\{P\}$ represents the set of Pauli operators acting on a single qubit, and $A_P^{(k'')}$ are Hermitian operators acting on $\pazocal{H}^{(k'')}$.The condition ${A^{(k'')}}^2 = \mathds{1}$ leads to the following relations: 
\begin{equation*}
\sum_P {A_P^{(k'')}}^2 = \mathds{1}, \quad \mathrm{and}\quad \left[A_P^{(k'')},A_{P'}^{(k'')}\right] = 0.
\end{equation*}
 
We now rewrite the key equations, Eqs.~\eqref{02}-\eqref{22}, using these expressions. We obtain the following system:
\begin{align}
\label{02a}
\cos(2\theta)\tr\left[\mathds{1}^{(j'')}\otimes A_\mathds{1}^{(k'')}\sigma_{junk}^{(j'',k'')}\right] + \tr\left[\mathds{1}^{(j'')}\otimes A_\mathrm{Z}^{(k'')}\sigma_{junk}^{(j'',k'')}\right] &= \alpha_z,\\ \label{12a}
\tr\left[\mathds{1}^{(j'')}\otimes A_\mathrm{X}^{(k'')}\sigma_{junk}^{(j'',k'')}\right] &= \alpha_x, \\
\tr\left[A_\mathrm{Y}^{(j'')}\otimes A_\mathrm{Y}^{(k'')}\sigma_{junk}^{(j''k'')}\right] &= \alpha_y, \label{eq: A2Ba}\\ \label{22a}
\tr\left[\mathds{1}^{(j'')}\otimes A_\mathds{1}^{(k'')}\sigma_{junk}^{(j''k'')}\right] + \cos(2\theta)\tr\left[\mathds{1}^{(j'')}\otimes A_\mathrm{Z}^{(k'')}\sigma_{junk}^{(j''k'')}\right] &= \alpha_z\,\cos{(2\theta)}\,.
\end{align}
From Eqs.~\eqref{02a} and \eqref{22a}, we can immediately deduce:
\begin{align}\nonumber
\tr\left[\mathds{1}^{(j'')}\otimes A_\mathds{1}^{(k'')}\sigma_{junk}^{(j''k'')}\right] &= 0,\\ \label{02b}
\tr\left[\mathds{1}^{(j'')}\otimes A_\mathrm{Z}^{(k'')}\sigma_{junk}^{(j''k'')}\right] &= \alpha_z.
\end{align}

Next, we introduce the following identity, derived from the normalization condition for $\alpha_x,\alpha_z,\alpha_y$
\begin{equation*}
  1 = \alpha_z^2 + \alpha_x^2 + \alpha_y^2.
\end{equation*}
This identity allows us to bound the terms involving the traces of products of operators. We can write:
\begin{equation*}
1 = \left(\tr\left[\mathds{1}^{(j'')}\otimes A_\mathrm{Z}^{(k'')}\sigma_{junk}^{(j''k'')}\right]\right)^2 + \left(\tr\left[\mathds{1}^{(j'')}\otimes A_\mathrm{X}^{(k'')}\sigma_{junk}^{(j''k'')}\right]\right)^2 + \left(\tr\left[A_Y^{(j'')}\otimes A_\mathrm{Y}^{(k'')}\sigma_{junk}^{(j''k'')}\right]\right)^2 \,.
\end{equation*}
By substituting Eqs.~\eqref{12a}, \eqref{eq: A2Ba}, and \eqref{02b} into the expression above we get:
\begin{equation*}
1 \leq
\tr\left[\mathds{1}^{(j'')}\otimes {A_\mathrm{Z}^{(k'')}}^2\sigma_{junk}^{(j''k'')}\right] 
+ \tr\left[\mathds{1}^{(j'')}\otimes {A_\mathrm{X}^{(k'')}}^2\sigma_{junk}^{(j''k'')} \right] 
+ \tr\left[\mathds{1}^{(j'')}\otimes {A_\mathrm{Y}^{(k'')}}^2\sigma_{junk}^{(j''k'')}\right] \,.
\end{equation*}
Using the Cauchy-Schwarz inequality for operators, we obtain:
\begin{equation} 
 1 \leq \tr\left[\mathds{1}^{(j'')}\otimes ({A_\mathrm{Z}^{(k'')}}^2+{A_\mathrm{X}^{(k'')}}^2+{A_\mathrm{Y}^{(k'')}}^2)\sigma_{junk}^{(j''k'')}\right] = 1,
\end{equation}
where equality holds because of the identity $\sum_P{A_P^{(k'')}}^2 = \mathds{1}$. As a result, the inequalities involving Cauchy-Schwarz become equalities, which implies that the following conditions are satisfied:
\begin{align*}
\left(\tr\left[\mathds{1}^{(j'')}\otimes {A_\mathrm{Z}^{(k'')}}\sigma_{junk}^{(j''k'')}\right]\right)^2 &= \tr\left[\mathds{1}^{(j'')}\otimes {A_\mathrm{Z}^{(k'')}}^2\sigma_{junk}^{(j''k'')}\right], \\
\left(\tr\left[\mathds{1}^{(j'')}\otimes {A_\mathrm{X}^{(k'')}}\sigma_{junk}^{(j''k'')}\right]\right)^2 &= \tr\left[\mathds{1}^{(j'')}\otimes {A_\mathrm{X}^{(k'')}}^2\sigma_{junk}^{(j''k'')} \right], \\
\left(\tr\left[A_Y^{(j'')}\otimes {A_\mathrm{Y}^{(k'')}}\sigma_{junk}^{(j''k'')}\right]\right)^2 &= \tr\left[\mathds{1}^{(j'')}\otimes {A_\mathrm{Y}^{(k'')}}^2\sigma_{junk}^{(j''k'')}\right].
\end{align*}
The saturation of the Cauchy-Schwarz inequality implies the existence of scalars $a_Z, a_X$ and $a_Y$ such that
\begin{align}\nonumber
\mathds{1}^{(j'')}\otimes A_\mathrm{Z}^{(k'')}\sqrt{\sigma_{junk}^{(j''k'')}} = a_Z\sqrt{\sigma_{junk}^{(j''k'')}} \quad &\Rightarrow \quad \mathds{1}^{(j'')}\otimes A_\mathrm{Z}^{(k'')}\sigma_{junk}^{(j''k'')} = a_Z\sigma_{junk}^{(j''k'')}, \\ \nonumber
\mathds{1}^{(j'')}\otimes A_\mathrm{X}^{(k'')}\sqrt{\sigma_{junk}^{(j''k'')}} = a_X\sqrt{\sigma_{junk}^{(j''k'')}} \quad &\Rightarrow \quad \mathds{1}^{(j'')}\otimes A_\mathrm{X}^{(k'')}\sigma_{junk}^{(j''k'')} = a_X\sigma_{junk}^{(j''k'')}, \\ \label{implication}
A_\mathrm{Y}^{(j'')}\otimes A_\mathrm{Y}^{(k'')}\sqrt{\sigma_{junk}^{(j''k'')}} = a_Y\sqrt{\sigma_{junk}^{(j''k'')}} \quad &\Rightarrow \quad A_\mathrm{Y}^{(j'')}\otimes A_\mathrm{Y}^{(k'')}\sigma_{junk}^{(j''k'')} = a_Y\sigma_{junk}^{(j''k'')} .
\end{align}
From Eqs.~\eqref{12a}, ~\eqref{eq: A2Ba} and ~\eqref{02b}, it follows that
\begin{equation*}
a_Z = \alpha_z, \quad a_X = \alpha_x, \quad a_Y = \alpha_y.
\end{equation*}
Thus, on the support of $\tr_{j''}(\sigma_{junk}^{(j''k'')})$, $A_\mathrm{Z}^{(k'')}$ and $A_\mathrm{X}^{(k'')}$ act as $\alpha_z\mathds{1}$ and $\alpha_x\mathds{1}$, respectively. From $({A_\mathrm{Z}^{(k'')}}^2+{A_\mathrm{X}^{(k'')}}^2+{A_\mathrm{Y}^{(k'')}}^2)\sigma_{junk}^{(j''k'')} = \sigma_{junk}^{(j''k'')}$, it follows that ${A_\mathrm{Y}^{(k'')}}^2\sigma_{junk}^{(j''k'')} = \alpha_y^2\sigma_{junk}^{(j''k'')}$ . By rescaling ${A_\mathrm{Y}^{(k'')}} = \alpha_y\bar{A}_\mathrm{Y}^{(k'')}$, we finally get that on the support of $\tr_{j''}(\sigma_{junk}^{(j''k'')})$ the operator $A^{(k)}$ can be written as
\begin{equation*}
U^{(k)}A^{(k)}{U^{(k)}}^\dagger = \alpha_z \sz^{(k_q)}\otimes \mathds{1}^{(k'')} + \alpha_x \sx^{(k_q)} \otimes \mathds{1}^{(k'')} + \alpha_y \sy^{(k_q)}\otimes \bar{A}_\mathrm{Y}^{(k'')}.
\end{equation*}
From \eqref{implication} we obtain 
\begin{equation*}
  A_\mathrm{Y}^{(j'')}\otimes {\bar{A}_\mathrm{Y}}^{(k'')}\sigma_{junk}^{(j''k'')} = \sigma_{junk}^{(j''k'')} .
\end{equation*}

\subsection{A different form of the Measurement Lemma}\label{altML}

In some cases, it is necessary to use the measurement lemma to self-test an unknown measurement $A^{(j)}$ applied by party $(j)$, given that the following self-testing statements about the state and the measurements of party $(k)$ are available.
\begin{lemma}[Alternative measurement lemma]\label{lemma:measurementlemmaAlt}
Let $\physstate$ be a bipartite state and $A_0^{(k)}, A_1^{(k)}, A_2^{(k)}$ and $A_3^{(k)}$ dichotomic measurement observables. Suppose there exist local unitaries $U_j$ and $U_k$ such that:
\begin{equation}\label{en}
\begin{aligned}
  &\left(U^{(j)}\otimes U^{(k)}\right)\rho\left(U^{(j)}\otimes U^{(k)}\right)^{\dagger} = \psi_\theta^{(j_q,k_q)}\tp \sigma_{junk}^{(j'',k'')},\\
&   U^{(k)} A_{0}^{(k)}{U^{(k)}}^{\dagger} = \cos\left(\frac{\mu}{2}\right)\sz^{(k_q)}\tp \mathds{1}^{(k'')} + \sin\left(\frac{\mu}{2}\right)\sx^{(k_q)}\tp \mathds{1}^{(k'')}, \\ 
  & U^{(k)} A_{1}^{(k)}{U^{(k)}}^{\dagger} = \cos\left(\frac{\mu}{2}\right)\sz^{(k_q)}\tp \mathds{1}^{(k'')} - \sin\left(\frac{\mu}{2}\right)\sx^{(k_q)}\tp \mathds{1}^{(k'')},\\
   & U^{(k)} A_{2}^{(k)}{U^{(k)}}^{\dagger} = \cos\left(\frac{\mu}{2}\right)\sz^{(k_q)}\tp \mathds{1}^{(k'')} + \sin\left(\frac{\mu}{2}\right)\sy^{(k_q)}\tp A_{\mathrm{Y}}^{(k'')}, \\ 
&U^{(k)} A_{3}^{(k)}{U^{(k)}}^{\dagger} = \cos\left(\frac{\mu}{2}\right)\sz^{(k_q)}\tp \mathds{1}^{(k'')} - \sin\left(\frac{\mu}{2}\right)\sy^{(k_q)}\tp A_{\mathrm{Y}}^{(k'')},
\end{aligned}
\end{equation}
where $A_{\mathrm{Y}}$ is a Hermitian $\pm$-eigenvalue operator. Suppose moreover that the following correlations are observed
\begin{subequations}
\begin{align}
\label{sm02}
\tr \left[A^{(j)}\otimes A_0^{(k)}\rho\right] &= \alpha_z\cos{\mu} + \alpha_x\sin\mu\sin{2\theta}\,,\\ \label{sm12}
\tr \left[A^{(j)}\otimes A_1^{(k)}\rho\right] &= \alpha_z\cos{\mu} - \alpha_x\sin\mu\sin{2\theta} \,,\\
\tr \left[A^{(j)}\otimes A_2^{(k)}\rho\right] &= \alpha_z\cos{\mu} + \alpha_y\sin\mu\sin{2\theta} \label{eq: smA2B}\,,\\ 
\tr \left[A^{(j)}\otimes A_3^{(k)}\rho\right] &= \alpha_z\cos{\mu} - \alpha_y\sin\mu\sin{2\theta} \label{eq: smA3B}\,,\\
\label{sm22}
\tr \left[ A^{(j)}\tp\mathds{1}^{(k)}\rho\right] &= \cos{2\theta}\alpha_z\,.
\end{align}
\end{subequations}
for some real numbers $\alpha_x,\alpha_y$ and $\alpha_z$ such that $\alpha_z^2 + \alpha_x^2 + \alpha_y^2 = 1$. Then we have
\begin{equation*}
  U^{(j)}A^{(j)}{U^{(j)}}^{\dagger} = \alpha_z \sz^{(j_q)}\otimes \mathds{1}^{(j'')} + \alpha_x \sx^{(j_q)} \otimes \mathds{1}^{(j'')} + \alpha_y \sy^{(j_q)}\otimes \bar{C}^{(j'')}_\mathrm{Y},
\end{equation*}
where $\bar{C}_\mathrm{Y}$ is a Hermitian $\pm1$-eigenvalue operator. 
\end{lemma}

Analogously to the proof of Lemma \ref{lemma:measurementlemma}, we start the proof by writing $A^{(j)}$ as follows:
\begin{equation*}
U^{(j)}A^{(j)}{U^{(j)}}^{\dagger} = \sum_{P}P^{(j_q)}\otimes C_P^{(j'')},
\end{equation*}
where $\{P\}$ is the basis of single-qubit Pauli operators acting on qubits, while $C_P$ are Hermitian operators acting on $\pazocal{H}_{j''}$. The fact that ${A^{(j)}}^2 = \mathds{1}$ implies
\begin{equation*}
\sum_P {C_P^{(j'')}}^2 = \mathds{1}.
\end{equation*}
Given Eqs.~\eqref{en}, the relations \eqref{sm02}-\eqref{sm22} can be rewritten in the following form 
\begin{subequations}
  \begin{align}
\nonumber
\cos{\mu}\cos{2\theta}\tr \left[C_{\mathds{1}}^{(j'')}\tp\mathds{1}^{(k'')}\sigma_{junk}^{(j'',k'')}\right] &+ \cos{\mu}\tr \left[C_{\mathrm{Z}}^{(j'')}\tp\mathds{1}^{(k'')}\sigma_{junk}^{(j'',k'')}\right] + \\ \label{tt}&+ \sin{\mu}\sin{2\theta}\tr \left[C_{\mathrm{X}}^{(j'')}\tp\mathds{1}^{(k'')}\sigma_{junk}^{(j'',k'')}\right] = \alpha_z\cos{\mu} + \alpha_x\sin\mu\sin{2\theta}\,,\\ \nonumber
\cos{\mu}\cos{2\theta}\tr \left[C_{\mathds{1}}^{(j'')}\tp\mathds{1}^{(k'')}\sigma_{junk}^{(j'',k'')}\right] &+ \cos{\mu}\tr \left[C_{\mathrm{Z}}^{(j'')}\tp\mathds{1}^{(k'')}\sigma_{junk}^{(j'',k'')}\right] - \\ \label{gg}&-\sin{\mu}\sin{2\theta}\tr \left[C_{\mathrm{X}}^{(j'')}\tp\mathds{1}^{(k'')}\sigma_{junk}^{(j'',k'')}\right] = \alpha_z\cos{\mu} - \alpha_x\sin\mu\sin{2\theta}\,, \\
\cos{\mu}\cos{2\theta}\tr \left[C_{\mathds{1}}^{(j'')}\tp\mathds{1}^{(k'')}\sigma_{junk}^{(j'',k'')}\right] &+ \cos{\mu}\tr \left[C_{\mathrm{Z}}^{(j'')}\tp\mathds{1}^{(k'')}\sigma_{junk}^{(j'',k'')}\right] + \nonumber \\ &+ \sin{\mu}\sin{2\theta}\tr\left[C_{\mathrm{Y}}^{(j'')}\tp A_\mathrm{Y}^{(k'')}\sigma_{junk}^{(j'',k'')}\right] = \alpha_z\cos{\mu} + \alpha_y\sin\mu\sin{2\theta}\,, \label{vv}\\ 
\cos{\mu}\cos{2\theta}\tr \left[C_{\mathds{1}}^{(j'')}\tp\mathds{1}^{(k'')}\sigma_{junk}^{(j'',k'')}\right] &+ \cos{\mu}\tr \left[C_{\mathrm{Z}}^{(j'')}\tp\mathds{1}^{(k'')}\sigma_{junk}^{(j'',k'')}\right] -\nonumber \\ &- \sin{\mu}\sin{2\theta}\tr\left[C_{\mathrm{Y}}^{(j'')}\tp A_\mathrm{Y}^{(k'')}\sigma_{junk}^{(j'',k'')}\right] = \alpha_z\cos{\mu} - \alpha_y\sin\mu\sin{2\theta} \,,\label{ll}\\
\label{ii}
\tr\left[C_{\mathds{1}}^{(j'')}\tp\mathds{1}^{(k'')}\sigma_{junk}^{(j'',k'')}\right] &+ \cos{2\theta}\tr \left[C_{\mathrm{Z}}^{(j'')}\tp\mathds{1}^{(k'')}\sigma_{junk}^{(j'',k'')}\right] = \cos{2\theta}\alpha_z\,.
\end{align}
\end{subequations}

From \eqref{tt} and \eqref{gg} we obtain
\begin{equation*}
  \tr\left[C_{\mathrm{X}}^{(j'')}\mathds{1}^{(k'')}\sigma_{junk}^{(j'',k'')}\right] = \alpha_x,
\end{equation*}
and, analogously, from \eqref{vv} and \eqref{ll} we get
\begin{equation*}
  \tr\left[C_{\mathrm{Y}}^{(j'')}A_\mathrm{Y}^{(k'')}\sigma_{junk}^{(j'',k'')}\right] = \alpha_y.
\end{equation*}
By manipulating the remaining equations we can obtain
\begin{align*}
  \tr\left[C_{\mathds{1}}^{(j'')}\mathds{1}^{(k'')}\sigma_{junk}^{(j'',k'')}\right] &= 0,\\
  \tr \left[C_{\mathrm{Z}}^{(j'')}\mathds{1}^{(k'')}\sigma_{junk}^{(j'',k'')}\right] &= \alpha_z.
\end{align*}
With the last four equations we can do exactly the same analysis as we did in the proof of Lemma \ref{lemma:measurementlemma}, and conclude that on the support of $\tr_{k''}(\sigma_{junk}^{(j''k'')})$, the operator $A^{(j)}$ can be written as
\begin{equation*}
  U^{(j)}A^{(j)}{U^{(j)}}^{\dagger} = \alpha_z \sz^{(j_q)}\otimes \mathds{1}^{(j'')} + \alpha_x \sx^{(j_q)} \otimes \mathds{1}^{(j'')} + \alpha_y \sy^{(j_q)}\otimes \bar{C}_\mathrm{Y}^{(j'')},
\end{equation*}
where $\bar{C}_\mathrm{Y}^{(j'')}\otimes {A}_\mathrm{Y}^{(k'')}\sigma_{junk}^{(j''k'')} = \sigma_{junk}^{(j''k'')} $.

\section{Tripartite scenario\label{app:tripartite}}

\setcounter{equation}{0}
\makeatletter
\renewcommand{\theequation}{C\arabic{equation}}

 In this appendix, we provide a proof of how to self-test any tripartite quantum state shared among Alice, Bob, and Charlie. The shared state is assumed to have the form:
\begin{equation*} \ket{\Psi} = \sum_{i,j,k = 0}^1 \lambda_{ijk}\ket{ijk}.\end{equation*}

To facilitate the self-testing procedure, we define the grouped correlations associated with the self-testing results presented in Appendices~\ref{app:proof1}, \ref{appML}, and~\ref{altML}. For $D \in \{A,B,C\}$, we introduce a measurement observable $D_\lozenge$ with its corresponding projective measurements given by $D_\lozenge^\pm = (\idd\pm D_\lozenge)/2$.

\begin{tcolorbox}[colback=violet!5!white,colframe=violet,title=\textbf{Substate self-testing correlations $\pazocal{C}_{st}$}] 
To streamline the presentation, we introduce the shorthand notation $\pazocal{C}_{st}\left(\cdot\right)$ for substate self-testing correlations. This notation has a single argument, representing the subnormalized post-measurement state. The correlations are derived from the self-testing procedure defined in Lemma~\eqref{lemma:selftesting}. In this framework, writing: \begin{equation*}
\pazocal{C}_{st}(D_\lozenge^{\pm}\rho) = (\Lambda,\alpha)
\end{equation*} is equivalent to the following conditions: \begin{align*}
\tr\left[D_\lozenge^{\pm}\rho^{\A\B\C}\right] = \Lambda,&\qquad
  I_{{\alpha}}\left(\frac{\tr_D\left[D_\lozenge^\pm\rho^{\A\B\C}\right]}{\tr\left[D_\lozenge^{\pm}\rho^{\A\B\C}\right]}\right) = 2\sqrt{2}\sqrt{1+{\alpha}^2/4}\,,\\ 
  J_{{\alpha}}\left(\frac{\tr_D\left[D_\lozenge^\pm\rho^{\A\B\C}\right]}{\tr\left[D_\lozenge^{\pm}\rho^{\A\B\C}\right]}\right) = 2\sqrt{2}\sqrt{1+{\alpha}^2/4},&\qquad \label{eq:d0pL}
  L\left(\frac{\tr_D\left[D_\lozenge^\pm\rho^{\A\B\C}\right]}{\tr\left[D_\lozenge^{\pm}\rho^{\A\B\C}\right]}\right) = 2\sqrt{2}\sin{\theta},
\end{align*}
where the parameter $\alpha$ is defined as $\alpha = 2\cos(2\theta)/\sqrt{1+\sin^2(2\theta)}$. 
  \end{tcolorbox}
This formulation explicitly links the norm of the post-measurement state, 
$\Lambda$, to its Schmidt decomposition and related metrics, such as $I_\alpha$, $J_\alpha$ and $L$.

To facilitate the application of Lemma~\ref{lemma:measurementlemma} in our proof, we define measurement self-testing correlations as follows:

\begin{tcolorbox}[colback=violet!5!white,colframe=violet,title=\textbf{Measurement self-testing correlations $\pazocal{C}_{mst}$}] 
The shorthand notation $\pazocal{C}_{mst}\left(\cdot\right)$ represents measurement self-testing correlations. It has six arguments: 
\begin{itemize}
  \item The projector defining the post-measurement subnormalized state,
  \item The unknown measurement to be self-tested,
  \item Three auxiliary measurements involved in Lemma~\ref{lemma:measurementlemma}, and
  \item The quantum state on which the correlations are evaluated.
\end{itemize}
The equation 
\begin{equation*} \pazocal{C}_{mst}\left(A_\lozenge^\pm,B_\lozenge,C_1,C_2,C_3,\rho\right) = (\phi,\alpha_z,\alpha_x,\alpha_y) 
\end{equation*}
is equivalent to the following set of correlations:
\begin{align*}
  \frac{\tr\left[A_\lozenge^\pm \tp B_\lozenge \rho^{\A\B\C}\right]}{\tr\left[A_\lozenge^\pm \rho^{\A\B\C}\right]} = \alpha_z\cos{2{\phi}}, &\qquad
  \frac{\tr\left[A_\lozenge^\pm\tp B_\lozenge\tp C_1\rho^{\A\B\C}\right]}{\tr\left[A_\lozenge^\pm \rho^{\A\B\C}\right]} = \alpha_z \,,\\ 
  \frac{\tr\left[A_\lozenge^\pm\tp B_\lozenge\tp C_2\rho^{\A\B\C}\right]}{\tr\left[A_\lozenge^\pm \rho^{\A\B\C}\right]} = \alpha_x\sin 2\phi, &\qquad 
  \frac{\tr\left[A_\lozenge^\pm\tp B_\lozenge\tp C_3\rho^{\A\B\C}\right]}{\tr\left[A_\lozenge^\pm \rho^{\A\B\C}\right]} = \alpha_y\sin 2\phi\,.
\end{align*}

\end{tcolorbox}
In this framework, the measurement self-testing correlations are used to verify Bob’s measurement $B_\lozenge$ by examining its correlations with Charlie’s three measurements, $C_1,C_2$ and $C_3$. These correlations are evaluated on the post-measurement state prepared by Alice through her measurement $A_\lozenge$. 
Similarly, by exchanging the roles of the parties, analogous correlations can be defined for other scenarios:
\begin{align*}
\pazocal{C}_{mst}\left(A_\lozenge^\pm,C_\lozenge,B_1,B_2,B_3,\rho\right) &= (\phi,\alpha_z,\alpha_x,\alpha_y)\,,\\
\pazocal{C}_{mst}\left(B_\lozenge^\pm,A_\lozenge,C_1,C_2,C_3,\rho\right) &= (\phi,\alpha_z,\alpha_x,\alpha_y)\,,\\
\pazocal{C}_{mst}\left(B_\lozenge^\pm,C_\lozenge,A_1,A_2,A_3,\rho\right) &= (\phi,\alpha_z,\alpha_x,\alpha_y)\,,\\
\pazocal{C}_{mst}\left(C_\lozenge^\pm,A_\lozenge,B_1,B_2,B_3,\rho\right) &= (\phi,\alpha_z,\alpha_x,\alpha_y)\,.
\end{align*}
In the equations above, the order of operators within the brackets is critical:
\begin{itemize}
\item The first operator (the projector) defines the post-measurement state.
\item The second operator represents the unknown measurement to be self-tested.
\item The final three operators correspond to auxiliary measurements, typically self-tested Pauli observables.
\end{itemize}
This structure ensures the self-testing procedure is robust and consistent across different scenarios.

To prepare for the use of Lemma~\ref{lemma:measurementlemmaAlt}, we define alternative measurement self-testing correlations as follows:
\begin{tcolorbox}[colback=violet!5!white,colframe=violet,title=\textbf{Alternative measurement self-testing correlations $\pazocal{C}_{amst}$}] 
The shorthand notation $\pazocal{C}_{amst}\left(\cdot\right)$ represents alternative measurement self-testing correlations. It includes seven arguments:\begin{itemize}
\item The projector defining the post-measurement subnormalized state,
\item The unknown measurement to be self-tested,
\item Four auxiliary measurements involved in Lemma~\ref{lemma:measurementlemmaAlt}, and
\item The quantum state on which the correlations are evaluated.
\end{itemize}
The equation
\begin{equation*}
\pazocal{C}_{amst}\left(A_\lozenge^\pm,C_\lozenge,B_1,B_2,B_3,B_4,\rho\right) = (\phi,\mu,\alpha_z,\alpha_x,\alpha_y)
\end{equation*}
is equivalent to the following set of correlations
\begin{align*}
  \frac{\tr\left[A_\lozenge^\pm\tp C_\lozenge\rho^{\A\B\C}\right]}{\tr\left[A_\lozenge^\pm \rho^{\A\B\C}\right]} &= \alpha_z\cos{2{\phi}}\,, \\ 
\frac{\tr\left[A_\lozenge^\pm\tp B_1\tp C_\lozenge\rho^{\A\B\C}\right]}{\tr\left[A_\lozenge^\pm \rho^{\A\B\C}\right]} &= \alpha_z\cos{\mu} + \alpha_x\sin{{\mu}}\sin{2{\phi}}\,,\\
  \frac{\tr\left[A_\lozenge^\pm\tp B_2\tp C_\lozenge\rho^{\A\B\C}\right]}{\tr\left[A_\lozenge^\pm \rho^{\A\B\C}\right]} &= \alpha_z\cos{\mu} - 
\alpha_x\sin{{\mu}}\sin{2{\phi}}\,,\\ 
  \frac{\tr\left[A_\lozenge^\pm\tp B_3\tp C_\lozenge\rho^{\A\B\C}\right]}{\tr\left[A_\lozenge^\pm \rho^{\A\B\C}\right]} &= \alpha_z\cos{\mu} + \alpha_y\sin{{\mu}}\sin{2{\phi}}\,,\\ 
  \frac{\tr\left[A_\lozenge^\pm\tp B_4\tp C_\lozenge\rho^{\A\B\C}\right]}{\tr\left[A_\lozenge^\pm \rho^{\A\B\C}\right]} &= \alpha_z\cos{\mu} - \alpha_y\sin{{\mu}}\sin{2{\phi}}\,.
\end{align*}
\end{tcolorbox}
In this framework, the alternative measurement self-testing correlations are used to self-test Charlie’s measurement $C_\lozenge$, by evaluating its correlations with four of Bob’s measurements, $B_1, B_2, B_3, B_4$. These correlations are determined on the state effectively prepared for Bob and Charlie by Alice’s measurement $A_\lozenge$. 
Analogous definitions can be constructed by exchanging the roles of the parties, as follows:
\begin{align*}
\pazocal{C}_{amst}\left(A_\lozenge^\pm,B_\lozenge,C_1,C_2,C_3,C_4,\rho\right) &= (\phi,\mu,\alpha_z,\alpha_x,\alpha_y)\,,\\
\pazocal{C}_{amst}\left(B_\lozenge^\pm,C_\lozenge,A_1,A_2,A_3,A_4,\rho\right) &= (\phi,\mu,\alpha_z,\alpha_x,\alpha_y)\,,\\
\pazocal{C}_{amst}\left(A_\lozenge^\pm,B_\lozenge,C_1,C_2,C_3,C_4,\rho\right) &= (\phi,\mu,\alpha_z,\alpha_x,\alpha_y)\,,\\
\pazocal{C}_{amst}\left(C_\lozenge^\pm,B_\lozenge,A_1,A_2,A_3,A_4,\rho\right) &= (\phi,\mu,\alpha_z,\alpha_x,\alpha_y)\,.
\end{align*}

\subsection{The first set of sub-tests}\label{sec:a0plus}

Consider a tripartite quantum state $\ket{\Psi} = \sum_{ijk}\lambda_{ijk}\ket{ijk}$ shared between three parties: Alice, Bob, and Charlie. Suppose Alice performs a measurement $A_\lozenge={\ketbra{0}{0},\ketbra{1}{1}}$, on her subsystem. The post-measurement states of Bob and Charlie, conditioned on Alice's measurement outcomes $0$ and $1$ are given by $\ket{\psi_{a^+}}^{\B'\C'} = \frac{A_\lozenge^+\ket{\psi}}{\bra{\psi}A_\lozenge^+\ket{\psi}}$ and $\ket{\psi_{a^-}}^{\B'\C'} = \frac{A_\lozenge^-\ket{\psi}}{\bra{\psi}A_\lozenge^-\ket{\psi}}$ where the operators $A_\lozenge^\pm$ are defined as $\frac{\mathds{1} \pm A_\lozenge}{2}$. Expressing these states in the corresponding Schmidt bases, we find their forms to be:
\begin{align}
  \label{SchmidtA0}
  \ket{{\Psi}_{a^+}}^{\B'\C'} &= \cos{\phi}_{a^+}\ket{{0}_{a^+}{0}_{a^+}} + \sin{\phi}_{a^+}\ket{{1}_{a^+}{1}_{a^+}}\,,\\
  \label{SchmidtA1}
  \ket{{\Psi}_{a^-}}^{\B'\C'} &= \cos{\phi}_{a^-}\ket{{0}_{a^-}{0}_{a^-}} + \sin{\phi}_{a^-}\ket{{1}_{a^-}{1}_{a^-}}\,,
\end{align}
where the Schmidt bases are related to the computational basis through the unitary operators 
$V_{a^\pm}$ and $W_{a^\pm}$, such that
\begin{align*} \ket{0_{a^+}}^{\B'} = V_{a^+}\ket{0}^{\B'}, \quad \ket{0_{a^+}}^{\C'} = W_{a^+}\ket{0}^{\C'}, \quad \ket{0_{a^-}}^{\B'} = V_{a^-}\ket{0}^{\B'}, \quad \ket{0_{a^-}}^{\C'} = W_{a^-}\ket{0}^{\C'}. \end{align*}
The unitary transformations $V_{a^\pm}$ and $W_{a^\pm}$ are determined by
\begin{align*} V_{a^+} \otimes W_{a^+} \left(\sum_{j,k}\lambda_{0jk}|jk\rangle\right) &= |\Psi_{a^+}\rangle^{\B'C'},\\ V_{a^-} \otimes W_{a^-} \left(\sum_{j,k}\lambda_{1jk}|jk\rangle\right) &= |\Psi_{a^-}\rangle^{\B'C'}. \end{align*}

The aim in the first set of sub-tests is to certify that by performing the measurement corresponding to the observable $A_\lozenge$ and obtaining the output $0$, the state is equivalent to \eqref{SchmidtA0}. The correlations necessary for the self-test are:
\begin{subequations}
\begin{align}\label{eq:a0pstate}
  \pazocal{C}_{st}(A_\lozenge^+,\rho) &= (\Lambda_0,\alpha_{a^+}),\\ \label{eq:a0pstmeas} 
  \pazocal{C}_{mst}(A_\lozenge^+,B_\lozenge,C_1,C_2,C_3) &= (\phi_{a^+},\upsilon^{a+}_z,\upsilon^{a+}_x,\upsilon^{a+}_y),\\ 
  \label{eq:a0pXstmeas} 
  \pazocal{C}_{mst}(A_\lozenge^+,B_\blacklozenge,C_1,C_2,C_3) &= (\phi_{a^+},\eta^{a+}_z,\eta^{a+}_x,\eta^{a+}_y),\\ 
  \label{eq:a0paltstmeas}
  \pazocal{C}_{qmst}(A_\lozenge^+,C_\lozenge,B_1,B_2,B_3,B_4) &= (\phi_{a^+},\mu_{a^+},\omega^{a+}_z,\omega^{a+}_x,\omega^{a+}_y),\\ 
  \label{eq:a0paltXstmeas}
  \pazocal{C}_{qmst}(A_\lozenge^+,C_\blacklozenge,B_1,B_2,B_3,B_4) &= (\phi_{a^+},\mu_{a^+},\delta^{a+}_z,\delta^{a+}_x,\delta^{a+}_y),
\end{align}
\end{subequations}
where $\Lambda_0 = \sqrt{\la_{000}^2 + \la_{001}^2 + \la_{010}^2 + \la_{011}^2}$, $\sin2{\phi}_{a^+} = \sqrt{(1-\alpha_{a^+}^2/4)/(1+\alpha_{a^+}^2/4)}$ 
and
$\sin{{\mu}_{a^+}/2} = \sqrt{(1-\alpha_{a^+}^2/4)/2}$. The correlations given in~\eqref{eq:a0pstate} self-test that Bob and Charlie have the correct state after Alice measures $A_\lozenge$ and obtains the outcome $0$. All remaining correlations are evaluated on the same state, conditioned on Alice's outcome $0$. Specifically:\begin{itemize}
  \item Correlations~\eqref{eq:a0pstmeas} and~\eqref{eq:a0pXstmeas} self-test Bob's measurements $B_\lozenge$ and $B_\blacklozenge$ which ideally correspond to $\sz$ and $\sx$ respectively.
  \item Correlations ~\eqref{eq:a0paltstmeas} and~\eqref{eq:a0paltXstmeas} self-test Charlie's measurements $C_\lozenge$ and $C_\blacklozenge$ which ideally correspond to $\sz$ and $\sx$ respectively.
\end{itemize}
The remaining parameters appearing in Eqs.~\eqref{eq:a0pstate}-\eqref{eq:a0paltXstmeas} are such that:
\begin{align}\label{eq:ub}
V_{a^+}^\dagger\sz V_{a^+} &= \upsilon^{a+}_z\sz + \upsilon^{a+}_x\sx + \upsilon^{a+}_y\sy,\\ \label{eq:ubx}
V_{a^+}^\dagger\sx V_{a^+} &= \eta^{a+}_z\sz + \eta^{a+}_x\sx + \eta^{a+}_y\sy,\\ \label{eq:uc}
  W_{a^+}^\dagger\sz W_{a^+} &= \omega^{a+}_z\sz + \omega^{a+}_x\sx + \omega^{a+}_y\sy,\\ \label{eq:ucX}
  W_{a^+}^\dagger\sx W_{a^+} &= \delta^{a+}_z\sz + \delta^{a+}_x\sx + \delta^{a+}_y\sy.
\end{align}
By taking the transpose of \eqref{eq:ub}-\eqref{eq:ucX} we find 
\begin{align}\label{eq:ub*}
  \left(V_{a^+}^*\right)^\dagger\sz V_{a^+}^* &= \upsilon^{a+}_z\sz + \upsilon^{a+}_x\sx - \upsilon^{a+}_y\sy,\\ \label{eq:ubX*}
  \left(V_{a^+}^*\right)^\dagger\sx V_{a^+}^* &= \eta^{a+}_z\sz + \eta^{a+}_x\sx - \eta^{a+}_y\sy,\\
\label{eq:uc*}
  \left(W_{a^+}^*\right)^\dagger\sz W_{a^+}^* &= \omega^{a+}_z\sz + \omega^{a+}_x\sx - \omega^{a+}_y\sy,\\
\label{eq:ucX*}
  \left(W_{a^+}^*\right)^\dagger\sx W_{a^+}^* &= \delta^{a+}_z\sz + \delta^{a+}_x\sx - \delta^{a+}_y\sy\,.
\end{align}

According to Lemma \ref{lemma:selftesting}, the correlations \eqref{eq:a0pstate} ensure the existence of unitaries $U^\B_{a_{+}}$ and $U^\C_{a_{+}}$ such that 
\begin{multline}\label{eq:state0}
 \left(U^\B_{a_{+}}\otimes U^\C_{a_{+}}\right)\tr_{A}(A_\lozenge^+\rho^{\A\B\C}) \left(U^\B_{a_{+}}\otimes U^\C_{a_{+}}\right)^{\dagger} \varpropto \\
 \varpropto \left(\cos{\phi}_{a^+}\ket{{0}_{a^+}{0}_{a^+}}^{\B_q\C_q} + \sin{\phi_{a^+}}\ket{{1_{a^+}}{1_{a^+}}}^{\B_q\C_q}\right)\left(\cos{\phi_{a^+}}\bra{{0_{a^+}}{0_{a^+}}}^{\B_q\C_q} + \sin{\phi_{a^+}}\bra{{1_{a^+}}{1_{a^+}}}^{\B_q\C_q}\right)\ot\xi_{a^+}^{\B''C''} \,,
\end{multline}
and 
\begin{align*}
  U^\C_{a_{+}}C_1{U^\C_{a_{+}}}^{\dagger} &= {\sigma}_{\textrm{z}{a^+}}^{\C_q} \otimes \mathds{1}^{\C''}\,, \\
  U^\C_{a_{+}}C_2{U^\C_{a_{+}}}^{\dagger} &= {\sigma}_{\textrm{x}{a^+}}^{\C_q} \otimes \mathds{1}^{\C''} \,,\\
  U^\C_{a_{+}}C_3{U^\C_{a_{+}}}^{\dagger} &= {\sigma}_{\textrm{y}{a^+}}^{\C_q} \otimes {C}_{Y_{a^+}}^{\C''},
\end{align*}
where notation ${\sigma}_{\textrm{z}{a^+}} = \ketbra{0_{a^+}}{0_{a^+}} - \ketbra{1_{a^+}}{1_{a^+}}$, ${\sigma}_{\textrm{x}{a^+}} = \ketbra{0_{a^+}}{1_{a^+}} + \ketbra{1_{a^+}}{0_{a^+}}$ and ${\sigma}_{\textrm{y}{a^+}} = i{\sigma}_{\textrm{x}{a^+}}{\sigma}_{\textrm{z}{a^+}}$ is introduced in the main text. It is important to note here that self-testing unitaries $U^\B_{a_+}$ and $U^\C_{a_+}$ act only on the support of $\tr_{\A\C}[A_\lozenge^+\rho^{\A\B\C}]$ and $\tr_{\A\B}[A_\lozenge^+\rho^{\A\B\C}]$, respectively. This will hold for all statements in the remaining appendices which deal with the way self-testing unitaries act on the measurement observables. By applying Lemma~\ref{lemma:measurementlemma} to correlations \eqref{eq:a0pstmeas} and \eqref{eq:a0paltXstmeas} we self-test Bob's measurements $B_\lozenge$ and $B_\blacklozenge$ on the support of $\tr_{\A\C}[A_\lozenge^+\rho^{\A\B\C}]$:
\begin{align*}
  U^\B_{a_{+}}B_\lozenge{U^\B_{a_{+}}}^{\dagger} &= \upsilon^{a+}_x{\sigma}_{\textrm{x}{a^+}}^{\B_q} \otimes \mathds{1}^{\B''} + \upsilon^{a+}_z{\sigma}_{\textrm{z}{a^+}}^{\B_q} \otimes \mathds{1}^{\B''} + \upsilon^{a+}_y{\sigma}_{\textrm{y}{a^+}}^{\B_q} \otimes {B}_{Ya^+}^{\B''}\,, \\
  U^\B_{a_{+}}B_\blacklozenge{U^\B_{a_{+}}}^{\dagger} &= \eta^{a+}_x{\sigma}_{\textrm{x}{a^+}}^{\B_q} \otimes \mathds{1}^{\B''} + \eta^{a+}_z{\sigma}_{\textrm{z}^{\B_q}{a^+}} \otimes \mathds{1}^{\B''} + \eta^{a+}_y{\sigma}_{\textrm{y}{a^+}}^{\B_q} \otimes {\tilde{B}_{Ya^+}}^{\B''}\,,
\end{align*}
where ${B}_{Ya^+}$ and $\tilde{B}_{Ya^+}$ are $\pm1$-eigenvalue operators, such that
\begin{align}\label{eq:bcy}
{B}^{\B''}_{Ya^+}{C}^{\C''}_{Ya^+}\xi_{a+}^{\B''\C''}= \xi_{a+}^{\B''\C''} &\qquad \Rightarrow \qquad {B}^{\B''}_{Ya^+}\xi_{a+}^{\B''\C''} = {C}^{\C''}_{Ya^+}\xi_{a+}^{\B''\C''},\\ \label{eq:bcy'}
\tilde{B}^{\B''}_{Ya^+}{C}^{\C''}_{Ya^+}\xi_{a+}^{\B''\C''} = \xi_{a+}^{\B''\C''}  &\qquad \Rightarrow \qquad \tilde{B}^{\B''}_{Ya^+}\xi_{a+}^{\B''\C''} = {C}^{\C''}_{Ya^+}\xi_{a+}^{\B''\C''}.
\end{align}
We can further detail the form of the measurements $B_\lozenge$ on the support of $\tr_{AC}[A_\lozenge^+\rho^{\A\B\C}]$. Using Eqs.~\eqref{eq:ub} and~\eqref{eq:ub*}, 
\begin{align}\nonumber
  U^\B_{a_{+}}B_\lozenge{U^\B_{a_{+}}}^{\dagger} &= (\upsilon^{a+}_x{\sigma}_{\textrm{x}{a^+}} + \upsilon^{a+}_z{\sigma}_{\textrm{z}{a^+}} + \upsilon^{a+}_y{\sigma}_{\textrm{y}{a^+}})^{\B_q} \otimes \left(\frac{\mathds{1}+{B}_{Ya^+}}{2}\right)^{\B''} + \\ \nonumber &\qquad\qquad\qquad\qquad + (\upsilon^{a+}_x{\sigma}_{\textrm{x}{a^+}} + \upsilon^{a+}_z{\sigma}_{\textrm{z}{a^+}} - \upsilon^{a+}_y{\sigma}_{\textrm{y}{a^+}})^{\B_q} \otimes \left(\frac{\mathds{1}-{B}_{Ya^+}}{2}\right)^{\B''}\\ \label{eq:b0final}
  &= \left(V_{a^+}^\dagger{\sigma}_{\textrm{Z}{a^+}}V_{a^+}\right)^{\B_q} \otimes \left(\frac{\mathds{1}+{B}_{Ya^+}}{2}\right)^{\B''} + \left({V_{a^+}^*}^\dagger{\sigma}_{\textrm{Z}{a^+}}V_{a^+}^*\right)^{\B_q} \otimes \left(\frac{\mathds{1}-{B}_{Ya^+}}{2}\right)^{\B''}\,.
  \end{align}
  Similarly using Eqs.~\eqref{eq:ubx} and \eqref{eq:ubX*} to express the observable $B_\blacklozenge$ on the support of $\tr_{AC}[A_\lozenge^+\rho^{\A\B\C}]$:
\begin{align}  \nonumber
U^\B_{a_{+}}B_\blacklozenge{U^\B_{a_{+}}}^{\dagger} &= (\eta^{a+}_x{\sigma}_{\textrm{x}{a^+}} + \eta^{a+}_z{\sigma}_{\textrm{z}{a^+}} + \eta^{a+}_y{\sigma}_{\textrm{y}{a^+}})^{\B_q} \otimes \left(\frac{\mathds{1}+\tilde{B}_{Ya^+}}{2}\right)^{\B''} + \\ \nonumber &\qquad\qquad\qquad\qquad + (\eta^{a+}_x{\sigma}_{\textrm{x}{a^+}} + \eta^{a+}_z{\sigma}_{\textrm{z}{a^+}} - \eta^{a+}_y{\sigma}_{\textrm{y}{a^+}})^{\B_q} \otimes \left(\frac{\mathds{1}-\tilde{B}_{Ya^+}}{2}\right)^{\B''}\\ \label{eq:b0Xfinal}
  &= \left(V_{a^+}^\dagger{\sigma}_{\textrm{x}{a^+}}V_{a^+}\right)^{\B_q} \otimes \left(\frac{\mathds{1}+\tilde{B}_{Ya^+}}{2}\right)^{\B''} + \left({V_{a^+}^*}^\dagger{\sigma}_{\textrm{x}{a^+}}V_{a^+}^*\right)^{\B_q} \otimes \left(\frac{\mathds{1}-\tilde{B}_{Ya^+}}{2}\right)^{\B''}\,.
\end{align}
Similarly, we can use correlations \eqref{eq:a0paltstmeas} to self-test the additional measurement on Charlie's side $C_\lozenge$ on the support of $\tr_{AB}(A_\lozenge^+\rho^{\A\B\C})$ by using Lemma~\ref{lemma:measurementlemmaAlt}:
\begin{align}\nonumber
  U^\C_{a_{+}}C_\lozenge{U^\C_{a_{+}}}^{\dagger} &= (\omega^{a+}_x{\sigma}_{\textrm{x}{a^+}} + \omega^{a+}_z{\sigma}_{\textrm{z}{a^+}} + \omega^{a+}_y{\sigma}_{\textrm{y}{a^+}})^{\C_q} \otimes \left(\frac{\mathds{1}+\tilde{C}_{Ya^+}}{2}\right)^{\C''} + \\ \nonumber &\qquad\qquad\qquad\qquad + (\omega^{a+}_x{\sigma}_{\textrm{x}{a^+}} + \omega^{a+}_z{\sigma}_{\textrm{z}{a^+}} - \omega^{a+}_y{\sigma}_{\textrm{y}{a^+}})^{\C_q} \otimes \left(\frac{\mathds{1}-\tilde{C}_{Ya^+}}{2}\right)^{\C''}\\ \label{eq:c6final}
  &= \left(W_{a^+}^\dagger{\sigma}_{\textrm{z}{a^+}}W_{a^+}\right)^{\C_q} \otimes \left(\frac{\mathds{1}+\tilde{C}_{Ya^+}}{2}\right)^{\C''} + \left({W_{a^+}^*}^\dagger{\sigma}_{\textrm{z}{a^+}}W_{a^+}^*\right)^{\C_q} \otimes \left(\frac{\mathds{1}-\tilde{C}_{Ya^+}}{2}\right)^{\C''}\,,\end{align}
and correlations \eqref{eq:a0paltXstmeas} to self-test $C_\blacklozenge$:
\begin{align}  
  \nonumber
U^\C_{a_{+}}C_\blacklozenge{U^\C_{a_{+}}}^{\dagger} &= (\delta^{a+}_x{\sigma}_{\textrm{x}{a^+}} + \delta^{a+}_z{\sigma}_{\textrm{z}{a^+}} + \delta^{a+}_y{\sigma}_{\textrm{y}{a^+}})^{\C_q} \otimes \left(\frac{\mathds{1}+\bar{C}_{Ya^+}}{2}\right)^{\C''} + \\ \nonumber &\qquad\qquad\qquad\qquad + (\delta^{a+}_x{\sigma}_{\textrm{x}{a^+}} + \delta^{a+}_z{\sigma}_{\textrm{z}{a^+}} - \delta^{a+}_y{\sigma}_{\textrm{y}{a^+}})^{\C_q} \otimes \left(\frac{\mathds{1}-\bar{C}_{Ya^+}}{2}\right)^{\C''}\\ \label{eq:c6Xfinal}
  &= \left(W_{a^+}^\dagger{\sigma}_{\textrm{x}{a^+}}W_{a^+}\right)^{\C_q} \otimes \left(\frac{\mathds{1}+\bar{C}_{Ya^+}}{2}\right)^{\C''} + \left({W_{a^+}^*}^\dagger{\sigma}_{\textrm{x}{a^+}}W_{a^+}^*\right)^{\C_q} \otimes \left(\frac{\mathds{1}-\bar{C}_{Ya^+}}{2}\right)^{\C''}\,,
\end{align}
where
\begin{align}
  \label{baker}{B}_{Ya^+}^{\B''}\tp\tilde{C}_{Ya^+}^{\C''}\xi_{Ya+}^{\B''\C''} = \xi_{Ya+}^{\B''\C''} \quad \Rightarrow \quad {B}_{Ya^+}^{\B''}\xi_{Ya+}^{\B''\C''} = \tilde{C}_{Ya^+}^{\C''}\xi_{Ya+}^{\B''\C''} \quad \Rightarrow \quad \tilde{C}_{Ya^+}^{\C''}\xi_{Ya+}^{\B''\C''} = {C}_{Ya^+}^{\C''}\xi_{Ya+}^{\B''\C''},\\ \label{kantemir}
{B}_{Ya^+}^{\B''}\tp\bar{C}_{Ya^+}^{\C''}\xi_{Ya+}^{\B''\C''} = \xi_{Ya+}^{\B''\C''} \quad \Rightarrow \quad {B}_{Ya^+}^{\B''}\xi_{Ya+}^{\B''\C''} = \bar{C}_{Ya^+}^{\C''}\xi_{Ya+}^{\B''\C''} \quad \Rightarrow \quad \bar{C}_{Ya^+}^{\C''}\xi_{Ya+}^{\B''\C''} = {C}_{Ya^+}^{\C''}\xi_{Ya+}^{\B''\C''}.
\end{align}
The first equalities in both lines come from Lemma~\ref{lemma:measurementlemmaAlt}, the second ones were obtained by multiplying the first one with $C'_{Ya^+}$ and $C''_{Ya^+}$, and in the third one we used Eqs.~\eqref{eq:bcy} and~\eqref{eq:bcy'}.
Finally, the self-tested substate \eqref{eq:state0}, expressed in the eigenbases of $B_\lozenge$ and $C_\lozenge$ has the form:
\begin{align}\label{eq:state0Comp}
\ketbra{{\Psi}_{a^+}}{{\Psi}_{a^+}}^{\B_q\C_q}\ot \xi_{a^+}^{\B''C''} &=
   \ketbra{{{\Psi}}_{a^+}}{{{\psi}}_{a^+}}^{\B_q\C_q}\ot{B}_{Ya^+}^+\xi_{a^+}^{\B''C''} + \ketbra{{\Psi}_{a^+}}{{\psi}_{a^+}}^{\B_q\C_q}\ot B_{Ya^+}^-\xi_{a^+}^{\B''C''} \nonumber \\ &= \ketbra{{\psi}_{a+}}{{\psi}_{+}}^{\B_q\C_q}\ot{C}_{Ya^+}^+\xi_{a^+}^{\B''C''} + \ketbra{\psi_{a+}^*}{\psi_{+}^*}^{\B_q\C_q}\ot{C}_{Ya^+}^-\xi_{a^+}^{\B''C''}\,,
\end{align}
where $\ket{\psi_+} = \la_{000}\ket{00}+\la_{001}\ket{01}+\la_{010}\ket{10}+\la_{011}\ket{11}$ and $\ket{\psi_+^*} = \la_{000}^*\ket{0^*0^*}+\la_{001}^*\ket{0^*1^*}+\la_{010}^*\ket{1^*0^*}+\la_{011}^*\ket{1^*1^*}$.

\subsection{The second set of sub-tests}\label{sec:a0minus}

Let us now move to the outcome $1$ of Alice's measurement $A_\lozenge$ and self-testing of the state \eqref{SchmidtA1}. The correlations necessary for the self-test are:
\begin{subequations}
\begin{align}\label{a0mst}
\pazocal{C}_{st}(A_\lozenge^-,\rho) &= (\Lambda_1,\alpha_{a^-}),\\ \label{eq:a0mstmeas} 
  \pazocal{C}_{mst}(A_\lozenge^-,C_\lozenge,B_7,B_8,B_9) &= (\phi_{a^-},\omega^{a-}_z,\omega^{a-}_x,\omega^{a-}_y)\\
  \label{eq:a0mXstmeas} 
  \pazocal{C}_{mst}(A_\lozenge^-,C_\blacklozenge,B_7,B_8,B_9) &= (\phi_{a^-},\delta^{a-}_z,\delta^{a-}_x,\delta^{a-}_y)\\
  \label{eq:a0maltstmeas}
  \pazocal{C}_{qmst}(A_\lozenge^-,B_\lozenge,C_4,C_5,C_6,C_7) &= (\phi_{a^-},\mu_{a^-},\upsilon^{a-}_z,\upsilon^{a-}_x,\upsilon^{a-}_y),\\
  \label{eq:a0maltXstmeas}
  \pazocal{C}_{qmst}(A_\lozenge^-,B_\blacklozenge,C_4,C_5,C_6,C_7) &= (\phi_{a^-},\mu_{a^-},\eta^{a-}_z,\eta^{a-}_x,\eta^{a-}_y),
\end{align}
\end{subequations}
where $\Lambda_1 = \sqrt{\la_{100}^2 + \la_{101}^2 + \la_{110}^2 + \la_{111}^2}$, $\sin2{\phi}_{a^-} = \sqrt{(1-{\alpha}_{a^-}^2/4)/(1+{\alpha}_{a^-}^2/4)}$ and
$\sin{{\mu}_{a^-}/2} = \sqrt{(1-{\alpha}_{a^-}^2/4)/2}$. The remaining parameters are such that
\begin{align}\label{eq:vb}
V_{a^-}^\dagger\sz V_{a^-} = \upsilon^{a-}_z\sz + \upsilon^{a-}_x\sx + \upsilon^{a-}_y\sy,\\
\label{eq:vbX}
V_{a^-}^\dagger\sx V_{a^-} = \eta^{a-}_z\sz + \eta^{a-}_x\sx + \eta^{a-}_y\sy,\\
\label{eq:vc}
  W_{a^-}^\dagger\sz W_{a^-} = \omega^{a-}_z\sz + \omega^{a-}_x\sx + \omega^{a-}_y\sy,\\
  \label{eq:vcX}
  W_{a^-}^\dagger\sx W_{a^-} = \delta^{a-}_z\sz + \delta^{a-}_x\sx + \delta^{a-}_y\sy.
\end{align}
By taking the transpose we obtain 
\begin{align}\label{eq:vb*}
  \left(V_{a^-}^*\right)^\dagger\sz V_{a^-}^* = \upsilon^{a-}_z\sz + \upsilon^{a-}_x\sx - \upsilon^{a-}_y\sy,\\
\label{eq:vbX*}
  \left(V_{a^-}^*\right)^\dagger\sx V_{a^-}^* = \eta^{a-}_z\sz + \eta^{a-}_x\sx - \eta^{a-}_y\sy,\\
\label{eq:vc*}
  \left(W_{a^-}^*\right)^\dagger\sz W_{a^-}^* = \omega^{a-}_z\sz + \omega^{a-}_x\sx - \omega^{a-}_y\sy,\\
  \label{eq:vcX*}
  \left(W_{a^-}^*\right)^\dagger\sx W_{a^-}^* = \delta^{a-}_z\sz + \delta^{a-}_x\sx - \delta^{a-}_y\sy.
\end{align}
According to Lemma \ref{lemma:selftesting}, correlations \eqref{a0mst} imply that there exist unitaries $U^\B_{a_{-}}$ and $U^\C_{a_{-}}$ such that
\begin{multline}\label{eq:state1}
 \left(U^\B_{a_{-}}\otimes U^\C_{a_{-}}\right)\tr_{A}(A_\lozenge^-\rho^{\A\B\C}) \left(U^\B_{a_{-}}\otimes U^\C_{a_{-}}\right)^{\dagger}\varpropto \\ \varpropto (\cos{\phi}_{a^-}\ket{{0_{a^-}}{0_{a^-}}}^{\B_q\C_q} + \sin{\phi_{a^-}}\ket{{1_{a^-}}{1_{a^-}}}^{\B_q\C_q})(\cos{\phi_{a^-}}\bra{{0_{a^-}}{0_{a^-}}}^{\B_q\C_q} + \sin{\phi_{a^-}}\bra{{1_{a^-}}{1_{a^-}}}^{\B_q\C_q})\ot\xi_{a^-}^{\B''C''} \,,
\end{multline}
 and 
\begin{align*}
  U^\B_{a_{-}}B_7{U^\B_{a_{-}}}^{\dagger} &= {\sigma}_{\textrm{z}{a^-}}^{\B_q} \otimes \mathds{1}^{\B''} \,,\\
  U^\B_{a_{-}}B_8{U^\B_{a_{-}}}^{\dagger} &= {\sigma}_{\textrm{x}{a^-}}^{\B_q} \otimes \mathds{1}^{\B''} \,,\\
  U^\B_{a_{-}}B_9{U^\B_{a_{-}}}^{\dagger} &= {\sigma}_{\textrm{y}{a^-}}^{\B_q} \otimes {B}_{\mathrm{Y}{a^-}}^{\B''},
\end{align*}
where ${\sigma}_{\textrm{z}{a^-}} = \ketbra{0_{a^-}}{0_{a^-}} - \ketbra{1_{a^-}}{1_{a^-}}$ and analogously for ${\sigma}_{\textrm{x}{a^-}}$ and ${\sigma}_{\textrm{y}{a^-}}$.
Now, we can use correlations \eqref{eq:a0mstmeas} and \eqref{eq:a0mXstmeas} to self-test Charlie's measurements $C_\lozenge$ and $C_\blacklozenge$ on the support of $\tr_{AB}[A_\lozenge^-\rho^{\A\B\C}]$, by using the measurement lemma (Lemma~\ref{lemma:measurementlemma}):
\begin{align*}
  U^\C_{a_{-}}C_\lozenge{U^\C_{a_{-}}}^{\dagger} &= \omega^{a-}_x{\sigma}_{\textrm{x}{a^-}}^{\C_q} \otimes \mathds{1}^{\C''} + \omega^{a-}_z{\sigma}_{\textrm{z}{a^-}}^{\C_q} \otimes \mathds{1}^{\C''} + \omega^{a-}_y{\sigma}_{\textrm{y}{a^-}}^{\C_q} \otimes {C}_{Ya^-}^{\C''} \,,\\
  U^\C_{a_{-}}C_\blacklozenge{U^\C_{a_{-}}}^{\dagger} &= \delta^{a-}_x{\sigma}_{\textrm{x}{a^-}}^{\C_q} \otimes \mathds{1}^{\C''} + \delta^{a-}_z{\sigma}_{\textrm{z}{a^-}}^{\C_q} \otimes \mathds{1}^{\C''} + \delta^{a-}_y{\sigma}_{\textrm{y}{a^-}}^{\C_q} \otimes \tilde{C}_{Ya^-}^{\C''}\,,
\end{align*}
where ${C}_{Ya^-}$ and $\tilde{C}_{Ya^-}$ are $\pm1$-eigenvalue operators, such that 
\begin{align}\label{eq:mbcy}
{B}_{Ya^-}^{\B''}\tp {C}_{Ya^-}^{\C''}\xi_{a-}^{\B''\C''} = \xi_{a-}^{\B''\C''} \qquad &\Rightarrow \qquad {B}_{Ya^-}^{\B''}\xi_{a-}^{\B''\C''} = {C}_{Ya^-}^{\C''}\xi_{a-}^{\B''\C''},\\
\label{eq:mbcy'}
{B}_{Ya^-}^{\B''}\tp\tilde{C}_{Ya^-}^{\C''}\xi_{a-}^{\B''\C''} =\xi_{a-}^{\B''\C''} \qquad &\Rightarrow \qquad {B}_{Ya^-}^{\B''}\xi_{a-}^{\B''\C''} = \tilde{C}_{Ya^-}^{\C''}\xi_{a-}^{\B''\C''}\,.
\end{align}
We can further write the form of $C_\lozenge$ on the support $\tr_{AC}(A_\lozenge^-\rho^{\A\B\C})$
\begin{align}\nonumber
   U^\C_{a_{-}}C_\lozenge{U^\C_{a_{-}}}^{\dagger} &= (\omega^{a-}_x{\sigma}_{\textrm{x}{a^-}} + \omega^{a-}_z{\sigma}_{\textrm{z}{a^-}} + \omega^{a-}_y{\sigma}_{\textrm{y}{a^-}})^{\C_q} \otimes \left(\frac{\mathds{1}+{C}_{Ya^-}}{2}\right)^{\C''} + \\ \nonumber &\qquad\qquad\qquad\qquad +(\omega^{a-}_x{\sigma}_{\textrm{x}{a^-}} + \omega^{a-}_z{\sigma}_{\textrm{z}{a^-}} - \omega^{a-}_y{\sigma}_{\textrm{y}{a^-}})^{\C_q} \otimes \left(\frac{\mathds{1}-{C}_{Ya^-}}{2}\right)^{\C''}\\ \label{eq:b0finall}
  &= \left(W_{a^-}^\dagger{\sigma}_{\textrm{z}{a^-}}W_{a^-}\right)^{\C_q} \otimes \left(\frac{\mathds{1}+{C}_{Ya^-}}{2}\right)^{\C''} + \left({W_{a^-}^*}^\dagger{\sigma}_{\textrm{z}{a^-}}W_{a^-}^*\right)^{\C_q} \otimes \left(\frac{\mathds{1}-{C}_{Ya^-}}{2}\right)^{\C''}\,,\end{align}
 and $C_\blacklozenge$: 
  \begin{align}
  \nonumber U^\C_{a_{-}}C_\blacklozenge{U^\C_{a_{-}}}^{\dagger} &= (\delta^{a-}_x{\sigma}_{\textrm{x}{a^-}} + \delta^{a-}_z{\sigma}_{\textrm{z}{a^-}} + \delta^{a-}_y{\sigma}_{\textrm{y}{a^-}})^{\C_q} \otimes \left(\frac{\mathds{1}+\tilde{C}_{Ya^-}}{2}\right)^{\C''} + \\ \nonumber &\qquad\qquad\qquad\qquad+(\delta^{a-}_x{\sigma}_{\textrm{x}{a^-}} + \delta^{a-}_z{\sigma}_{\textrm{z}{a^-}} - \delta^{a-}_y{\sigma}_{\textrm{y}{a^-}})^{\C_q} \otimes \left(\frac{\mathds{1}-\tilde{C}_{Ya^-}}{2}\right)^{\C''}\\ \label{eq:b0finallX}
  &= \left(W_{a^-}^\dagger{\sigma}_{\textrm{x}{a^-}}W_{a^-}\right)^{\C_q} \otimes \left(\frac{\mathds{1}+\tilde{C}_{Ya^-}}{2}\right)^{\C''} + \left({W_{a^-}^*}^\dagger{\sigma}_{\textrm{x}{a^-}}W_{a^-}^*\right)^{\C_q} \otimes \left(\frac{\mathds{1}-\tilde{C}_{Ya^-}}{2}\right)^{\C''}\,,
\end{align}
where we used \eqref{eq:vb}, \eqref{eq:vbX}, \eqref{eq:vb*}, and \eqref{eq:vbX*}. In a similar manner, we can use \eqref{eq:a0maltstmeas} to self-test the additional measurement on Bob's side $B_\lozenge$ on the support of $\tr_{AC}(A_\lozenge^-\rho^{\A\B\C})$
\begin{align}\nonumber
U^\B_{a_{-}}B_\lozenge{U^\B_{a_{-}}}^{\dagger} &= (\upsilon^{a-}_x{\sigma}_{\textrm{x}{a^-}} + \upsilon^{a-}_z{\sigma}_{\textrm{z}{a^-}} + \upsilon^{a-}_y{\sigma}_{\textrm{y}{a^-}})^{\B_q} \otimes \left(\frac{\mathds{1}+\tilde{B}_{Ya^-}}{2}\right)^{\B''} + 
\\ \nonumber &\qquad\qquad\qquad\qquad +(\upsilon^{a-}_x{\sigma}_{\textrm{x}{a^-}} + \upsilon^{a-}_z{\sigma}_{\textrm{z}{a^-}} - \upsilon^{a-}_y{\sigma}_{\textrm{y}{a^-}})^{\B_q} \otimes \left(\frac{\mathds{1}-\tilde{B}_{Ya^-}}{2}\right)^{\B''}\\ 
  &= \left(V_{a^-}^\dagger{\sigma}_{\textrm{z}{a^-}}V_{a^-}\right)^{\B_q} \otimes \left(\frac{\mathds{1}+\tilde{B}_{Ya^-}}{2}\right)^{\B''} + \left({V_{a^-}^*}^\dagger{\sigma}_{\textrm{z}{a^-}}V_{a^-}^*\right)^{\B_q} \otimes \left(\frac{\mathds{1}-\tilde{B}_{Ya^-}}{2}\right)^{\B''}\,,
\label{eq:B0suppA-}\end{align}
and \eqref{eq:a0maltXstmeas} to self-test $B_\blacklozenge$:
\begin{align}
\nonumber 
U^\B_{a_{-}}B_\blacklozenge{U^\B_{a_{-}}}^{\dagger} &= (\eta^{a-}_x{\sigma}_{\textrm{x}{a^-}} + \eta^{a-}_z{\sigma}_{\textrm{z}{a^-}} + \eta^{a-}_y{\sigma}_{\textrm{y}{a^-}})^{\B_q} \otimes \left(\frac{\mathds{1}+\bar{B}_{Ya^-}}{2}\right)^{\B''} + \\ \nonumber &\qquad\qquad\qquad\qquad +(\eta^{a-}_x{\sigma}_{\textrm{x}{a^-}} + \eta^{a-}_z{\sigma}_{\textrm{z}{a^-}} - \eta^{a-}_y{\sigma}_{\textrm{y}{a^-}})^{\B_q} \otimes \left(\frac{\mathds{1}-\bar{B}_{Ya^-}}{2}\right)^{\B''}\\ 
  &= \left(V_{a^-}^\dagger{\sigma}_{\textrm{x}{a^-}}V_{a^-}\right)^{\B_q} \otimes \left(\frac{\mathds{1}+\bar{B}_{Ya^-}}{2}\right)^{\B''} + \left({V_{a^-}^*}^\dagger{\sigma}_{\textrm{x}{a^-}}V_{a^-}^*\right)^{\B_q} \otimes \left(\frac{\mathds{1}-\bar{B}_{Ya^-}}{2}\right)^{\B''}\,,
\label{eq:BXsuppA-}
\end{align}
where
\begin{align}
  {C}_{Ya^-}^{\C''}\tp\tilde{B}_{Ya^-}^{\B''}\xi_{a-}^{\B''\C''} = \xi_{a-}^{\B''\C''} \quad \Rightarrow \quad {C}_{Ya^-}^{\C''}\xi_{a-}^{\B''\C''} = \tilde{B}_{Ya^-}^{\B''}\xi_{a-}^{\B''\C''} \quad \Rightarrow \quad \tilde{B}_{Ya^-}^{\B''}\xi_{a-}^{\B''\C''} = {B}_{Ya^-}^{\B''}\xi_{a-}^{\B''\C''},\\
  {C}_{Ya^-}^{\C''}\tp\bar{B}_{Ya^-}^{\B''}\xi_{a-}^{\B''\C''} =\xi_{a-}^{\B''\C''} \quad \Rightarrow \quad {C}_{Ya^-}^{\C''}\xi_{a-}^{\B''\C''}= \bar{B}_{Ya^-}^{\B''}\xi_{a-}^{\B''\C''} \quad \Rightarrow \quad \bar{B}_{Ya^-}^{\B''}\xi_{a-}^{\B''\C''} = {B}_{Ya^-}^{\B''}\xi_{a-}^{\B''\C''}.
\end{align}
Therefore, the self-tested substate \eqref{eq:state1} in the eigenbases of $B_\lozenge$ and $C_\lozenge$ has the following form: 
\begin{align}\label{eq:state1Comp}
\ketbra{{{\Psi}}_{a^-}}{{{\Psi}}_{a^-}}^{\B_q\C_q}\ot {\xi_{a^-}}^{\B''\C''} &=
   \ketbra{\Psi_{a^-}}{\Psi_{a^-}}^{\B_q\C_q}\ot{B}_{Ya^-}^+{\xi}_{a^-}^{\B''\C''} + \ketbra{\Psi_{a^-}}{\Psi_{a^-}}^{\B_q\C_q}\ot{B}_{Ya^-}^-{\xi}_{a^-}^{\B''\C''} \nonumber \\ &= \ketbra{{\psi}_{a-}}{{{\psi}}_{a-}}^{\B_q\C_q}\ot{C}_{Ya^-}^+\xi_{a^-}^{\B''\C''} + \proj{{{\psi}}^*_{a-}}^{\B_q\C_q}\ot{C}_{Ya^-}^-\xi_{a^-}^{\B''\C''}\,,
\end{align}
where $\ket{{\psi}_{a-}} = \la_{100}\ket{{0}{0}}+\la_{101}\ket{{0}{1}}+\la_{110}\ket{{1}{0}}+\la_{111}\ket{{1}{1}}$ and $\ket{{\psi}^*_{a-}} = \la_{100}^*\ket{{0}^*{0}^*}+\la_{101}^*\ket{{0}^*{1}^*}+\la_{110}^*\ket{{1}^*{0}^*}+\la_{111}^*\ket{{1}^*{1}^*}$.

\subsection{The third set of sub-tests}\label{sec:b0plus}

The third sub-tests aim to certify that the state of Alice and Charlie when Bob measures $B_\lozenge$ and obtains outcome $0$ is 
\be\label{psiC0}
\Lambda_2\ket{\psi_{b+}} = \la_{000}\ket{00}+\la_{001}\ket{01} + \la_{100}\ket{10} + \la_{101}\ket{11}\,,
\ee
where $\Lambda_2 = \sqrt{\la_{000}^2 + \la_{001}^2 + \la_{100}^2 + \la_{101}^2}$. The state $\ket{{\psi}_{b+}}$ has Schmidt decomposition:
\be\label{Schmidt2}
\ket{{\Psi}_{b^+}}^{\A'\C'} = \cos{{\phi}_{b^+}}\ket{{0}_{b^+}{0}_{b^+}} + \sin{{\phi}_{b^+}}\ket{{1}_{b^+}{1}_{b^+}}.
\ee
where $\ket{{0}_{b^+}}^{\A'} = T_{b^+}\ket{0}^{\A'}$, $\ket{{1}_{b^+}}^{\A'} = T_{b^+}\ket{1}^{\A'}$ and $\ket{{0}_{b^+}}^{\C'} = W_{b^+}\ket{0}^{\C'}$, $\ket{{1}_{b^+}}^{\C'} = W_{b^+}\ket{1}^{\C'}$.
Unitaries $T_{b^+}$ and $W_{b^+}$ are such that
\begin{align}\label{eq:tbpa}
T_{b^+}^\dagger\sz T_{b^+} &= \tau^{b+}_z\sz + \tau^{b+}_x\sx + \tau^{b+}_y\sy,\\
\label{eq:tbpaX}
T_{b^+}^\dagger\sx T_{b^+} &= \gamma^{b+}_z\sz + \gamma^{b+}_x\sx + \gamma^{b+}_y\sy,\\ \label{eq:tbpc}
  W_{b^+}^\dagger\sz W_{b^+} &= \omega^{b+}_z\sz + \omega^{b+}_x\sx + \omega^{b+}_y\sy,\\
  \label{eq:tbpcX}
  W_{b^+}^\dagger\sx W_{b^+} &= \delta^{b+}_z\sz + \delta^{b+}_x\sx + \delta^{b+}_y\sy.
\end{align}
By taking the transpose of \eqref{eq:tbpa} and~\eqref{eq:tbpc} we obtain 
\begin{align}\label{eq:tbpa*}
\left(T_{b^+}^*\right)^\dagger\sz T_{b^+}^* &= \tau^{b+}_z\sz + \tau^{b+}_x\sx - \tau^{b+}_y\sy,\\ \label{eq:tbpaX*}
\left(T_{b^+}^*\right)^\dagger\sx T_{b^+}^* &= \gamma^{b+}_z\sz + \gamma^{b+}_x\sx - \gamma^{b+}_y\sy,\\ 
\left(W_{b^+}^*\right)^\dagger\sz W_{b^+}^* &= \omega^{b+}_z\sz + \omega^{b+}_x\sx - \omega^{b+}_y\sy,\\
\left(W_{b^+}^*\right)^\dagger\sx W_{b^+}^* &= \delta^{b+}_z\sz + \delta^{b+}_x\sx - \delta^{b+}_y\sy.
\end{align}

Let us now summarize all the correlations that have to be observed to perform the self-test of the conditional state \eqref{psiC0}:
\begin{subequations} 
\begin{align}
\label{eq:b0pstate}
  \pazocal{C}_{st}(B_\lozenge^+,\rho) &= (\Lambda_2,\alpha_{b^+}),\\ \label{eq:b0pstmeas} 
  \pazocal{C}_{mst}(B_\lozenge^+,A_\lozenge,C_{10},C_{11},C_{12}) &= (\phi_{b^+},\tau^{b+}_z,\tau^{b+}_x,\tau^{b+}_y),\\ \label{eq:b0Xpstmeas} 
  \pazocal{C}_{mst}(B_\lozenge^+,A_\blacklozenge,C_{10},C_{11},C_{12}) &= (\phi_{b^+},\gamma^{b+}_z,\gamma^{b+}_x,\gamma^{b+}_y),\\ \label{eq:b0paltstmeas}
  \pazocal{C}_{amst}(B_\lozenge^+,C_\lozenge,A_1,A_2,A_3,A_4) &= (\phi_{b^+},\mu_{b^+},\omega^{b+}_z,\omega^{b+}_x,\omega^{b+}_y),\\ \label{eq:b0paltXstmeas}
  \pazocal{C}_{amst}(B_\lozenge^+,C_\blacklozenge,A_1,A_2,A_3,A_4) &= (\phi_{b^+},\mu_{b^+},\delta^{b+}_z,\delta^{b+}_x,\delta^{b+}_y),
\end{align}
\end{subequations}
where $\sin2{\phi}_{b^+} = \sqrt{(1-\alpha_{b^+}^2/4)/(1+\alpha_{b^+}^2/4)}$ 
and
$\sin{{\mu}_{b^+}/2} = \sqrt{(1-\alpha_{b^+}^2/4)/2}$.

Correlations \eqref{eq:b0pstate} self-test the state \eqref{Schmidt2} and three Pauli measurements in the corresponding Schmidt basis:
\begin{align}\label{state2}
  \left(U^{\A}_{b_+}\otimes U^\C_{b_+}\right)\tr_B(B_\lozenge^+\rho^{\A\B\C}) \left(U^{\A}_{b_+}\otimes U^\C_{b_+}\right)^{\dagger} &\varpropto \ketbra{{\Psi}_{b^+}}{{\Psi}_{b^+}}^{\A_q\C_q}\ot\xi_{b^+}^{\A''\C''}\,,\\ \nonumber
  U^\C_{b_+}C_{10}{U^\C_{b_+}}^{\dagger} &= {\sigma}_{\textrm{z}{b^+}}^{\C_q} \otimes \mathds{1}^{\C''}\,, \\ \nonumber
  U^\C_{b_+}C_{11}{U^\C_{b_+}}^{\dagger} &= {\sigma}_{\textrm{x}{b^+}}^{\C_q} \otimes \mathds{1}^{\C''}\,, \\ \nonumber
  U^\C_{b_+}C_{12}{U^\C_{b_+}}^{\dagger} &= {\sigma}_{\textrm{y}{b^+}}^{\C_q} \otimes {C}_{Yb^+}^{\C''}.
\end{align}
Following that step, correlations \eqref{eq:b0paltstmeas} and \eqref{eq:b0Xpstmeas} can be used to self-test $A_\lozenge$ and $A_\blacklozenge$ on the support of $\tr_{BC}(B_\lozenge^+\rho^{\A\B\C})$:
\begin{align*}
  U^{\A}_{b_+}A_\lozenge{U^{\A}_{b_+}}^{\dagger} = \tau^{b+}_x{\sigma}_{\textrm{x}{b^+}}^{\A_q} \otimes \mathds{1}^{\A''} + \tau^{b+}_z{\sigma}_{\textrm{z}{b^+}}^{\A_q} \otimes \mathds{1}^{\A''} + \tau^{b+}_y{\sigma}_{\textrm{y}{b^+}}^{\A_q} \otimes {A}_{Yb^+}^{\A''} \,,\\ U^{\A}_{b_+}A_\blacklozenge{U^{\A}_{b_+}}^{\dagger} = \gamma^{b+}_x{\sigma}_{\textrm{x}{b^+}}^{\A_q} \otimes \mathds{1}^{\A''} + \gamma^{b+}_z{\sigma}_{\textrm{z}{b^+}}^{\A_q} \otimes \mathds{1}^{\A''} + \gamma^{b+}_y{\sigma}_{\textrm{y}{b^+}}^{\A_q} \otimes \tilde{A}_{Yb^+}^{\A''} \,,
\end{align*}
or differently written for $A_\lozenge$
\begin{align}\nonumber
   U^{\A}_{b_+}A_\lozenge{U^{\A}_{b_+}}^{\dagger} &= (\tau^{b+}_x{\sigma}_{\textrm{x}{b^+}} + \tau^{b+}_z{\sigma}_{\textrm{z}{b^+}} + \tau^{b+}_y{\sigma}_{\textrm{y}{b^+}})^{\A_q} \otimes \left(\frac{\mathds{1}+{A}_{Yb^+}}{2}\right)^{\A''} + \\ &\qquad\qquad\qquad\qquad + \nonumber (\tau^{b+}_x{\sigma}_{\textrm{x}{b^+}} + \tau^{b+}_z{\sigma}_{\textrm{z}{b^+}} - \tau^{b+}_y{\sigma}_{\textrm{y}{b^+}})^{\A_q} \otimes \left(\frac{\mathds{1}-{A}_{Yb^+}}{2}\right)^{\A''}  \\ 
  &= \left(T_{b^+}^\dagger{\sigma}_{\textrm{z}{b^+}}T_{b^+}\right)^{\A_q} \otimes \left(\frac{\mathds{1}+{A}_{Yb^+}}{2}\right)^{\A''} + \left({T_{b^+}^*}^\dagger{\sigma}_{\textrm{z}{b^+}}T_{b^+}^*\right)^{\A_q} \otimes \left(\frac{\mathds{1}-{A}_{Yb^+}}{2}\right)^{\A''}\,,
\label{eq:A0suppB+}\end{align}
and $A_\blacklozenge$
\begin{align}
  \nonumber U^{\A}_{b_+}A_\blacklozenge{U^{\A}_{b_+}}^{\dagger} &= (\gamma^{b+}_x{\sigma}_{\textrm{x}{b^+}} + \gamma^{b+}_z{\sigma}_{\textrm{z}{b^+}} + \gamma^{b+}_y{\sigma}_{\textrm{y}{b^+}})^{\A_q} \otimes \left(\frac{\mathds{1}+\tilde{A}_{Yb^+}}{2}\right)^{\A''} + \\ &\qquad\qquad\qquad\qquad + \nonumber(\gamma^{b+}_x{\sigma}_{\textrm{x}{b^+}} + \gamma^{b+}_z{\sigma}_{\textrm{z}{b^+}} - \gamma^{b+}_y{\sigma}_{\textrm{y}{b^+}})^{\A_q} \otimes \left(\frac{\mathds{1}-\tilde{A}_{Yb^+}}{2}\right)^{\A''}  \\ 
  &= \left(T_{b^+}^\dagger{\sigma}_{\textrm{x}{b^+}}T_{b^+}\right)^{\A_q} \otimes \left(\frac{\mathds{1}+\tilde{A}_{Yb^+}}{2}\right)^{\A''} + \left({T_{b^+}^*}^\dagger{\sigma}_{\textrm{x}{b^+}}T_{b^+}^*\right)^{\A_q} \otimes \left(\frac{\mathds{1}-\tilde{A}_{Yb^+}}{2}\right)^{\A''}\,,
\label{eq:AXsuppB+}
\end{align}
where the Hermitian operators ${A}_{Yb^+}$ and $\tilde{A}_{Yb^+}$ are such that ${A}_{Yb^+}{C}_{Yb^+}\xi_{b+}^{\A\B\C} = \xi_{b+}$ and $\tilde{A}_{Yb^+}\otimes {C}_{Yb^+}\xi_{b+} = \xi_{b+}$. Similarly, correlations \eqref{eq:b0paltstmeas}, using Lemma~\ref{lemma:measurementlemmaAlt} self-test measurement $C_\lozenge$ on the support of $\tr_{AB}(B_\lozenge^+\rho^{\A\B\C})$:
\begin{align}\nonumber
   U^\C_{b_+}C_\lozenge{U^\C_{b_+}}^{\dagger} &= (\omega^{b+}_x{\sigma}_{\textrm{x}{b^+}} + \omega^{b+}_z{\sigma}_{\textrm{z}{b^+}} + \omega^{b+}_y{\sigma}_{\textrm{y}{b^+}})^{\C_q} \otimes \left(\frac{\mathds{1}+\tilde{C}_{Yb^+}}{2}\right)^{\C''} + \\ &\qquad\qquad\qquad\qquad +\nonumber (\omega^{b+}_x{\sigma}_{\textrm{x}{b^+}} + \omega^{b+}_z{\sigma}_{\textrm{z}{b^+}} - \omega^{b+}_y{\sigma}_{\textrm{y}{b^+}})^{\C_q} \otimes \left(\frac{\mathds{1}-\tilde{C}_{Yb^+}}{2}\right)^{\C''}\\ \label{c6finalA}
  &= \left(W_{b^+}^\dagger{\sigma}_{\textrm{z}{b^+}}W_{b^+}\right)^{\C_q} \otimes\left( \frac{\mathds{1}+\tilde{C}_{Yb^+}}{2}\right)^{\C''} + \left({W_{b^+}^*}^\dagger{\sigma}_{\textrm{z}{b^+}}W_{b^+}^*\right)^{\C_q} \otimes \left(\frac{\mathds{1}-\tilde{C}_{Yb^+}}{2}\right)^{\C''}\,,\end{align}
  and correlations \eqref{eq:b0paltXstmeas} self-test measurement $C_\blacklozenge$:
  
  \begin{align}
  \nonumber U^\C_{b_+}C_\blacklozenge{U^\C_{b_+}}^{\dagger} &= (\delta^{b+}_x{\sigma}_{\textrm{x}{b^+}} + \delta^{b+}_z{\sigma}_{\textrm{z}{b^+}} + \delta^{b+}_y{\sigma}_{\textrm{y}{b^+}})^{\C_q} \otimes \left(\frac{\mathds{1}+\bar{C}_{Yb^+}}{2}\right)^{\C''} + \\ &\qquad\qquad\qquad\qquad + \nonumber (\delta^{b+}_x{\sigma}_{\textrm{x}{b^+}} + \delta^{b+}_z{\sigma}_{\textrm{z}{b^+}} - \delta^{b+}_y{\sigma}_{\textrm{y}{b^+}})^{\C_q} \otimes \left(\frac{\mathds{1}-\bar{C}_{Yb^+}}{2}\right)^{\C''}\\ \label{c6finalAX}
  &= \left(W_{b^+}^\dagger{\sigma}_{\textrm{x}{b^+}}W_{b^+}\right)^{\C_q} \otimes \left(\frac{\mathds{1}+\bar{C}_{Yb^+}}{2}\right)^{\C''} + \left({W_{b^+}^*}^\dagger{\sigma}_{\textrm{x}{b^+}}W_{b^+}^*\right)^{\C_q} \otimes \left(\frac{\mathds{1}-\bar{C}_{Yb^+}}{2}\right)^{\C''}\,,
\end{align}
where 
\begin{equation}\label{b0+ac}\tilde{C}_{Yb^+}^{\C''}\xi_{b+}^{\A''\C''} = \xi_{b+}^{\A''\C''}\,,\qquad
{A}_{Yb^+}^{\A''}\tp\bar{C}_{Yb^+}^{\C''}\xi_{b+}^{\A''\C''} = \xi_{b+}^{\A''\C''}, 
\end{equation}
which in turn imply 
\begin{equation}\label{allc}\tilde{C}_{Yb^+}^{\C''}\xi_{b+}^{\A''\C''} = {C}_{Yb^+}^{\C''}\xi_{b+}^{\A''\C''}\,, \qquad 
\bar{C}_{Yb^+}^{\C''}\xi_{b+}^{\A''\C''} = {C}_{Yb^+}^{\C''}\xi_{b+}^{\A''\C''}.
\end{equation}
 The self-tested substate \eqref{state2} in the eigenbasis of $A_\lozenge$ and $C_\lozenge$, has the form:
\begin{align}\label{state2Comp}
\ketbra{{\Psi}_{b^+}}{{\Psi}_{b^+}}^{\A_q\C_q}\ot {\xi_{b^+}}^{\A''\C''} &= \ketbra{{\Psi}_{b^+}}{{\Psi}_{b^+}}^{\A_q\C_q}\ot{C}_{Yb^+}^+\xi_{b^+}^{\A''\C''} + \ketbra{{\Psi}_{b^+}}{{\Psi}_{b^+}}^{\A_q\C_q}\ot{C}_{Yb^+}^-\xi_{b^+}^{\A''\C''} \nonumber \\ &= \ketbra{{\psi}_{b+}}{{\psi}_{b+}}^{\A_q\C_q}\ot{C}_{Yb^+}^+\xi_{b^+}^{\A''\C''} + \ketbra{{\psi}^*_{b+}}{{\psi}^*_{b+}}^{\A_q\C_q}\ot\hat{C}_{Yb^+}^-\xi_{b^+}^{\A''\C''}\,,
\end{align}
where $\ket{{\psi}_{b+}} = \la_{000}\ket{{0}{0}}+\la_{001}\ket{{0}r{1}} + \la_{100}\ket{{1}{0}} + \la_{101}\ket{{1}{1}}$ and $\ket{{\psi}_{b+}^*} = \la_{000}^*\ket{{0}^*{0}^*}+\la_{001}^*\ket{{0}^*{1}^*} + \la_{100}^*\ket{{1}^*{0}^*} + \la_{101}^*\ket{{1}^*{1}^*}$.

\subsection{Self-testing with the SWAP isometry}\label{sec:SWAP}

In the previous sections, we demonstrated the existence of unitaries that map the postmeasurement states of Bob and Charlie, following Alice's measurement $A_\lozenge$, to the corresponding reference states. Similarly, we established that there are unitaries mapping the postmeasurement states of Alice and Charlie, following Bob's measurement $B_\lozenge$, to the reference states. However, these unitaries, derived from separate sets of self-tests, are not guaranteed to be identical. This discrepancy prevents consistent unification of the self-tested structures across all parties.

Nevertheless, the self-testing results provide crucial insights. For each party, there exists a pair of observables (associated with the inputs $\lozenge$ and $\blacklozenge$) that anticommute. Moreover, the correlations of observables corresponding to $\lozenge$ faithfully reproduce the correlations of computational basis measurements on the reference states, albeit potentially up to complex conjugation.

To address these challenges, we propose utilizing the SWAP isometry, depicted in Figure~\ref{isoTel}, as a tool for constructing the desired global structure. This approach enables us to effectively reconcile the local self-testing results and establish a unified description.

By applying the SWAP unitary to the tripartite state $\pur$ we get:
\begin{subequations}
\begin{align} \label{eq:000}
  \Phi\left(\pur\otimes\ket{+++}^{\A'\B'\C'}\right) &= \ket{000}^{\A'\B'\C'}\otimes\frac{\idd + A_\lozenge}{2}\otimes\frac{\idd + B_\lozenge}{2}\otimes\frac{\idd + C_\lozenge}{2}\pur + \\ \label{eq:001}
  & + \ket{001}^{\A'\B'\C'}\otimes\frac{\idd + A_\lozenge}{2}\otimes\frac{\idd + B_\lozenge}{2}\otimes C_\blacklozenge\frac{\idd - C_\lozenge}{2}\pur + \\ \label{eq:010} & +
  \ket{010}^{\A'\B'\C'}\otimes\frac{\idd + A_\lozenge}{2}\otimes B_\blacklozenge\frac{\idd - B_\lozenge}{2}\otimes\frac{\idd + C_\lozenge}{2}\pur + \\ \label{eq:011} & +\ket{011}^{\A'\B'\C'}\otimes\frac{\idd + A_\lozenge}{2}\otimes B_\blacklozenge\frac{\idd - B_\lozenge}{2}\otimes C_\blacklozenge\frac{\idd - C_\lozenge}{2}\pur + \\ \label{eq:100} &+
  \ket{100}^{\A'\B'\C'}\otimes A_\blacklozenge\frac{\idd - A_\lozenge}{2}\otimes\frac{\idd + B_\lozenge}{2}\otimes\frac{\idd + C_\lozenge}{2}\pur + \\ \label{eq:101}
  &+ \ket{101}^{\A'\B'\C'}\otimes A_\blacklozenge\frac{\idd - A_\lozenge}{2}\otimes\frac{\idd + B_\lozenge}{2}\otimes C_\blacklozenge\frac{\idd - C_\lozenge}{2}\pur + \\ \label{eq:110} &+
  \ket{110}^{\A'\B'\C'}\otimes A_\blacklozenge\frac{\idd - A_\lozenge}{2}\otimes B_\blacklozenge\frac{\idd - B_\lozenge}{2}\otimes\frac{\idd + C_\lozenge}{2}\pur + \\ \label{eq:111} & +\ket{111}^{\A'\B'\C'}\otimes A_\blacklozenge\frac{\idd - A_\lozenge}{2}\otimes B_\blacklozenge\frac{\idd - B_\lozenge}{2}\otimes C_\blacklozenge\frac{\idd - C_\lozenge}{2}\pur.
\end{align}
\end{subequations}
Expression above can be simplified by taking into account the results from subsection~\ref{sec:a0plus}.
Let us recall Eq.~\eqref{eq:state0Comp}:
\begin{align*}
  U^\B_{a^+}\otimes U^\C_{a^+} &\left(\frac{\tr_A\left(A_\lozenge^+\proj{\psi}^{\A\B\C\Pp}\right)}{\tr\left(A_\lozenge^+\proj{\psi}^{\A\B\C\Pp}\right)}\right) \left({U^\B_{a^+}}\otimes {U^\C_{a^+}}\right)^\dagger = \\ &= \ketbra{{\psi}_{a+}}{{\psi}_{a+}}^{\B_q\C_q}\ot\left(\frac{\idd+{C}_{Ya^+}}{2}\right)^{\C''}\xi_{a^+}^{\B''\C''} + \ketbra{\psi_{a+}^*}{\psi_{a+}^*}^{\B_q\C_q}\ot\left(\frac{\idd-{C}_{Ya^+}}{2}\right)^{\C''}\xi_{a^+}^{\B''\C''}\,.
\end{align*}
It implies that state $A_\lozenge^+\pur$ has the following form:
\begin{align}
  A_\lozenge^+\pur = {U^\B_{a^+}}^\dagger\otimes {U^\C_{a^+}}^\dagger\left(\ket{\psi_{a+}}^{\B_q\C_q}\otimes\ket{\xi_{a^+}^+}^{\A\B''\C''\Pp} + \ket{\psi_{a+}^*}^{\B_q\C_q}\otimes\ket{\xi_{a^+}^-}^{\A\B''\C''\Pp}\right),
\end{align}
where we introduced a notation that makes explicit the fact that the Hilbert spaces of Bob and Charlie can be written as tensor products of their qubit components, denoted with $\B_q$ and $\C_q$ respectively, and the rest denoted with $\B''$ and $\C''$. By denoting $p(0|x=\lozenge) = \tr\left(A_\lozenge^+\proj{\psi}^{\A\B\C\Pp}\right)$ the states $\ket{\xi_{a^+}^+}$ and $\ket{\xi_{a^+}^-}$ are purifications of $p(0|x=\lozenge){C}_{Ya^+}^+\xi_{a^+}$ and $p(0|x=\lozenge){C}_{Ya^+}^-\xi_{a^+}$ respectively, such that 
\begin{align*}
  \tr_A\left[\ket{\xi_{a^+}^+}\bra{\xi_{a^+}^-}\right] = 0.
\end{align*}
Eqs.~\eqref{eq:bcy} and~\eqref{baker} imply the following relations:
\begin{align*}
  \left(\frac{\idd+{B}_{Ya^+}}{2}\right)^{\B''}\xi_{a^+}^{\B''\C''} = \left(\frac{\idd+{C}_{Ya^+}}{2}\right)^{\C''}\xi_{a^+}^{\B''\C''}, \qquad \left(\frac{\idd-{B}_{Ya^+}}{2}\right)^{\B''}\xi_{a^+}^{\B''\C''} = \left(\frac{\idd-{C}_{Ya^+}}{2}\right)^{\C''}\xi_{a^+}^{\B''\C''},\\
  \left(\frac{\idd+\tilde{C}_{Ya^+}}{2}\right)^{\C''}\xi_{a^+}^{\B''\C''} = \left(\frac{\idd+{C}_{Ya^+}}{2}\right)^{\C''}\xi_{a^+}^{\B''\C''}, \qquad \left(\frac{\idd-\tilde{C}_{Ya^+}}{2}\right)^{\C''}\xi_{a^+}^{\B''\C''} = \left(\frac{\idd-{C}_{Ya^+}}{2}\right)^{\C''}\xi_{a^+}^{\B''\C''},
\end{align*}
which further implies:
\begin{align*}
  \left(\frac{\idd+{B}_{Ya^+}}{2}\right)^{\B''}\otimes \left(\frac{\idd+\tilde{C}_{Ya^+}}{2}\right)^{\C''}\left(\frac{\idd+{C}_{Ya^+}}{2}\right)^{\C''}\xi_{a+}^{\B''\C''} &= \left(\frac{\idd+{C}_{Ya^+}}{2}\right)^{\C''}\xi_{a+}^{\B''\C''},\\
  \left(\frac{\idd-{B}_{Ya^+}}{2}\right)^{\B''}\otimes \left(\frac{\idd-\tilde{C}_{Ya^+}}{2}\right)^{\C''}\left(\frac{\idd-{C}_{Ya^+}}{2}\right)^{\C''}\xi_{a+}^{\B''\C''} &= \left(\frac{\idd-{C}_{Ya^+}}{2}\right)^{\C''}\xi_{a+}^{\B''\C''},
\end{align*}
or on the level of purifications $\ket{\xi_{a+}^\pm}$:
\begin{align}\label{puri1}
  \left(\frac{\idd+{B}_{Ya^+}}{2}\right)^{\B''}\otimes \left(\frac{\idd+\tilde{C}_{Ya^+}}{2}\right)^{\C''}\ket{\xi_{a+}^+}^{\A\B''\C''\Pp} &= \ket{\xi_{a+}^+}^{\A\B''\C''\Pp}\\ \label{puri2}
  \left(\frac{\idd-{B}_{Ya^+}}{2}\right)^{\B''}\otimes \left(\frac{\idd-\tilde{C}_{Ya^+}}{2}\right)^{\C''}\ket{\xi_{a+}^-}^{\A\B''\C''\Pp} &= \ket{\xi_{a+}^-}^{\A\B''\C''\Pp}.
\end{align}

Let us now consider the state from~\eqref{eq:000}:
\begin{align}\nonumber
  A_\lozenge^+\otimes& B_\lozenge^+ \otimes C_\lozenge^+\pur = \left(B_\lozenge^+ \otimes C_\lozenge^+\right) \left({U^\B_{a^+}}^\dagger\otimes {U^\C_{a^+}}^\dagger\right)\left(\ket{\psi_{a+}}^{\B_q\C_q}\otimes\ket{\xi_{a^+}^+}^{\A\B''\C''\Pp} + \ket{\psi_{a+}^*}^{\B_q\C_q}\otimes\ket{\xi_{a^+}^-}^{\A\B''\C''\Pp}\right)\\ \nonumber
  &= \left({U^\B_{a^+}}^\dagger\otimes {U^\C_{a^+}}^\dagger\right) \left({U^\B_{a^+}}B_\lozenge^+{U^\B_{a^+}}^\dagger\otimes {U^\C_{a^+}}C_\lozenge^+ {U^\C_{a^+}}^\dagger\right)\left(\ket{\psi_{a+}}^{\B_q\C_q}\otimes\ket{\xi_{a^+}^+}^{\A\B''\C''\Pp} + \ket{\psi_{a+}^*}^{\B_q\C_q}\otimes\ket{\xi_{a^+}^-}^{\A\B''\C''\Pp}\right)\,.
\end{align}
Now we can use Eqs.~\eqref{eq:b0final} and~\eqref{eq:c6final} to get:
\begin{align*}
   A_\lozenge^+\otimes &B_\lozenge^+ \otimes C_\lozenge^+\pur =\\&= {U^\B_{a^+}}^\dagger\otimes {U^\C_{a^+}}^\dagger\left(\lambda_{000}\ket{00}^{\B_q\C_q}\otimes {B}_{Ya^+}^+\tilde{C}_{Ya^+}^+\ket{\xi_{a^+}^+}^{\A\B''\C''\Pp} + \lambda_{000}^*\ket{0^*0^*}^{\B_q\C_q}\otimes {B}_{Ya^+}^-\tilde{C}_{Ya^+}^-\ket{\xi_{a^+}^-}^{\A\B''\C''\Pp}\right),\end{align*}
and taking into account~\eqref{puri1} and~\eqref{puri2}
   
   \begin{equation*}
   A_\lozenge^+\otimes B_\lozenge^+ \otimes C_\lozenge^+\pur= {U^\B_{a^+}}^\dagger\otimes {U^\C_{a^+}}^\dagger\left(\lambda_{000}\ket{00}^{\B_q\C_q}\otimes \ket{\xi_{a^+}^+}^{\A\B''\C''\Pp} + \lambda_{000}^*\ket{0^*0^*}^{\B_q\C_q}\otimes \ket{\xi_{a^+}^-}^{\A\B''\C''\Pp}\right).
\end{equation*}

Taking notation $\ket{\xi_0}^{\A\B\C\Pp} = {U^\B_{a^+}}^\dagger\otimes {U^\C_{a^+}}^\dagger\ket{00}^{\B_q\C_q}\otimes \ket{\xi_{a^+}^+}^{\A\B''\C''\Pp}$ and $\ket{\xi_1}^{\A\B\C\Pp} = {U^\B_{a^+}}^\dagger\otimes {U^\C_{a^+}}^\dagger\ket{0^*0^*}^{\B_q\C_q}\otimes \ket{\xi_{a^+}^-}^{\A\B''\C''\Pp}$ the equation above becomes
\begin{equation}\label{johnwoo}
   A_\lozenge^+\otimes B_\lozenge^+ \otimes C_\lozenge^+\pur = \lambda_{000} \ket{\xi_0} + \lambda_{000}^*\ket{\xi_1},
\end{equation}
where $\braket{\xi_0}{\xi_1} = 0$.

Using a similar argumentation, the state from Eq.~\eqref{eq:001} becomes:
\begin{align}\nonumber
  & A_\lozenge^+\otimes B_\lozenge^+ \otimes C_\blacklozenge C_\lozenge^-\pur = \\ \nonumber&= {U^\B_{a^+}}^\dagger\otimes {U^\C_{a^+}}^\dagger \left({U^\C_{a^+}}C_\blacklozenge {U^\C_{a^+}}^\dagger\right) \left(\lambda_{001}\ket{01}^{\B_q\C_q}\otimes \ket{\xi_{a^+}^+}^{\A\B''\C''\Pp} + \lambda_{001}^*\ket{0^*1^*}^{\B_q\C_q}\otimes \ket{\xi_{a^+}^-}^{\A\B''\C''\Pp}\right)\\ \nonumber
  &= {U^\B_{a^+}}^\dagger\otimes {U^\C_{a^+}}^\dagger \left(\lambda_{001}\ket{00}^{\B_q\C_q}\otimes \bar{C}_{Ya^+}^+\ket{\xi_{a^+}^+}^{\A\B''\C''\Pp} + \lambda_{001}^*\ket{0^*0^*}^{\B_q\C_q}\otimes \bar{C}_{Ya^+}^-\ket{\xi_{a^+}^-}^{\A\B''\C''\Pp}\right)\\ \nonumber
  &= \lambda_{001} \ket{\xi_0} + \lambda_{001}^*\ket{\xi_1}\,.
\end{align}
The second line is obtained in the same way as in the case discussed above, while the third line is obtained by using~\eqref{eq:c6Xfinal}. In the fourth line we used relations~\eqref{kantemir}.
By using exactly the same argumentation we obtain:
\begin{align}\label{hk}
  A_\lozenge^+\otimes B_\blacklozenge B_\lozenge^- \otimes C_\lozenge^+\pur = \lambda_{010} \ket{\xi_0} + \lambda_{010}^*\ket{\xi_1}\,, \\ \nonumber
  A_\lozenge^+\otimes B_\blacklozenge B_\lozenge^- \otimes C_\blacklozenge C_\lozenge^-\pur = \lambda_{011} \ket{\xi_0} + \lambda_{011}^*\ket{\xi_1}\,.
\end{align}
Now we analyze the state from line~\eqref{eq:100}. The result~\eqref{eq:state1Comp} implies that state $A_\lozenge^-\pur$ has the form:
\begin{equation*}
A_\lozenge^-\pur = {U^\B_{a^-}}^\dagger\otimes {U^\C_{a^-}}^\dagger\left(\ket{{\psi}_{a-}}^{\B_q\C_q}\otimes\ket{\xi_{a^-}^+}^{\A\B''\C''\Pp} + \ket{{\psi}_{a-}^*}^{\B_q\C_q}\otimes\ket{\xi_{a^-}^-}^{\A\B''\C''\Pp}\right).
\end{equation*}
By denoting $p(1|x=\lozenge) = \tr\left(A_\lozenge^-\proj{\psi}^{\A\B\C\Pp}\right)$, the states $\ket{\xi_{a^-}^+}$ and $\ket{\xi_{a^-}^-}$ are purifications of $p(1|x=\lozenge){C}_{Ya^-}^+\xi_{a^-}$ and $p(1|x=\lozenge){C}_{Ya^-}^-\xi_{a^-}$, respectively, such that 
\begin{align*}
  \tr_A\left[\ket{\xi_{a^-}^+}\bra{\xi_{a^-}^-}\right] = 0.
\end{align*}
Using these expressions we can write
\begin{align}\nonumber
  &A_\blacklozenge A_\lozenge^-\otimes B_\lozenge^+ \otimes C_\lozenge^+\pur = \\ \nonumber &= \left(B_\lozenge^+ \otimes C_\lozenge^+\right) \left({U_\B^{a^-}}^\dagger\otimes {U_\C^{a^-}}^\dagger\right)\left(\ket{{\psi}_{a-}}^{\B_q\C_q}\otimes A_\blacklozenge\ket{\xi_{a^-}^+}^{\A\B''\C''\Pp} + \ket{{\psi}_{a-}^*}^{\B_q\C_q}\otimes A_\blacklozenge\ket{\xi_{a^-}^-}^{\A\B''\C''\Pp}\right)\\ \nonumber
  &= \left({U^\B_{a^-}}^\dagger\otimes {U^\C_{a^-}}^\dagger\right) \left({U^\B_{a^-}}B_\lozenge^+{U^\B_{a^-}}^\dagger\otimes {U^\C_{a^-}}C_\lozenge^+ {U^\C_{a^-}}^\dagger\right)\left(\ket{{\psi}_{a-}}^{\B_q\C_q}\otimes A_\blacklozenge\ket{\xi_{a^-}^+}^{\A\B''\C''\Pp} + \ket{{\psi}_{a-}^*}^{\B_q\C_q}\otimes A_\blacklozenge\ket{\xi_{a^-}^-}^{\A\B''\C''\Pp}\right)\,.
\end{align}
Taking into account Eqs.~\eqref{eq:b0finall} and~\eqref{eq:B0suppA-} to get:
\begin{align*}
   &A_\blacklozenge A_\lozenge^-\otimes B_\lozenge^+ \otimes C_\lozenge^+\pur =\\&= {U^\B_{a^-}}^\dagger\otimes {U^\C_{a^-}}^\dagger\left(\lambda_{100}\ket{{0}{0}}^{\B_q\C_q}\otimes A_\blacklozenge{B}_{Ya^-}^+{C}_{Ya^-}^+\ket{\xi_{a^-}^+}^{\A\B''\C''\Pp} + \lambda_{100}^*\ket{{0}^*{0}^*}^{\B_q\C_q}\otimes A_\blacklozenge{B}_{Ya^-}^-{C}_{Ya^+-}^-\ket{\xi_{a^-}^-}^{\A\B''\C''\Pp}\right)\\
   &= {U^\B_{a^-}}^\dagger\otimes {U^\C_{a^-}}^\dagger\left(\lambda_{100}\ket{{0}{0}}^{\B_q\C_q}\otimes A_\blacklozenge\ket{\xi_{a^-}^+}^{\A\B''\C''\Pp} + \lambda_{100}^*\ket{{0}^*{0}^*}^{\B_q\C_q}\otimes A_\blacklozenge\ket{\xi_{a^-}^-}^{\A\B''\C''\Pp}\right)\,.
\end{align*}
Taking notation $\ket{\xi_0'}^{\A\B\C\Pp} = {U^\B_{a^-}}^\dagger\otimes {U^\C_{a^-}}^\dagger\ket{{0}{0}}^{\B_q\C_q}\otimes A_\blacklozenge\ket{\xi_{a^-}^+}^{\A\B''\C''\Pp}$ and $\ket{\xi'_1}^{\A\B\C\Pp} = {U^\B_{a^-}}^\dagger\otimes {U^\C_{a^-}}^\dagger\ket{{0}^*{0}^*}^{\B_q\C_q}\otimes A_\blacklozenge\ket{\xi_{a^-}^-}^{\A\B''\C''\Pp}$, the equation above becomes
\begin{equation}\label{johnwoo1}
   A_\blacklozenge A_\lozenge^-\otimes B_\lozenge^+ \otimes C_\lozenge^+\pur = \lambda_{100} \ket{\xi'_0}^{\A\B\C\Pp} + \lambda_{100}^*\ket{\xi'_1}^{\A\B\C\Pp},
\end{equation}
where $\braket{\xi'_0}{\xi'_1} = 0$.
By using the same argumentation the states from lines~\eqref{eq:101}-\eqref{eq:111} have the form:
\begin{align}\label{ct}
   A_\blacklozenge A_\lozenge^-\otimes B_\lozenge^+ \otimes 
C_\blacklozenge C_\lozenge^-\pur &= \lambda_{101}\ket{\xi'_0}^{\A\B\C\Pp} + \lambda_{101}^*\ket{\xi'_1}^{\A\B\C\Pp},\\ \nonumber
A_\blacklozenge A_\lozenge^-\otimes B_\blacklozenge B_\lozenge^- \otimes C_\lozenge^+\pur &= \lambda_{110}\ket{\xi'_0}^{\A\B\C\Pp} + \lambda_{110}^*\ket{\xi'_1}^{\A\B\C\Pp},\\ \nonumber
A_\blacklozenge A_\lozenge^-\otimes B_\blacklozenge B_\lozenge^- \otimes 
C_\blacklozenge C_\lozenge^-\pur &= \lambda_{111}\ket{\xi'_0}^{\A\B\C\Pp} + \lambda_{111}^*\ket{\xi'_1}^{\A\B\C\Pp}.
\end{align}
The states $\ket{\xi'_0}$ and $\ket{\xi'_1}$ satisfy:
\begin{align*}
\braket{\xi'_0}{\xi'_1} = 0, \qquad  \braket{\xi'_0}{\xi'_0} + \braket{\xi'_1}{\xi'_1} = 1.  
\end{align*}
Using these results, the action of the SWAP isometry reduces to:
\begin{align}\nonumber
    \Phi\left(\pur\otimes\ket{+++}^{\A'\B'\C'}\right) &= \left(\sum_{jk}\lambda_{0jk}\ket{0jk}^{\A'\B'\C'}\right)\otimes\ket{\xi_0}^{\A\B\C\Pp} + \left(\sum_{jk}\lambda^*_{0jk}\ket{0jk}^{\A'\B'\C'}\right)\otimes\ket{\xi_1}^{\A\B\C\Pp} \,+ \\ &+ \left(\sum_{jk}\lambda_{1jk}\ket{1jk}^{\A'\B'\C'}\right)\otimes\ket{\xi'_0}^{\A\B\C\Pp} + \left(\sum_{jk}\lambda^*_{1jk}\ket{1jk}^{\A'\B'\C'}\right)\otimes\ket{\xi'_1}^{\A\B\C\Pp}\,.\label{swapalmost}
\end{align}
Further structure is obtained when we incorporate the results of the third set of sub-tests given in subsection~\ref{sec:b0plus}. In particular relations~\eqref{eq:A0suppB+}-\eqref{state2Comp} allow rewriting the states in lines~\eqref{eq:000},~\eqref{eq:001},~\eqref{eq:100} and~\eqref{eq:101} in the form:
\begin{subequations}
\begin{align}\label{johnwoo2}
   A_\lozenge^+\otimes B_\lozenge^+ \otimes C_\lozenge^+\pur &= \lambda_{000}\ket{\xi_0''}^{\A\B\C\Pp} + \lambda_{000}^*\ket{\xi_1''}^{\A\B\C\Pp},\\ \label{tw}
  A_\lozenge^+\otimes B_\lozenge^+ \otimes C_\blacklozenge C_\lozenge^-\pur &= \lambda_{001}\ket{\xi_0''}^{\A\B\C\Pp} + \lambda_{001}^*\ket{\xi_1''}^{\A\B\C\Pp},\\ \label{bd}
  A_\blacklozenge A_\lozenge^-\otimes B_\lozenge^+ \otimes C_\lozenge^+\pur &= \lambda_{100}\ket{\xi_0''}^{\A\B\C\Pp} + \lambda_{001}^*\ket{\xi_1''}^{\A\B\C\Pp},\\
  A_\blacklozenge A_\lozenge^-\otimes B_\lozenge^+ \otimes C_\blacklozenge C_\lozenge^-\pur &= \lambda_{101}\ket{\xi_0''}^{\A\B\C\Pp} + \lambda_{101}^*\ket{\xi_1''}^{\A\B\C\Pp},\label{ko}
\end{align}
\end{subequations}
where states $\ket{\xi_0''}$ and $\ket{\xi_1''}$ satisfy:
\begin{align*}
\braket{\xi_0''}{\xi_1''} = 0, \qquad  \braket{\xi_0''}{\xi_0''} + \braket{\xi_1''}{\xi_1''} = 1.  
\end{align*}

Eqs.~\eqref{johnwoo} and~\eqref{johnwoo2} imply:
\begin{align}\label{niksic}
  \lambda_{000}\ket{\xi_0} + \lambda_{000}^*\ket{\xi_1} = \lambda_{000}\ket{\xi_0''} + \lambda_{000}^*\ket{\xi_1''}\,,
\end{align}
and similarly Eqs.~\eqref{hk} and~\eqref{tw} imply:
\begin{align*}
  \lambda_{001}\ket{\xi_0} + \lambda_{001}^*\ket{\xi_1} = \lambda_{001}\ket{\xi_0''} + \lambda_{001}^*\ket{\xi_1''}.
\end{align*}
If we define $\lambda_{ijk} = \left|\lambda_{ijk}\right|\exp(i\phi_{ijk})$, the last two equations become:
\begin{align*}
  \exp(i\phi_{000})\ket{\xi_0} +\exp(-i\phi_{000})\ket{\xi_1} &= \exp(i\phi_{000})\ket{\xi_0''} +\exp(-i\phi_{000})\ket{\xi_1''}\,,\\
  \exp(i\phi_{001})\ket{\xi_0} +\exp(-i\phi_{001})\ket{\xi_1} &= \exp(i\phi_{001})\ket{\xi_0''} +\exp(-i\phi_{001})\ket{\xi_1''}\,,
\end{align*}
where both $\lambda_{000}$ and $\lambda_{001}$ are nonzero by the choice of the computational basis described at the beginning of Section~\ref{sec:tripartite} in the main text. By multiplying the first equation above with $\exp(i\phi_{001})$ and the second one with $-\exp(i\phi_{000})$ and adding them, we obtain
\begin{align*}
  2i\sin(\phi_{001}-\phi_{000})\ket{\xi_1} = 2i\sin(\phi_{001}-\phi_{000})\ket{\xi_1''}\,.
\end{align*}
Given that our choice of Charlie's local computational basis, explained at the beginning of Section~\ref{sec:tripartite} in the main text $\phi_{001} \neq \phi_{000}$, the last equation implies 
\begin{equation*}
  \ket{\xi_1} = \ket{\xi_1''}\,,
\end{equation*}
which when plugged into~\eqref{niksic} gives
\begin{equation*}
  \ket{\xi_0} = \ket{\xi_0''}\,.
\end{equation*}
In a similar way equations~\eqref{johnwoo1},~\eqref{ct},~\eqref{bd} and~\eqref{ko}, under the assumption $\phi_{100} = \phi_{101}$, can be used to obtain:
\begin{align*}
  \ket{\xi_1'} &= \ket{\xi_1''},\qquad \ket{\xi_0'} = \ket{\xi_0''},
\end{align*}
which by transitivity gives
\begin{align*}
  \ket{\xi_1'} &= \ket{\xi_1},\qquad \ket{\xi_0'} = \ket{\xi_0}.
\end{align*}
Returning to Eq.~\eqref{swapalmost} we obtain
\begin{align*}
    \Phi\left(\pur\otimes\ket{+++}^{\A'\B'\C'}\right) &= \left(\sum_{ijk}\lambda_{ijk}\ket{ijk}^{\A'\B'\C'}\right)\otimes\ket{\xi_0}^{\A\B\C\Pp} + \left(\sum_{ijk}\lambda^*_{ijk}\ket{ijk}^{\A'\B'\C'}\right)\otimes\ket{\xi_1}^{\A\B\C\Pp}\\
    &=\ket{\Psi}^{\A'\B'\C'}\otimes \ket{\xi_0}^{\A\B\C\Pp} + \ket{\Psi^*}^{\A'\B'\C'}\otimes \ket{\xi_1}^{\A\B\C\Pp},
\end{align*}
which proves Theorem~\ref{thmTri}.

\section{The proof of Theorem~\ref{theoremMulti}}\label{app:thmMulti}

\setcounter{equation}{0}
\makeatletter
\renewcommand{\theequation}{D\arabic{equation}}

As outlined in the main text, the self-testing procedure for $n$-partite states consists of $n-1$ sub-tests. In the $j$-th sub-test, all parties except for parties $1$ and $j$ receive only the input $\lozenge$; these are referred to as the projecting parties. The states of parties $1$ and $j$ are self-tested for every local output of parties $j+1,\cdots,n$, but only for the output $0$ of parties $2,\cdots,j-1$. Before demonstrating how the full state can be self-tested by combining these $n-1$ sub-tests, let us introduce some notation. The vector $\vec{a}^j$ has $n-2$ components: the first $j-2$ are always $0$, and the remaining can take value $0$ or $1$:
\begin{equation}
  \vec{a}^j = \left(\underbrace{0,\cdots,0}_{j-2},a^j_{j+1},\cdots,a^j_n\right).
\end{equation}
This implies that there are $2^{n-j}$ possible assignments to the vector $\vec{a}^j$. This vector encodes the outcomes of the $\lozenge$-measurements of the projecting parties. This explains the notation where $j-1$-st entry of vector $\vec{a}^j$ is denoted with $a^j_{j+1}$, and similarly for the entries following it. For each valid vector $\vec{a}^j$, there corresponds a specific state of parties $1$ and $j$, denoted as $\ket{\psi_{\vec{a}^j}^{(1,j)}}$, which is the normalized state to which parties $1$ and $j$ are projected when the remaining parties obtain input $\lozenge$ and obtain the global output described by vector $\vec{a}^j$.
Let us furthermore define $n$-dimensional vector
\begin{equation}
  \vec{a}^j_{kl} = \left(k,\underbrace{0,\cdots,0}_{j-2},l,a^j_{j+1},\cdots,a^j_n\right),
\end{equation}
where $k$ and $l$ can take values $0$ or $1$.
Ideally, when all parties but $1$ and $j$ measure the reference state $\ket{\Psi}$ in the computational basis, the remaining state of parties $1$ and $j$ is projected to
\begin{align}
  \ket{\psi_{\vec{a}^j}^{(1',j')}}&= \sum_{k,l=0}^1\lambda_{\vec{a}^j_{kl}}\ket{k}^{(1')}\ket{l}^{(j')}
\\ &= \cos\theta_{\aj}\ket{0_{\aj}}^{(1')}\ket{0_{aj}}^{(j')} + \sin\theta_{\aj}\ket{1_{\aj}}^{(1')}\ket{1_{aj}}^{(j')},
\end{align}
where $\{\ket{0_{\aj}},\ket{0_{\aj}}\}$ and $\{\ket{0_{\aj}},\ket{0_{\aj}}\}$ are the local Schmidt bases. As in the case of tripartite states, we can define Pauli matrices in the Schmidt bases: $\sigma_{j,\textrm{z}\aj} = \proj{0_{\aj}}-\proj{1_{\aj}}$, $\sigma_{j,\textrm{x}\aj} = \proj{+_{\aj}}-\proj{-_{\aj}}$, where $\ket{\pm_{aj}} = (\ket{0_{\aj}} \pm \ket{1_{\aj}})/\sqrt{2}$, and $\sigma_{j,\textrm{y}\aj} = \sqrt{-1}\sigma_{j,\textrm{z}\aj}\sigma_{j,\textrm{x}\aj}$. We also define $W_{j,\aj}\ket{0_{\aj}} = \ket{0}$ and $W_{j,\aj}\ket{1_{\aj}} = \ket{1}$.

In the $j$-th sub-test we use Lemma~\ref{lemma:selftesting} to self-test state $\ket{\psi_{\aj}}$ for every value of $\aj$. In half of those self-tests party $1$ needs $3$ measurements and party $j$ $6$ measurements and in the other half party $1$ needs $6$ and party $j$ 3 measurements. Furthermore, Lemma~\ref{lemma:measurementlemma} and Lemma~\ref{lemma:measurementlemmaAlt} are used to self-test measurement observables $A_\lozenge^{(1)}$, $A_\blacklozenge^{(1)}$, $A_\lozenge^{(j)}$ and $A_\blacklozenge^{(1)}$. For succinctness we avoid writing all the correlations necessary to reproduce the conditions of the corresponding lemma. Let us define the global projector $A_\lozenge^{\aj}$ as
\begin{equation}
  A_\lozenge^{\aj} = \idd^{(1)}\otimes \frac{\idd + A_\lozenge^{(2)}}{2}\otimes \cdots \otimes \frac{\idd + A_\lozenge^{(j-1)}}{2}\otimes \idd^{(j)}\otimes \frac{\idd + (-1)^{a^j_{j+1}}A_\lozenge^{(j+1)}}{2}\otimes\cdots\otimes \frac{\idd + (-1)^{a^j_n}A_\lozenge^{(n)}}{2},
\end{equation}
and we introduce shortened notation $n\setminus j = (2,\cdots,j-1,j+1,\cdots,n)$.

The successful self-tests result in the following statements:
\begin{align}\label{prvamulti}
  U_{\aj}^{(1)}\otimes U_{\aj}^{(j)}\tr_{n\setminus j}\left[A_\lozenge^{\aj}\rho\right]\left({U_{\aj}^{(1)}}\otimes {U_{\aj}^{(j)}}\right)^\dagger &\propto \proj{\psi_{\aj}}^{(1_q,j_q)}\otimes {A_{\aj}^+}^{(j'')}\xi_{\aj}^{(1'',j'')} + \proj{{\psi_{\aj}^*}}^{(1_q,j_q)}\otimes {A_{\aj}^-}^{(j'')}\xi_{\aj}^{(1'',j'')}\\
U_{\aj}^{(1)}A_\lozenge^{(1)}{U_{\aj}^{(1)}}^{\dagger} & = {W_{1,\aj}^{\dagger}} \sigma_{1,\textrm{z}\aj}{W_{1,\aj}^{(1)}}\otimes {A_{\aj}^+}^{(1'')} + {W_{1,\aj}^{T}} \sigma_{1,\textrm{z}\aj}{W_{1,\aj}^{*}}\otimes {A_{\aj}^-}^{(1'')}
\\  U_{\aj}^{(1)}A_\blacklozenge^{(1)}{U_{\aj}^{(1)}}^{\dagger} & = {W_{1,\aj}^{\dagger}} \sigma_{1,\textrm{x}\aj}{W_{1,\aj}}\otimes \left({\tilde{A}}_{\aj}^+\right)^{(1'')} + {W_{1,\aj}^{T}} \sigma_{1,\textrm{x}\aj}{W_{1,\aj}^{*}}\otimes \left(\tilde{A}_{\aj}^-\right)^{(1'')}
\\
U_{\aj}^{(j)}A_\lozenge^{(j)}{U_{\aj}^{(1)}}^{\dagger} & = {W_{j,\aj}^{\dagger}} \sigma_{j,\textrm{z}\aj}{W_{\aj}}\otimes \left({\tilde{A}}_{\aj}^+\right)^{(j'')} + {W_{j,\aj}^{T}} \sigma_{j,\textrm{z}\aj}{W_{j,\aj}^{*}}\otimes \left({\tilde{A}}_{\aj}^-\right)^{(j'')} \\
U_{\aj}^{(j)}A_\blacklozenge^{(j)}{U_{\aj}^{(1)}}^{\dagger} & = {W_{j,\aj}^{\dagger}} \sigma_{j,\textrm{x}\aj}W_{j,\aj}\otimes \left({\bar{A}}_{\aj}^+\right)^{(j'')} + {W_{j,\aj}^{T}} \sigma_{j,\textrm{x}\aj}{W_{j,\aj}^{*}}\left({\bar{A}}_{\aj}^-\right)^{(j'')},\label{zadnjamulti}
\end{align}
where operators ${A_{\aj}^\pm}^{(1'')}$, $\left(\tilde{A}_{\aj}^\pm\right)^{(1'')}$, $\left(\tilde{A}_{\aj}^\pm\right)^{(j'')}$ and $\left(\bar{A}_{\aj}^\pm\right)^{(j'')}$ come from applying Lemma~\ref{lemma:measurementlemma} and satisfy:
\begin{align*}
{A_{\aj}^\pm}^{(j'')}\xi_{\aj}^{(1'',j'')} &= {A_{\aj}^\pm}^{(1'')}\xi_{\aj}^{(1'',j'')},\\
{A_{\aj}^\pm}^{(j'')}\xi_{\aj}^{(1'',j'')} &= \left(\tilde{A}_{\aj}^\pm\right)^{(1'')}\xi_{\aj}^{(1'',j'')},\\
{A_{\aj}^\pm}^{(j'')}\xi_{\aj}^{(1'',j'')} &= \left(\tilde{A}_{\aj}^\pm\right)^{(j'')}\xi_{\aj}^{(1'',j'')},\\
{A_{\aj}^\pm}^{(j'')}\xi_{\aj}^{(1'',j'')} &= \left(\bar{A}_{\aj}^\pm\right)^{(j'')}\xi_{\aj}^{(1'',j'')}.
\end{align*}
The SWAP isometry is created for every party from self-tested operators $A_\lozenge^{(j)}$ and $A_\blacklozenge^{(j)}$, and the output of the isometry has form: 
\begin{align}
  \Phi\left(\ket{\psi}^{[n],P}\otimes_{k = 1}^n\ket{+}^{k'}\right) = \sum_{\vec{a}}\ket{\xi_{\vec{a}}}^{([n],P)}\otimes\ket{\vec{a}}^{1',\cdots,n'},
\end{align}
where 
\begin{equation}
  \ket{\xi_{\vec{a}}}^{([n],P)} = \bigotimes_{j=1}^n\left(A_\blacklozenge^{(j)}\right)^{a_j}\frac{\idd + (-1)^{a_j}A_\lozenge^{(j)}}{2}\ket{\psi}^{([n],P)}.
\end{equation}
Relations~\eqref{prvamulti}-\eqref{zadnjamulti} for $j=2$ allow simplifying the form of the states $\ket{\xi_{\vec{a}}}$\footnote{To obtain this, repeat the procedure explained in detail for deriving Eq.~\eqref{johnwoo}}:
\begin{equation}
  \ket{\xi_{\vec{a}}}^{([n],P)} = \lambda_{\vec{a}^2_{a_1a_2}}\ket{\xi_{\vec{a}^2}}^{([n],P)} + \lambda^*_{\vec{a}^2_{a_1a_2}}\ket{\xi'_{\vec{a}^2}}^{([n],P)},
\end{equation}
where $\braket{\xi_{\vec{a}^2}}{\xi'_{\vec{a}^2}} = 0$. By plugging these states into the output of the isometry we get
\begin{multline}
  \Phi\left(\ket{\psi}^{[n],P}\otimes_{k = 1}^n\ket{+}^{k'}\right) = \sum_{\vec{a}^2}\left(\lambda_{\vec{a}^2_{00}}\ket{\vec{a}^2_{00}} + \lambda_{\vec{a}^2_{01}}\ket{\vec{a}^2_{01}} + \lambda_{\vec{a}^2_{10}}\ket{\vec{a}^2_{10}} + \lambda_{\vec{a}^2_{11}}\ket{\vec{a}^2_{11}}\right)^{[n']}\otimes\ket{\xi_{\vec{a}^2}}^{([n],P)} +\\ + \sum_{\vec{a}^2}\left(\lambda^*_{\vec{a}^2_{00}}\ket{\vec{a}^2_{00}} + \lambda^*_{\vec{a}^2_{01}}\ket{\vec{a}^2_{01}} + \lambda^*_{\vec{a}^2_{10}}\ket{\vec{a}^2_{10}} + \lambda^*_{\vec{a}^2_{11}}\ket{\vec{a}^2_{11}}\right)^{[n']}\otimes\ket{\xi'_{\vec{a}^2}}^{([n],P)},
\end{multline}
Relations~\eqref{prvamulti}-\eqref{zadnjamulti} for $j=3$ provide us another way to write the states $\ket{\xi_{\vec{a}}}$ for all $\vec{a}$ such that $a_2 = 0$:
\begin{equation}
  \ket{\xi_{\vec{a}}}^{([n],P)} = \lambda_{\vec{a}^2_{a_10}}\ket{\xi_{\vec{a}^3}}^{([n],P)} + \lambda^*_{\vec{a}^2_{a_10}}\ket{\xi'_{\vec{a}^3}}^{([n],P)}.
\end{equation}
Here we note that the choice of the local computational bases is such that for every $\vec{a}^2$ coefficients $\lambda_{\vec{a}^2_{00}}$ and $\lambda_{\vec{a}^2_{10}}$ are both different than $0$ and with different complex phases. By reproducing the argumentation from Appendix~\ref{sec:SWAP}\footnote{It is the argumentation used after eq.~\eqref{niksic}}, we have that for all $\vec{a}^2$ and $\vec{a}^3$ having the same last $n-3$ components it holds
\begin{align}
  \ket{\xi_{\vec{a}^2}}^{([n],P)} = \ket{\xi_{\vec{a}^3}}^{([n],P)}, \qquad \ket{\xi'_{\vec{a}^2}}^{([n],P)} = \ket{\xi'_{\vec{a}^3}}^{([n],P)}.
\end{align}
Taking these relations into account, the output of the SWAP isometry reduces to:
\begin{multline}
  \Phi\left(\ket{\psi}^{[n],P}\otimes_{k = 1}^n\ket{+}^{k'}\right) = \sum_{\vec{a}^3}\left(\sum_{a_1,a_2,a_3}\lambda_{a_1,a_2,a_3,a^3_4,\cdots,a^3_n}\ket{a_1,a_2,a_3,a^3_4,\cdots,a^3_n}\right)^{[n']}\otimes\ket{\xi_{\vec{a}^3}}^{([n],P)} + \\ + \sum_{\vec{a}^3}\left(\sum_{a_1,a_2,a_3}\lambda^*_{a_1,a_2,a_3,a^3_4,\cdots,a^3_n}\ket{a_1,a_2,a_3,a^3_4,\cdots,a^3_n}\right)^{[n']}\otimes\ket{\xi'_{\vec{a}^3}}^{([n],P)}.
\end{multline}
By repeating the process $n-4$ times more, after factoring in relations~\eqref{prvamulti}-\eqref{zadnjamulti} for $j=n-1$ the output of the SWAP isometry becomes:
\begin{multline}
  \Phi\left(\ket{\psi}^{[n],P}\otimes_{k = 1}^n\ket{+}^{k'}\right) = \sum_{\vec{a}^{n-1}}\left(\sum_{a_1,\cdots,a_{n-1}}\lambda_{a_1,\cdots,a_{n-1},a^{n-1}_n}\ket{a_1,\cdots,a_{n-1},a^{n-1}_n}\right)^{[n']}\otimes\ket{\xi_{\vec{a}^{n-1}}}^{([n],P)} + \\ + \sum_{\vec{a}^{n-1}}\left(\sum_{a_1,\cdots,a_{n-1}}\lambda^*_{a_1,\cdots,a_{n-1},a^{n-1}_n}\ket{a_1,\cdots,a_{n-1},a^{n-1}_n}\right)^{[n']}\otimes\ket{\xi'_{\vec{a}^{n-1}}}^{([n],P)},
\end{multline}
Note that first $n-3$ components of vector $\vec{a}^{n-1}$ are by definition equal to $0$, meaning there are two different valid vectors $\vec{a}^{n-1}$. By taking into account relations~\eqref{prvamulti}-\eqref{zadnjamulti} we reach expressions involving $\vec{a}^n$, which by definition has $0$-s for all components. Thus, by denoting $\ket{\xi}^{([n],P)} = \ket{\xi_{\vec{a}^n}}^{([n],P)}$ and $\ket{\xi'}^{([n],P)} = \ket{\xi'_{\vec{a}^n}}^{([n],P)}$ we get:
\begin{align}
\Phi\left(\ket{\psi}^{[n],P}\otimes_{k = 1}^n\ket{+}^{k'}\right) &= \left(\sum_{\vec{a}}\lambda_{\vec{a}}\ket{\vec{a}}\right)^{[n']}\otimes\ket{\xi}^{([n],P)} + \left(\sum_{\vec{a}}\lambda^*_{\vec{a}}\ket{\vec{a}}\right)^{[n']}\otimes\ket{\xi'}^{([n],P)}\\
 &= \ket{\Psi}^{[n']}\otimes\ket{\xi}^{([n],P)} + \ket{\Psi^*}^{[n']}\otimes\ket{\xi'}^{([n],P)},
\end{align}
where $\braket{\xi}{\xi'} = 0$.

\end{document}